\documentclass[aps,final,notitlepage,oneside,twocolumn,nobibnotes,nofootinbib,
superscriptaddress,noshowpacs,centertags]{revtex4-1}

\usepackage[english]{babel}
\usepackage{graphicx}
\usepackage{latexsym}
\usepackage{amssymb}
\usepackage{amsmath}

\begin{document}

\author{V.S. Berezinsky}\thanks{e-mail: berezinsky@lngs.infn.it}
\affiliation{Gran Sasso Science Institute (GSSI)
and Laboratori Nazionali del Gran Sasso, INFN,
I-67010 Assergi (AQ), Italy}
\title{Small-scale clumps of dark matter}
\author{V.I. Dokuchaev}\thanks{e-mail: dokuchaev@lngs.infn.it}
\author{Yu.N. Eroshenko}\thanks{e-mail: eroshenko@inr.ac.ru}
\affiliation{Institute for Nuclear Research,
Russian Academy of Sciences,
prosp. 60-letiya Oktyabrya 7a, 117312 Moscow, Russian Federation}

\date{\today}

\begin{abstract}Small-scale clumps of dark matter are gravitationally
bounded structures that have masses comparable to or lower
than stellar masses and consist of noninteracting or weakly
interacting dark matter particles. In this paper, the current
knowledge about the formation and evolution of such structures is reviewed, various types of spectra of primordial cosmological perturbations are considered, and various dark matter
models are discussed. Depending on the particular spectrum
type, dark matter clumps may differ considerably in their formation processes and ultimate characteristics. The role of
clumps in experiments on indirect detection of dark matter
particles via their annihilation products is discussed. A number
of astrophysical problems and phenomena that are related to
dark matter clumps are examined.
\end{abstract}

\maketitle

\tableofcontents

\bigskip

\section{Introduction}

Modern cosmology (see monographs [1-5], review [6], and
historical overview [7]) studies the most fundamental problems of the origin of our Universe by combining the physics
of elementary particles and quantum field theory, especially
as regards studying the very early epochs. The inflationary
paradigm as part of quantum cosmology should be specially
mentioned here. This paradigm offers elegant solutions to
several cosmological problems simultaneously, but the
specific model of inflation remains to be chosen.
The study of the nature of dark matter and dark energy
that are present in the Universe is another important field of
modern cosmology. The importance of these substances of an
unknown nature is dictated by their dominance, 95\% of the
mass in the modern Universe: dark energy and dark matter
(also called ``hidden mass'') respectively provide about 68.3\%
and 26.8\% of the total mass, while ordinary baryonic matter
(stars, interstellar and intergalactic gas) contributes only
about 4.9\%. The nature of fields that form dark energy and
dark matter particles remains unknown. As in the case of the
inflationary paradigm, many theoretical models have been
offered for dark energy and dark matter, but it is unclear
which of them will be viable in the future and whether there is
a true model among them that is realized in ``our Universe''.

The energy-dominating substance at present, dark energy,
was discovered by analysing the accelerating expansion of the
Universe inferred from remote type-Ia supernova observations
[8-10]. Later, its existence was confirmed by independent measurements of cosmic microwave background (CMB)
fluctuations and other effects. Some scalar fields that remain
from early epochs or the energy of quantum fluctuations of
the vacuum could provide an explanation of dark energy. In
this review, we focus on the cosmological epochs in which
dark energy effects were negligible.

The problem of dark matter, which is the main topic of
this review, has good chances to be solved in the nearest
future. Dark matter particles, incident on Earth from space,
can be registered by the existing and future detectors, or can
be produced in collider experiments (at the Large Hadron
Collider, first of all).
There is little doubt that dark matter does exist in the
Universe. Dark matter explains the shape of the rotation
curves of spiral galaxies, the dynamics of stars in elliptical
galaxies, high gas temperatures in the halos of galaxies and
galaxy clusters, and the motion of galaxies in small groups and
in clusters. The presence of dark matter reconciles the results of
primordial nucleosynthesis calculations with the observed
elemental abundances by taking CMB anisotropy measurements into account. Only with dark matter is it possible to
explain the stability of galactic discs and their spiral structure.
The distribution of dark matter can be ``mapped'' using strong
and weak gravitational lensing observations: the appearance of multiple images of galaxies and their shape distortions
[11], including in colliding galaxy clusters (the Bullet Cluster, etc.). Although the nature of dark matter is unknown, the
inflationary paradigm offers a natural model for generating
density perturbations from which dark matter halos formed
at later stages; ordinary baryonic matter cooled and streamed
toward the centers of these halos. This process led to the
formation of galaxies, stars, planets, and ultimately life.

Reviews of possible particle candidates for dark matter
and methods for searching fort hem can be found, e.g., in [1,12].
Weakly interacting massive particles (WIMPs) are most
frequently considered to be the likely dark matter candidates. They could have formed in the early Universe in a
number suitable to explain the observed dark matter.
Among these particles, the neutralino --- the lightest supersymmetric particle?is very popular [13,14]. Other candidates include sterile neutrinos [15,16], axions [17], gravitinos
[18,19], superheavy particles [20-23], and primordial black
holes (PBHs) [24, 25].

The discovery of the light Higgs boson with the mass
$m_{\rm H}\simeq 125$~GeV [26-28] has strengthened the status of the
neutralino as the dark matter particle. The existence of a light
Higgs boson, heavy Grand Unification particles, and supergravity are contradictory in principle because radiation
diagrams (loops) predict the mass of the light boson $\delta m^2_{\rm H}\sim(\alpha/\pi)\Lambda^2$, where $\alpha$ is the coupling constant and $\Lambda\sim10^{16}$~GeV
is the Grand Unification scale. Supersymmetry elegantly
solves this problem by providing compensation to radiation
diagrams, and no simpler and more elegant solution is known
in the theory. If the light Higgs boson with a mass
$m_{\rm H}< 130$~GeV had not been discovered, the existence of
supersymmetry would have been in doubt. The discovery of
the light Higgs boson with the mass $m_{\rm H}< 130$~GeV involves
supersymmetry as the simplest and most natural explanation
for its small mass. Supersymmetry is assumed to be broken
using so-called soft terms in the Lagrangian, which are
described by at least five free parameters. The fixed mass of
the Higgs boson imposes constraints on these parameters and
those of the Grand Unification, as well as on the masses of
supersymmetric scalar particles (sparticles). The supersymmetric particles and the lightest particle among them, the
neutralino, have not been discovered so far in the ATLAS
(A Tororidal LHC Apparatus) and CMS (Compact Muon
Solenoid) experiments at LHC, possibly not because these
particles are too heavy but due to the weakness of their
interaction. Future measurements with increasing fluxes of
these particles will likely uncover supersymmetric particles
(see the discussion of Higgs particles and supersymmetry after
the discovery of the Higgs boson in [29]).

The direct registration [30,31] or production of dark
matter particles in accelerator experiments can provide the
most reliable information on their nature; however, indirect
methods of searching for them (for example, using their
annihilation products and other effects) should also be
appreciated, for two reasons. First, there is no guarantee as
yet that dark matter particles will be directly detected in the
foreseeable future. Second, if dark matter particles are
directly detected successfully, many ``applied'' problems will
arise regarding the role of these particles in astrophysics, the
influence of these particles and their annihilation and decay
products on the composition of cosmic rays, the state of
interstellar matter in the past and now, and other astrophysical processes. This means that many questions that are
currently being considered in connection with the indirect
detection of dark matter particles will be treated at a new level
(with the already known dark matter particles).

From the standpoint of indirect detection, searches for
signals (first of all, photons) from dark matter particle
annihilation seem to be the most promising. If dark matter
particles can annihilate, the efficiency of this process is mainly
determined by the dark matter density. The local annihilation
rate is proportional to the square of the particle number
density, and therefore the signal from dense clumps can be
higher than from the diffusive dark matter component. For
example, dark halos of galaxies, which were formed from
primordial density perturbations several billion years ago,
can have a higher density than the average matter density in
the Universe. At the time of writing this review, the most
distant galaxy was discovered by the Hubble Space Telescope
at the redshift $z\approx11.9$ (without spectroscopic confirmation),
which means that the Galaxy was formed when the age of the
Universe was 380~mln years (after the Big Bang, or since the
end of the inflation stage), when the reionization of the
Universe had not yet completed. But even at earlier times,
protogalaxies should have been formed: the `building
blocks' of galaxies, which are similar to present-day dwarf
galaxies. The character of dark matter clustering, the mass
and formation time of these `blocks', and the efficiency of
their merging and mass increase due to the accretion of
external matter depend primarily on the density perturbation
spectrum. According to the most reliable model, perturbations were generated at the inflation stage from quantum
fluctuations. Recently, the fluctuation spectrum was measured with high accuracy by the WMAP (Wilkinson Microwave Anisotropy Probe) [32] mission and the Planck space
observatory [33]. However, the telescopes from these space
missions can measure the perturbation spectrum only on very
large scales. According to the inflationary paradigm, the
spectrum of primordial perturbations extends to much
smaller scales, which are bounded from below by the
cosmological horizon scale at the end of inflation (the mass
of dark matter on this scale is about $\sim10^{-17}$~g). This means
that already before the formation of protogalaxies, clumps of
dark matter with masses much smaller than the Solar mass
could have been formed. These clumps are the first structures
that emerged in the Universe. Small-scale clumps of dark
matter and their role in astrophysics are considered in detail
in this review. For brevity in what follows, we refer to these
small-scale clumps of dark matter simply as clumps.

We can visualize a clump as a sufficiently loose ``cloud'' of
moving dark matter particles that are bound by gravitation
forces but do not fall onto a common center because of their
angular momenta. The clump has an almost spherical shape
and is in a quasistationary state if tidal forces from other
objects or density perturbations are absent. The slow
evolution of the internal structure of clumps can be due to
weak gravitational perturbations, the binary scattering of
particles, or annihilations of particles. The slow accretion of
dark matter particles onto clumps from the surrounding
diffuse component tends mainly to increase the effective
radius of the clumps. Rapid and strong restructuring of
clumps can be due to gravitational impacts of interacting
clumps or collisions with stars. In certain cases, a clump can
capture another clump or, just the opposite, can be captured
by a larger clump. The boundary between the slow accretion
of dark matter particles and the process of gravitational
capture of small clumps (coalescence of clumps with
strongly different masses) is somewhat conventional, but
can be described by the generalized Press--Schechter
formalism [34].

We briefly discuss the terminology. In the English--language literature, dark matter clumps are usually referred
to as ``clumps'', ``DM objects'' (dark-matter objects), ``halos'',
``minihalos'', or ``microhalos''. The last three terms stress that in many respects, clumps are diminished analogs of large dark
matter halos in galaxies. Indeed, by neglecting gas-dynamic
processes with baryons occurring in galaxies, the equations of
gravitational dynamics and principal formation processes of
small-scale clumps and galactic halos are basically the same
and differ only by the characteristic scales and the form of the
perturbation spectrum at these scales. At small scales, the
spectrum flattens, leading to rapid aggregation of clumps:
the formation of large-scale clumps, with the characteristic
time of this clustering being of the order of the formation time
of the internal density profile in the clumps. By contrast,
during the hierarchical clustering on galactic scales, the
objects with density profiles already formed coalesce more
frequently.

The formation and evolution of small-scale clumps have
been studied in many papers [35-51] (see [52] for a short
review of some aspects of the clump problem as of 2009). In
some papers (see, e.g., [53]), a phenomenological approach
has been used, which postulates the presence of clumps in
some fraction of dark matter with a certain density profile and
other free parameters, but unrelated to the specific density
perturbation spectrum.

The mass spectrum of clumps is bounded by some
minimum mass $M_{\min}$ depending on the properties of dark
matter particles and their interactions with cosmic plasma in
the early Universe at the radiation-dominated stage [54]. The
formation of clumps with minimal masses occurs not due to
the coalescence of smaller-scale objects but from individual
density fluctuations without significant internal inhomogeneities (substructures). Hence, clumps with minimal masses, as
well as those emerging from the perturbation spectrum with a
high peak, can be segregated into a separate class of clumps to
which the analytic model in [55-57] can be applied.

The mass $M_{\min}$ is model dependent; the value of $M_{\min}$ for
the neutralino in the Minimal Supersymmetric Standard
Model (MSSM) can range from $10^{-11}M_{\odot}$ to $10^{-3}
M_{\odot}$ [41,42,46,58,59]. The minimal clump mass (the mass spectrum
cutoff) $M_{\min}$ is determined by collisional and collisionless
decay processes (see, e.g., [41] and the references therein).
Additionally, the mass spectrum cutoff can be affected by
acoustic waves [60] during the kinetic decoupling of dark
matter particles [61], as well as by perturbation modes with
the horizon scale [62]. Calculations [46] showed that to flatten
small-scale perturbations, friction between dark matter
particles and cosmic plasma, similar to the Silk effect, is
important. Uncertainties in estimates of the minimal clump
mass are due to uncertainties in the neutralino models
involving free parameters.
A clump wit a minimal mass $M\sim10^{-6}M_{\odot}$, which is
obtained for the most likely supersymmetry parameters for
the primordial power-law perturbation spectrum with the
exponent $n_s=0.96$ derived from the CMB data, has the
following characteristic parameters. The first dark matter
objects, which in this case are formed at the redshift $z\sim60$
(for $2\sigma$-perturbations) with the mean density
$\bar\rho=2.6\times10^{-23}$~g~cm$^{-3}$, have the virial radius
$R=8.6\times10^{-3}$~pc and the internal velocity dispersion $v=71$~cm~s$^{-1}$.

The evolution of clumps in general should be studied in
the hierarchical framework, which takes coalescence and
aggregation of clump into account. Tidal interactions play a
fundamentally important role in these processes, because
tidal forces destroy most clumps at the early stages of
hierarchical structure formation [39]. Clumps that survive
the early hierarchical stage are further destroyed by collisions
with stars in galactic bulges, halos, and discs. In this review,
we show that the collective gravitational field of the disc has a
much stronger effect than the total contribution of individual
nearby stars.

In the foreseeable future, there is hope to directly register
gamma-ray fluxes from particle annihilations in clumps. It is
virtually impossible to observe small-scale clumps (they
collectively contribute to the diffuse gamma-ray background), but sufficiently large clumps can be observed, in
principle, as individual sources: they can form bright spots in
the X-ray and gamma-ray sky [63-65]. The contribution of
clumps to the annihilation signal is characterized by the
enhancement coefficient, or the boost factor, which is by
definition equal to the ratio of the signal with account for
dark matter clumping to that from diffusively distributed
dark matter.

Theoretical studies of dark matter halo formation
[55-57,76] predict power-law internal density profiles
$\rho_{\rm int}(r)\propto r^{-\beta}$
with $\beta\simeq1.8-2$. The density profiles obtained in numerical
simulations (Navarro-Frenk-White (NFW) profiles and
Moore~et~al's profiles [77]) have a similar divergent central
density with $\beta=1$. A result close to the isothermal power-law
profile with $\beta\simeq2$ was obtained in numerical simulations of
small-scale clumps [77]. In [51], it was found that if a clump
results from hierarchical clustering, then its density profile
coincides with the NFW one. If a clump is formed from an
isolated density perturbation (including minimal-mass
clumps), its density profile has a central core with $\beta\simeq1.4$.
This means that the singular profiles that have not experienced collisions have central profiles closer to those of Moore
et al and Gurevich--Zybin [55-57].

The increase in the density $\rho_{\rm int}(r)$ in the clump center as
$r\to0$ is limited by the total (integral) clump mass and
especially by the finite value of the annihilation signal:
\begin{equation}
Q_{\rm ann}(r)\propto\int dr\,r^2\rho_{\rm int}^2(r).
 \label{Qann}
\end{equation}
For density distributions with $\beta>1.5$, the presence of the
central core is physically necessary, i.e., the core must exist in
a wide class ofmodels. Most frequently, the core is postulated
in the form rint
$\rho_{\rm int}(r)=const$ for $r<r_c$. Different physical
mechanisms can be responsible for the formation of the
central part (core) of clumps during their formation and
evolution. It cannot be ruled out that these mechanisms are
different for clumps with different masses formed under
different conditions. The structure of the clump core is
possibly determined by small-scale density and velocity
inhomogeneities inside the forming clump; this problem was
studied for galaxies in the framework of the entropy theory
[78,79]. A theoretical estimate of the clump core radius can be
obtained using the energy criterion: $x_c\equiv R_c/R\simeq\delta_{\rm eq}^3$, where $\delta_{\rm eq}$ is the density perturbation amplitude at the beginning of
the matter-dominated cosmological stage [57]. In application
to clumps with a minimal mass $\sim10^{-6}M_\odot$ formed from
$2\sigma$-peaks ($\delta_{\rm eq}\simeq0.017$) in the spectrum, this estimate yields
$x_c\simeq1.8\times10^{-5}$. It cannot be ruled out either that the core
size or, at least, the characteristic scale of the density profile
change $\beta$ is also determined by tidal forces [39]. There is an
alternative core formation mechanism in models with
``metacold dark matter'', in which the core appears due to the
late decay of dark matter particles into light nonrelativistic
particles with low phase density [80,81].

The low spatial resolution of numerical simulations
currently prevents the determination of the core radius. The
only example pointing to the presence of a core with a radius
$x_c\simeq10^{-2}$ was obtained in calculations in [77]. However, in
the simulations in [82], the core was not resolved down to the
relative radius $x_c\simeq10^{-3}$, to which the power-law density
profile with $\beta=1.5$ extends. New, dedicated high-resolution
modeling is required to solve this problem.

In [40], the conclusion was made that almost all small-scale clumps in the Galaxy are destroyed by tidal interactions
with stars and are transformed into ``ministreams'' of dark
matter. The properties of ministreams can be important for
the direct detection of dark matter particles, because dark
matter particles in the streams move anisotropically in
different discrete directions. However, as shown in [83],
cores of clumps (or clump remnants) generally survive in
tidal interactions with galactic stars. Although the external
layers of clumps are ``stripped off'' and produce dark matter
ministreams, their cores are shielded by the adiabatic
invariant conservation and continue to produce the annihilation signal. Most of the annihilation signal power is generated
in the central parts of the clumps, and therefore the total
annihilation flux is weakly affected by the tidal destruction
process [37,43]. This statement is critically dependent on the
unknown core radius: cores with smaller sizes are more stable,
because dark matter particles there have higher orbital
frequencies and, accordingly, higher values of the adiabatic
parameter. The clump remnants form a low-mass ``tail'' of the
standard mass function for masses below $M_{\rm min}$.

In addition to the ``standard'' clump formation scenario,
hypothetical models have been proposed in which super-dense clumps are formed at the radiation-dominated stage
from entropy [84-87] and adiabatic [88] density perturbations, taking restrictions from primordial black holes into
account. Entropy perturbations, for example, can be produced by loops of cosmic strings. Superdense clumps are
interesting from the standpoint of annihilation of superheavy
dark matter particles, because their small annihilation cross
section can be compensated by the high density of the clumps.
Therefore, ultra-high energy annihilation signals become
possible. In this case, new effects can appear that determine
the central density core formation in superdense clumps, such
as Fermi degeneration of superheavy dark matter particles in
the cores of such clumps. The superdense clumps can be
registered by the tidal effects they produce on gravitational
wave detectors. In this review, we consider the formation of
superdense clumps in much detail, because this problem has
not been analysed sufficiently in the literature.

%%%***************************************************************

\section{Small-scale spectrum of density perturbations}

The presence of matter inhomogeneities at small scales in
early epochs is a necessary condition for the formation of
small-scale clumps. Galaxies and other structures form from
density perturbations $\delta(\vec x,t)=\delta\rho/\bar\rho=[\rho(\vec x,t)-\bar\rho]/\bar\rho$,
which generally can be adiabatic perturbations, entropy
perturbations, or a mixture of both. Adiabatic perturbations
are also referred to as curvature perturbations because the
local curvature in such a perturbation differs from the mean
curvature of the Universe. Entropy perturbations represent
another possible type of perturbation. These include dark
matter perturbations on top of a homogeneous radiation
background. The effect of entropy perturbations on the space
curvature is typically insignificant, and they are usually called
isocurvature perturbations. According to CMB observations,
the statistical distribution of adiabatic perturbations is
Gaussian with high accuracy, and the contribution of
entropy perturbations to the total perturbation amplitude, if
present, does not exceed a few percent.

We briefly discuss the perturbation statistics. The Fourier
transform
\begin{equation}
\delta_{\vec k}=\int\delta(\vec r)e^{i\vec k\vec x}\,d^3x \label{delfur}
\end{equation}
is characterized by a power spectrum $P(k)$,
\begin{equation}
\langle\delta^*_{\vec k}\delta_{\vec k'} \rangle=(2\pi)^3P(k)\delta_{\rm D}^{(3)}(\vec k-\vec
k'), \label{pdef}
\end{equation}
where $*$ denotes complex conjugation, $k$ is the wave vector,
$\delta_{\rm D}^{(3)}(\vec k-\vec k')$ is Dirac's delta function, and the angular
brackets mean ensemble averaging, i.e., averaging over a
large number of spatial volumes. The power spectrum
$P(t,k)$ at an arbitrary time instant $t$ is related to the initial
power spectrum $P_p(k)$ (at superhorizon scales) as $P(t,k)=P_p(k)T^2(k)D^2(t)$, where $T(k)$ is the transfer function and
$D(t)$ is the linear growth factor (see, e.g., [89, 90]). For
small-scale modes, we have $T(k)\simeq(k_{\rm eq}/k)^2$ for $k_{\rm
eq}/k\ll1$
with the logarithmic accuracy. The rms density perturbation $\sigma(R)=\langle\delta^2\rangle$ on a scale $R$ is expressed in terms of $P(k)$
as
\begin{equation}
 \sigma(R)=\frac{1}{2\pi^2}\int\limits_0^\infty k^2\,dk\,P(k)W^2(k,R),
 \label{sig0}
\end{equation}
where $W(k,R)$ is the Fourier transform of the window
function [89].

The statistical properties of forming structures are
considered using two basic approaches: (1) by counting the
number of peaks with certain characteristics [89]; (2) in the
framework of the Press--Schechter formalism and its generalizations (see [34, 91, 92]). The second approach deals with the
fraction of the dark matter mass contained in objects with a
given mass scale. A generalization of these two approaches,
and in some sense their synthesis, the so-called ``peak-patch
picture'', was developed in [93].

%%%

\subsection{Generation of adiabatic perturbations at the inflation stage}

In the inflation theory, curvature perturbations are generated
from quantum fluctuations of a scalar field. The amplitude of
the latter, $|\delta\phi|=H(\phi)/2\pi$, is connected with the Hubble
parameter $H(\phi)=\dot a/a$ during the inflation stage [2], where
$a(t)$ is the scale factor of the Universe. The density perturbations at the horizon crossing time are given by
\begin{equation}
\delta_{\mathrm{\rm H}}\sim M_{\mathrm{Pl}}^{-3}V^{3/2}/(dV/d\phi)\label{delinfl},
\end{equation}
where $M_{\rm Pl}$ is the Planck mass and $V(\phi)$ is the potential of the
scalar field (inflaton) that is responsible for the inflation.
The simplest inflation models give a nearly flat perturbation power spectrum $P(k)\equiv\delta^2_k\propto k^{n_s}$ and $n_s\simeq1$. However,
some tilt of the spectrum is usually predicted, which is
fundamentally important for small-scale clump formation,
because even a small tilt can significantly change the
perturbation amplitudes on low mass scales. The difference
between $n_s$ and unity is expressed in terms of the slow-roll
inflation parameters $\varepsilon=(V'/V)^2/(16\pi G)$ and $\eta=(V''/V)/(8\pi G)$, where $G$ is the gravitational constant, as
follows:
\begin{equation}
n_s=1-6\varepsilon+2\eta. \label{nsvarepseta}
\end{equation}
In the $R^2$-model by Starobinsky, $n_s\simeq1-2/N$, where $N$ is the
number of e-foldings (increases in the scale factor by $e$ times)
since the beginning of the generation of these perturbations
until the end of inflation. The value of $n_s$ in the inflation
model has not been precisely fixed yet; as well as the common
normalization constant, $n_s$ is determined from the CMB
anisotropy measurements and the inhomogeneities of matter
distribution in the Universe. Observations thus constrain
possible inflation parameters [94].

Immediately after the end of inflation, the perturbations (on the scales considered here) have a size greatly
exceeding the cosmological horizon, but as the horizon
increases $\sim ct$, they enter under the horizon at some
instant of time. To investigate perturbations with scales
larger than the horizon size, it is necessary to solve the
linearized Einstein equations [2]. Using them, it is possible to
match the solutions (and the initial conditions) on subhorizon
scales. The perturbations that are deep inside the horizon can
be studied using the Newtonian equations (taking the
potential of a homogeneous relativistic background into
account), but setting the initial conditions does require
solving the linearized Einstein equations. At the radiation-dominated stage, small adiabatic perturbations (those on the
linear stage), $\delta\ll1$, grow slowly on subhorizon scales,
$\delta_k\propto\ln(t/t_i)+const$, and after the transition to the dust-like stage, $t>t_{\rm eq}$, they start increasing rapidly as $\delta_k\propto t^{2/3}$.

%%%

\subsection{Normalization of the perturbation spectrum
from observational data}

The normalization constant in perturbation spectra is
frequently chosen from the requirement that the relative
mass fluctuations on the scale $8h^{-1}$~Mpc correspond to the
value $\sigma_8\simeq0.82$, which is directly obtained from galaxy and
galaxy cluster counts. An alternative method of spectrum
normalization, which is nonetheless consistent with the
above, is based on referencing to CMB fluctuations, because
they are closely related to dark matter density fluctuations.

According to the CMB anisotropy data, it is convenient to
normalize the curvature perturbation spectrum [2] as
\begin{equation}
{\mathcal P}_{\mathcal R}=A_{\mathcal R}\left(\frac{k}{k_*}\right)^{n_s-1}, \label{prspectrum}
\end{equation}
where $k_*/a_0=0.002$~Mpc$^{-1}$, $a_0$ is the present-day value
of the scale factor, $A_{\mathcal R}=(2.46\pm0.09)\times10^{-9}$, and $n_s=0.9608\pm0.0080$ according to the WMAP data [32] and
$n_s=0.9608\pm0.0054$ according to the Planck data [33].
Here, the typical perturbation amplitude is $\Delta_{\mathcal R}\simeq5\times10^{-5}$.
With only low multipoles used in the CMB power spectrum,
the WMAP data [32] correspond to the constant $n_s$ with an
accuracy $dn_s/d\ln k=-0.019\pm 0.025$, but taking higher
multipoles into account shows a (presently statistically
insignificant) tendency of decreasing $n_s$ toward small scales,
$dn_s/d\ln k=-0.022^{+0.012}_{-0.011}$. The Planck data [94] also point to a
statistically insignificant (at a level of $1.5\sigma$) decrease in $n_s$,
$dn_s/d\ln k=-0.0134\pm 0.0090$. If such a decrease actually
takes place, small-scale clumps form with a smaller efficiency
than for the constant $n_s\approx0.96$.

The rms amplitude of density perturbations normalized to
the CMB data on the horizon scale at the radiation-
dominated stage has the form [95]
\begin{equation}
\sigma_H(M)\simeq9.5\times10^{-5}\left(\frac{M}{10^{56}
\mbox{~g}}\right)^{\frac{1-n_s}{4}}.\label{normsigmah}
\end{equation}
For $n_s<1$, the mean amplitude of perturbations decreases
with decreasing the mass. Nevertheless, clumps can be formed
before galaxies due to the transfer function $T(k)$, which for
not too small $n_s$ leads to an rms perturbation increase with
decreasing the mass. For estimates, it is useful to write the
mean value of perturbations on scales $M\leq M_\odot$ at the time of
transition to the dust-like stage $t_{\rm eq}$:
\begin{eqnarray}
\sigma_{\rm eq}(M)&\simeq &8.2\times 10^{3.7(n_s-1)-3}
 \left(\frac{M}{M_{\odot}}\right)^{\frac{1-n_s}{6}}
 \nonumber
\\
&\times & \left[1-0.06\log\left(\frac{M}{M_{\odot}}\right)\right]^{\frac{3}{2}}. \label{A5}
\end{eqnarray}
It can be seen that for $n_s\approx1$, the logarithmic term in the
transfer function is very important.

It should be borne in mind that the use of the spectrum
normalized to the CMB data assumes a huge extrapolation,
by more than 15 orders of magnitude. Such an extrapolation
is justified for inflation models that give power-law spectra in
a wide range of scales. However, strictly speaking, the
predictions of inflation models for small scales have not
been confirmed by observations so far. Therefore, to a large
extent, the shape of the spectrum remains a ``free parameter''.
If the existence of small-scale clumps is confirmed (for
example, by observations of dark matter particle annihilation), the clump properties can allow determining the shape of
the perturbation spectrum on small scales and studying the
processes of perturbation generation at the inflation stage.
For example, it will be possible to fix parameters in the
Lagrangians of the specific field models.

%%%

\subsection{Perturbation spectra with peaks}

Although the perturbation spectrum is sufficiently well
known on the scales of galaxies and galaxy clusters, a
nonstandard form of the spectrum cannot be ruled out on
small scales, deviating from the simple power-law dependence
and possibly containing local maxima or narrow peaks. As we
can see from (5), an increase in $\delta_{\rm H}$ is possible if the inflation
potential $V(\phi)$ has a flat part [24, 96], i.e., if $dV(\phi)/d\phi\to0$ at
some value of the scalar field $\phi$. The increase in $\delta_{\rm H}$ can occur
inside a wide region or in several regions. The clumps are
formed in a broad mass range if the primordial perturbation
spectrum has a power-law shape or there is a broad maximum
in the spectrum. On the other hand, if the flat part of the
potential is local, the clumps are formed near the corresponding mass. We note that outside these regions, the spectrum
can remain close to the Harrison-Zeldovich one (Fig.~1) and
lead to the formation of galaxies and galaxy clusters in the
standard way.

\begin{figure}[t]
\begin{center}
\includegraphics[angle=0,width=0.45\textwidth]{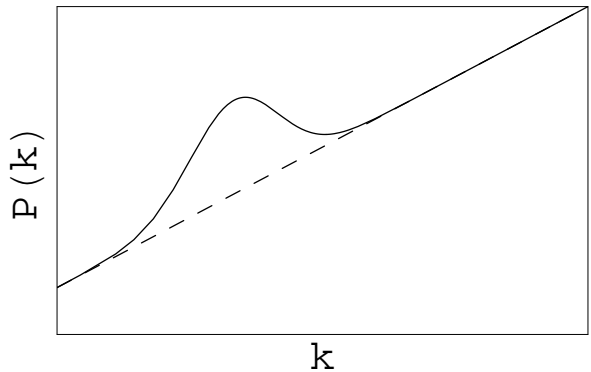}
\end{center}
\caption{Schematic view of a local maximum in the primordial density
perturbation spectrum.} \label{grpeak}
\end{figure}

A peak at the minimal scale produces clumps with the
highest dark matter density in the Universe. The discovery of
such clumps could provide invaluable information on the
inflation potential. The presence of peaks can also lead to a
more effective formation of PBHs with narrow mass distribution [24, 97].
Inflation models with several scalar peaks represent
another class of models producing spectra with peaks [98,
99]. Theoretical models of non-standard spectra were discussed, in particular, in [100-102].
An interesting mechanism in which the generation and
amplification of perturbations is possible, including the
appearance of spectra with peaks, is provided by cosmological phase transitions, for example, the quantum chromodynamics (QCD) transition at a temperature $\sim150-200$~MeV
(the time instant $\sim10^{-5}$~s) [35, 37]. During the QCD
transition, the equation of state of matter becomes softer
(the ratio $w=p/\rho$ decreases), and perturbations grow more
efficiently. However, in the model with a standard
($m\sim100$~GeV) neutralino, the peak corresponding to the
QCD transition is somewhat flattened in the subsequent
evolution due to free-streaming and the corresponding
spectral cutoff [62].

%%%%

\subsection{Entropy perturbations}

Entropy perturbations can be generated, in particular, in
axion dark matter and lead to the formation of axion miniclusters [84, 85, 104].
Primordial black holes and various topological defects
can also generate entropy perturbations. The nature of
these perturbations can be illustrated as follows. We
consider a point-like object with a mass $M_c$ on the
background of homogeneously distributed dark matter.
We encircle this object by a sphere containing dark matter
with a mass $M$. Then the effective value of perturbations
within the volume of the sphere is $\delta = M_c/M$, where $\delta$ is
obviously dependent on the sphere radius and decreases as
the radius increases. For the motion of test particles at the
spherical boundary, only the mean value of $\delta$ is important,
and the specific density distribution of dark matter inside
the sphere does not matter, according to the Kirchhoff
theorem, if the density distribution is spherically symmetric.
Perturbations produced by a seed mass evolve according to
the `secondary accretion' mechanism [76, 105, 106]. For
example, PBHs or their clusters or cosmic string loops can
serve as seed masses.

Strongly nonspherical perturbations, which have an
entropy nature, can be generated by infinite cosmic strings
or textures [107]. From the standpoint of the evolution of
perturbations, the difference between entropy perturbations
and adiabatic perturbations is in the absence of relatively
large initial peculiar velocities of dark matter particles in the
entropy perturbations.

%%%

\subsection{Constraints from primordial black holes}

The adiabatic perturbation spectrum is bounded by the effect
of PBH formation in the early Universe, because, for a
sufficiently large number of perturbations, an inadmissibly
large number of PBHs would be created, in contradiction to
observational constraints.

The possibility of a collapse of a small mass in general
relativity (GR) and the creation of BPHs was first discovered
in [108], and later PBH formation at the radiation-dominated
stage was studied in many papers (see, e.g., [109-112]). In fact,
photons and a mixture of relativistic particles collapse into
PBHs. The formation of PBHs is also possible at the early
dust-like stages [113]. PBH studies became especially relevant
after the Hawking quantum evaporation was suggested,
which can lead to high-energy gamma-ray emission at the
final stages of evaporation. The observational absence of such
sources strongly constrains the density perturbation spectrum
at a mass scale corresponding to evaporating PBHs [114].

The threshold perturbation value $\delta_{\rm th}=1/3$ leading to the
formation of a PBH was found analytically in [110], and
numerical simulations [111, 112] approximately confirmed
this value, although they indicate a more complicated
character of the gravitational collapse than was assumed in
the simple model in [110]. Subsequent numerical experiments
discovered the phenomenon of critical gravitational collapse,
when the mass of the forming PBH is $M_{\rm BH}=AM_{\rm H}(\delta_{\rm H}- \delta_{\rm th})^{\gamma}$
[115, 116], where $A\sim3$, $\gamma\simeq0.36$, $\delta_{\rm th}\simeq(0.65\div0.7)$, and $M_{\rm H}$ is
the mass of matter inside the horizon. The mass MBH can be
significantly smaller than $M_{\rm H}$; however, the PBH mass
distribution is concentrated near $M_{\rm BH}\sim M_{\rm H}$ [117].

We set $\Delta_{\rm H}\equiv\langle\delta_{\rm H}^2\rangle^{1/2}$. The fraction of the mass of radiation
transformed into a PBH at the time instant $t_{\rm H}$ is expressed
as [110]
\begin{equation}
\beta= \int\limits_{\delta_{\rm th}}^{1} \frac{\displaystyle
d\delta_{\mathrm{H}}}{\displaystyle\sqrt{2\pi}\Delta_ {\mathrm{H}}}
\exp(-\frac{\displaystyle\delta_{\mathrm{H}}^2}{\displaystyle2\Delta_{\mathrm {H}}^2}) \simeq
\frac{\Delta_{\mathrm{H}}}{\delta_{\rm th}\sqrt{2\pi}}\exp(-\frac{\delta_ {\rm
th}^2}{2\Delta_{\mathrm{H}}^2}), \label{bet2}
\end{equation}
and the present-day PBH density parameter is $\Omega_{\mathrm{BH}}\simeq\beta
a(t_{\mathrm{\rm eq}})/a(t_{\mathrm{H}})$. The above relations allow imposing bounds on $\Delta_{\rm H}$ from observational constraints on PBHs [118].
Constraints on the perturbation spectrum, in turn, impose
bounds on the parameters of the clumps being formed (see
Section~3.3).

%%%%

\section{Clump formation scenarios and models}

There are two basic methods to study the process of
formation and clustering and the internal structure of
clumps: numerical modelling and the use of approximate
analytic models. Analytic approaches, in historical
sequence, were developed in [55,56,76,106,119], and the
fundamentally important results in numerical modelling
were obtained in [105, 120-123], where the formation of
galaxies and large-scale structures was mainly studied.
Numerical methods are permanently improving by increasing the spatial resolution and the time-scale range. It is
expected that in the nearest future, numerical modelling will
improve our knowledge of processes involving clumps.
However, analytic calculations, which remain (and will
remain in the future) the necessary link in formulating
numerical problems, are important for a qualitative understanding of both physical processes and the results of
numerical simulations.
In this section, we discuss some analytic approaches, in
the order of increasing complexity and detail. For a more
detailed discussion of the theory of structure formation in the
Universe, without an accent on the small-scale clumps of
interest here, we can recommend concise introductory courses
[124,125].

%%%

\subsection{Spherical model of the evolution of perturbations}

Let there be an isolated positive density perturbation $\delta(\vec r)$, a
protohalo. The clump is detached from the cosmological
expansion at the time of transition of fluctuations to the
nonlinear stage, growing to $\delta\geq1$. Starting from that instant,
the clump is compressed by its own gravity with a small
correction due to tidal forces from nearby perturbations. In
the first approximation, the protohalo can be considered a
spherically symmetric object [34,91]. Because of a fairly
simple evolution, this approximation proves to be very
useful, it suffices to understand basic processes and to
obtain quantitative estimates.

We first write the very general equation for the evolution
of a spherical layer with a radius $r$ on scales much smaller than
the horizon scale, $r\ll ct$. Let the mass of dark matter inside
this layer be $M$. The contribution of the pressure of relativistic
density components to the energy-momentum tensor can be
taken into account in the Newtonian equations by substituting $\rho\to\rho+3p/c^2$ [126,127]. Then the evolution of the layer
radius is described by the equation
\begin{equation}
\frac{d^2r}{dt^2}=-\frac{G(M_{\rm BH}+M)}{r^2}-\frac{8\pi G\rho_r r}{3}+\frac{8\pi G\rho_{\Lambda}
r}{3} \label{d2rdt1}
\end{equation}
Equation (11) takes into account that $\varepsilon_r+3p_r=2\varepsilon_r$ for
radiation and $\varepsilon_{\Lambda}+3p_{\Lambda}=-2\varepsilon_{\Lambda}$ for the cosmological constant, where $\varepsilon_r$ and $p_r$ are the energy density and pressure of
radiation, and $\varepsilon_{\Lambda}$ and $p_{\Lambda}$ are the energy density and pressure
of the cosmological constant. For generality, the possibility of
the presence of a seed mass $M_{\rm BH}$, for example, the mass of a
central black hole, is taken into account.

We first consider the scenario of clump formation at the
matter-dominated stage. The small-scale perturbations we
consider are formed at $z\gg1$ ($t\ll t_\Lambda$), when the dark energy
contribution to the total density of the Universe can be
neglected. The initial conditions for (11) are set from the
linear theory of perturbation growth, according to which
dark matter perturbations grow as [2]
\begin{equation}
\delta(k,z)\simeq\frac{27}{2}\Phi_i(k)\frac{1+z_{\rm eq}}{1+z}\ln(0,2k\eta_{\rm eq}),
\label{deltalog}
\end{equation}
where $\Phi_i$ is the gravitational potential at the time the
perturbation scale was much larger than the cosmological
horizon. The distribution of $\Phi_i$ can be obtained from (7). The
power spectrum of the potential is described by a formula
similar to Eqn~(7), but with $A_\Phi=(4/9)A_{\mathcal R}$ [2]. Although it is
possible to use $\delta\simeq\delta_{\rm eq}(t/t_{\rm eq})^{2/3}$ for estimates, expression (12)
gives a more precise result because the transition to the dust-like stage with the equation of state $p=0$ takes some time.

At $t\gg t_{\rm eq}$, the nonlinear stage of the evolution of
perturbations is described by Eqn (11) without the last two
terms in the right-hand side and with $M_{\rm BH}=0$:
\begin{equation}
\frac{d^2r}{dt^2}=-\frac{GM}{r^2}. \label{d2rdt2}
\end{equation}

\begin{figure}[t]
\begin{center}
\includegraphics[angle=0,width=0.45\textwidth]{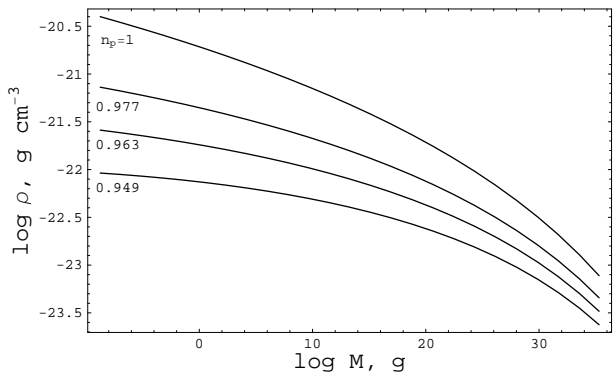}
\end{center}
\caption{The mean clump density $\rho$ as a function of its mass $M$ for the
spectral power-law exponents $n_s$ (marking the curves), calculated in the
framework of the simple spherical model at the dust-like stage.} \label{gr1}
\end{figure}

The initial expansion rate of the layer is smaller than that of
the Hubble expansion due to the presence of peculiar
velocities directed toward the center of the perturbation.
These velocities can be specified using linear theory [128],
according to which
\begin{equation}
\vec v=\frac{Ha}{4\pi}\nabla_x\int\frac{d^3x'\delta(\vec x')}{|\vec x'-\vec x|}. \label{vnabla}
\end{equation}
in a flat universe. Then the initial expansion rate of a spherical
layer at some instant of time $t_i\gg t_{\rm
eq}$ (but as long as  $\delta(t_i)\ll1$)
takes the form $dr/dt(t_i)=H(t_i)(1-\delta_i/3)$. The solution of
Eqn~(13) with the initial conditions as above can be
parametrically expressed as [128]
\begin{equation}
r=r_s\cos^2p,~~~ p+\frac{1}{2}\sin(2p) =\frac{2}{3} \left(\frac{5\delta_i}{3}
\right)^{3/2}\frac{t-t_s}{t_i}\label{param}
\end{equation}
where $t_s$
is the time of the maximum expansion, $p$ is a parameter, and
\begin{equation}
t_s=t_i\left[ 1+\frac{3\pi}{4} \left(\frac{5\delta_i}{3} \right)^{-3/2} \right], \label{tssph}
\end{equation}
and
\begin{equation}
r_s=r_i\left(\frac{3}{5\delta_i}\right), \label{rssph}
\end{equation}
is the radius of the layer at the time of maximum expansion. If
we assume that the Hubble flow is not perturbed (this is
possible for entropy perturbations at an instant close to $t_{\rm eq}$),
i.e., the correction $\delta_i/3$ is absent in $dr/dt(t_i)$, then the
substitution $5\delta_i/3\to \delta_i$ should be made in Eqns (15)-(17)
[76, p. 40].

At the instant $t_c=2t_s$, the collapse of the layer to a point
should formally occur, but in the simple model under
consideration we assume that prior to this time, when the
layer is compressed from the radius $r=r_s$ to $R\equiv r_s/2$, the
spherical layer is virialized and the compression stops. The
virialization, which represents the mixing and relaxation of
different elements inside the layer, occurs due to radial
oscillations, the presence of inhomogeneities in the compressing layer, and nonradial motions of matter. These processes
take some time; therefore, the virialization is mainly completed by the time $t\simeq 3t_s$ [125].

If perturbations are also present on larger scales, a further
increase in the protohalo mass occurs until its coalescence
with other objects due to the segregation and virialization of
new spherical layers. As a rule, large-scale perturbations
decrease with time, which leads to collapses at later times.
When the zero-energy boundary is reached, where $\delta_i=0$, the
object growth stops. After the virialization, the system
becomes highly nonlinear, $\delta\gg1$, but if the initial density
perturbation can be formally extrapolated to the instant $t_c$
using a linear law, we obtain the perturbation amplitude
$\delta_c=\delta_i(t_c/t_i)^{2/3}=3(12\pi)^{2/3}/20\simeq1,686$. The criterion of
object formation with $\delta(t_c)=\delta_c$ is used in the Press--Schechter formalism; the collapse time $t_c$ can be found from
here. At this time, the density of the virialized object is
$\kappa=18\pi^2\simeq178$ times larger than the mean density of the
universe, $\bar\rho_{\rm int}=\kappa\bar\rho(z_{c})$, and its radius is
\begin{equation}
R=\left(\frac{3M}{4\pi\bar\rho_{\rm int}}\right)^{1/3}. \label{rad1}
\end{equation}
The perturbation amplitude is expressed in units of the rms
perturbations $\nu=\delta/ \sigma$; typical objects are formed from
perturbations with $\nu\sim1-2$. Frequently, the virial radius is
assumed to equal $r_{200}$, inside which the dark matter density is
200 times larger than the mean density of the universe.

Figure~2 shows the mean clump density $\rho$ obtained using
the above formalism as a function of the clump mass $M$ for
different primordial perturbation spectral exponents $n_s$; we
formally assume an arbitrary clump mass $M$.

%%%%

\subsection{Spherical model for entropy perturbations}

The evolution of entropy perturbations, including at the
radiation-dominated (RD) stage, is studied in detail in [85].
The main difference between entropy perturbations and
adiabatic ones appears at the radiation-dominated stage of
cosmological evolution. At this stage, adiabatic perturbations
inside the horizon, as long as they are small, increase
logarithmically, while small entropy perturbations are
``frozen'' and, according to the Meszaros solution, increase
only by a factor of $5/2$ by the time $t_{\rm eq}$. However, sufficiently
large entropy perturbations can evolve and lead to the
formation of clumps already at the RD stage.

Equation (11) for a spherical layer can be conveniently
rewritten in the form [85, 129]
\begin{equation}
 y(y+1)\frac{d^2b}{dy^2} + \left[1 + \frac{3}{2}y\right]
 \frac{db}{dy} + \frac{1}{2} \left[ \frac{1+\delta_i}{b^2}-b \right]
 = 0 \,,
 \label{bigeq}
\end{equation}
where $y=a(\eta)/a_{\rm eq}$, $d\eta=dt/a(t)$ is the conformal time,
$\eta_{\rm eq}$, and $\delta_i = \delta\rho_{\rm DM}/\rho_{\rm DM}$ is the initial relative dark matter density perturbation. Here, the perturbed region
radius is parameterized as
\begin{equation}
 r=a(\eta)b(\eta)\xi \,. \label{abxi}
\end{equation}
where $\xi$ is the initial comoving coordinate and the function
$b(\eta)$ takes the expansion deceleration into account.

The cosmological expansion of the clumps stops at the
time when $dr/dt=0$, which is equivalent to the condition
$db/dy=-b/y$ [85]. We let $b_{\rm max}$ and $y_{\rm max}$ denote $b$ and $y$ at the
time of the expansion stopping. At that time, the cold dark
matter density in the clump is
\begin{equation}
\rho_{\mathrm{max}}=\rho_{\mathrm{eq}}y_{\mathrm{max}}^{- 3}b_{\mathrm{max}}^{-3}
\end{equation}
and the clump radius is
\begin{equation}
R_{\mathrm{max}}=\left(\frac{3M_x}{4\pi\rho_{\mathrm{max}}} \right)^{1/3}.\label{rmax}
\end{equation}
At a later time, the object is virialized when compressed by a
factor of two in radius.

For entropy perturbations, the initial conditions are
$\delta_i=\delta\rho_{\rm
DM}/\rho_{\rm DM}$ and $db/dt=0$ [85]. According to numerical simulations [85], the clump density can be approximated with
good accuracy as
\begin{equation}
 \rho\simeq140\delta_i^3(\delta_i+1)\rho_{\rm eq}.
 \label{rho140}
\end{equation}
For example, $\delta_i\simeq 1\div10^4$ for axion dark matter, and
masses of emerging ``axion miniclusters'' [84] fall within the
range $\sim(10^{-13}\div0,1)M_{\odot}$. Possible observational appearances of the axion miniclusters in the galactic halo were
discussed in [85, 104].

%%%

\subsection{Spherical model for adiabatic perturbations}

at the radiation-dominated stage
In order to consider the evolution of adiabatic perturbations
at the RD stage in a similar way, the initial conditions for (19)
should be chosen in accordance with the linear solution for
$\delta\ll1$. On subhorizon scales [37],
\begin{equation}
 \delta=\frac{3A_{\rm in}}{2}\left[
 \ln\left(\frac{k\eta}{\sqrt{3}}\right)
+\gamma_E-\frac{1}{2} \right] \,,
 \label{dgame}
\end{equation}
where $\gamma_E-1/2\approx0,077$, $A_{\rm
in}=\delta_{\rm H}/\phi$, $\phi\simeq0,817$, and $\delta_{\rm H}$ is the
radiation density perturbation at the horizon crossing time.
For $k\eta\gg1$ and $y\ll1$, we have the relation [88]
\begin{equation}
k\eta=\frac{\pi}{2^{2/3}}\left(\frac{3}{2\pi}\right)^{1/6}
 \frac{yc}{G^{1/2}M^{1/3}\rho_{\rm eq}^{1/6}} \,.
 \label{xyconn}
\end{equation}

The value of $b$ in (20) can be expressed through $\delta$ as [129]
\begin{equation}
b=(1+\delta)^{-1/3}.\label{delb}
\end{equation}
This relation means the transition from the Euler description,
Eqn~(24), to the Lagrange description, Eqn (19), of the
perturbation evolution. For adiabatic perturbations, we can
set $\delta_i=0$, but the initial velocity $db/dt$ is nonzero and is
determined by the linear stage of evolution.

It is convenient to connect analytic solution (24), obtained
in the linear theory, to a numerical solution of nonlinear
equation (19) at the time corresponding to the ``transitional''
perturbation amplitude $\delta=0.2$ (see [88]). At that time, we
determine the initial velocity of the spherical layer from
\begin{equation}
 \frac{db}{dy}=-\frac{\delta_{\rm H}b^4}{2y\phi} \,.
\end{equation}
The example of the evolution of $\delta=b^{-3}-1$ is shown in
Fig.~3.

\begin{figure}[t]
\begin{center}
\includegraphics[angle=0,width=0.45\textwidth]{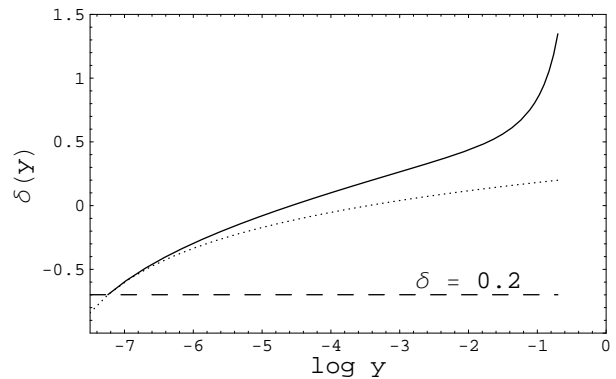}
\end{center}
\caption{Example of the evolution of a cold dark matter density
perturbation $\delta$ for $\delta_{\mathrm{H}}=0.04$, $M=0.1M_{\odot}$, $\Omega_{m}=0.3$. Up to the point
($y_{\mathrm{i}}=4\cdot10^{-6}$, $\delta_i=0.2$), the solid curve is obtained using formula (24),
and for $y>y_i$, from a numerical solution of Eqn~(19). The clump
decouples from the cosmological expansion at the radiation-dominated
stage at $y=a/a_{\rm eq}\simeq0.5$. The dashed curve shows the evolution of $\delta$
according to the linear theory, Eqn~(24).\label{gr1un}}
\end{figure}

The linear version, for $\delta\ll1$, of the formalism described
above (including the solution of the linear limit of Eqn~(19) at
the RD stage) is typically used to calculate the transition
function and perturbation spectrum at the dust-like stage [90].
However, the clumps can already be formed at the RD stage
from sufficiently large perturbations.

The characteristic clump densities r (cross section of the
surface in Fig.~4) are shown in Fig.~5. We see that at small $\delta_{\rm H}$,
the curves converge to $\rho\sim\rho_{\rm eq}\sim10^{-19}$~g~cm$^{-3}$. This corresponds to solution (15), according to which the perturbation
evolution at the dust-like stage does not depend on mass and
is determined by the initial perturbation amplitude (at
$t=t_{\rm eq}$).

\begin{figure}[t]
\begin{center}
\includegraphics[angle=0,width=0.45\textwidth]{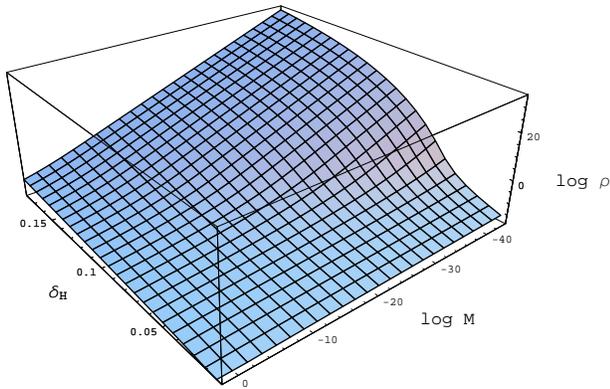}
\end{center}
\caption{Mean clump density $\rho$ [g~cm$^{-3}$] as a function of the clump mass $M$ (in units of $M_{\odot}$) and the radiation density perturbation dh on the horizon scale.} \label{gr5}
\end{figure}

\begin{figure}[t]
\begin{center}
\includegraphics[angle=0,width=0.45\textwidth]{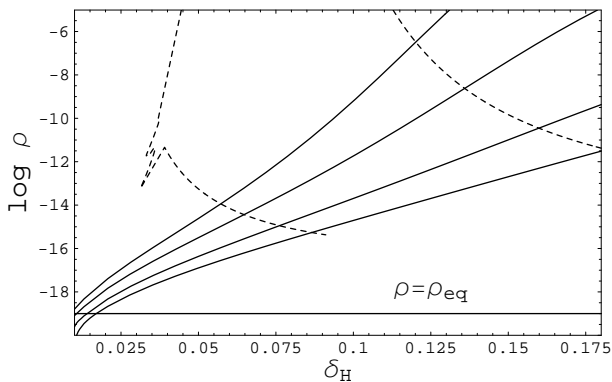}
\end{center}
\caption{Mean clump density $\rho$ [g~cm$^{-3}$] as a function of the radiation
density perturbation amplitude $\delta_{\rm H}$ on the horizon scale. The solid curves (from top down) respectively correspond to the clump masses $M=10^{-11}$, $10^{-6}$, $10^{-1}$, $10^{2}M_\odot$. The dashed curves correspond to
the clump density limits from primordial black hole overproduction for
the formation thresholds $\delta_{\rm th}=1/3$ and $\delta_{\rm th}=0.7$ (curves 1 and 2 respectively).\label{gr6}}
\end{figure}

To produce superdense clumps, excessive small-scale
perturbations are needed, e.g., in the form of peaks. It is
clear from (8) that perturbations with a simple power-law
spectrum and $n_s<1$ are too small to form clumps at the RD
stage. We note that superdense clumps formed from the
spectrum maximum do not aggregate into hierarchical
structures for a long time and are therefore almost unaffected
by tidal forces, and their mass function is concentrated
around some mass corresponding to the spectral maximum.

The clump formation scenario from adiabatic perturbations at the RD stage is constrained by the PBH formation
effect. In the model of ``neutralino stars'' with masses close to
$\sim0.1M_\odot $, this constraint was obtained in [88]. Following,
e.g., [24,96,98], we assume that there is a sufficiently high
maximum on some comoving scale $\xi=r/a(t)$ in the cosmological perturbation spectrum. Fluctuations with
$\delta_{\rm H}>\delta_{\rm th}\sim1/3$ collapse to form PBHs, and smaller perturbations turn into clumps. The mass fraction of such dark matter
clumps is expressed, as in (10), by the integral [103]
\begin{equation}
\beta_{\rm cl}= \int\limits_0^{\delta_{\rm th}}\frac{d\delta_{\rm H}}{\sqrt{2\pi}\Delta_ {\rm
H}}\exp\left(-\frac{\delta_{\rm H}^2}{2\Delta_{\rm H}^2}\right).  \label{bet3}
\end{equation}
As noted in Section~2.5, to avoid overproduction of PBHs, the
condition $\Delta_{\rm H}\ll\delta_{\rm th}$ should be satisfied; therefore, $\beta_{\rm cl}\approx1/2$. That is, half of the dark matter is in the region of positive
perturbations. However, not every perturbation from this
region can evolve into clumps. In Section~3.4, we precisely
determine the fraction of superdense dark matter clumps,
taking the nonsphericity of perturbations into account.

The relation of the PBH mass, which is equal to the total
dark matter mass under the horizon $M_{\rm H}$ by an order of
magnitude, to the dark matter mass $M$ in fluctuations of the
same comoving scale at the horizon crossing time
$t_{\rm H}\simeq GM_{\rm H}/c^3$ has the form [88]
\begin{eqnarray}
M_{\rm H}&=&\frac{1}{2^{2/3}} \left(\frac{3}{2\pi}\right)^{1/6}
\frac{M^{2/3}c}{G^{1/2}\rho_{\rm eq}^{1/6}} \label{mhmx}
\\
&\simeq& 1.5\times10^5\left(\frac{M}{0.1M_{\odot}}\right)^{2/3}M_{\odot}.
\end{eqnarray}
Using (10) and (29), we can express the cosmological PBH
density parameter $\Omega_{\rm BH}$ at the present time through dark
matter perturbations. Thus, PBHs put bounds on the clump
parameters shown in Fig.~5. The cusp on curve 1 corresponds
to PBHs that are evaporating at the present time via Hawking
radiation.

We recall that PBH formation occurs on the tail of the
Gaussian distribution of fluctuations, and most of the clumps
are formed from fluctuations with the rms amplitude $\delta\sim\Delta_{\rm H}$.
Therefore, we repeat, not every fluctuation that produced a
clump could collapse into a PBH at the instant $t_{\rm H}$. In other
words, due to the high PBH formation threshold, a significant
fraction of fluctuations did not collapse into PBHs and
continued to evolve (Fig.~6).

\begin{figure}[t]
\begin{center}
\includegraphics[angle=0,width=0.45\textwidth]{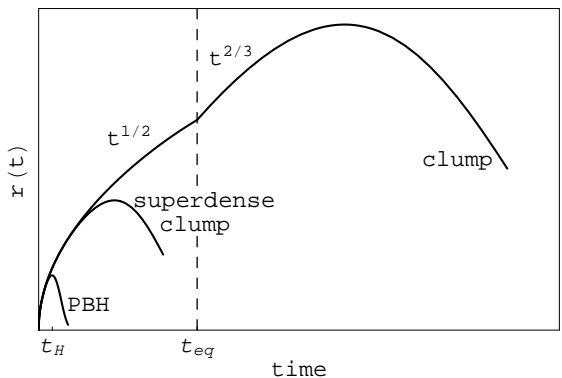}
\end{center}
\caption{Different evolutions of a density perturbation as a function of its
amplitude. Perturbations above the threshold collapse to form PBHs
immediately after the horizon crossing time $t_{\rm H}$. If the perturbation
amplitude is insufficient to form a PBH, but the perturbed region
becomes self-gravitating at the radiation-dominated stage at $t<t_{\rm eq}$, a
superdense clump is formed. Clumps with lower densities are produced at
$t>t_{\rm eq}$. \label{grevol}}
\end{figure}

Because the clumps and PBHs originate from the same
perturbation spectrum, the mass and radius clump distributions can, in theory, restore the form of the perturbation
spectrum and the PBH mass functions can be obtained using a
formalism similar to the Press--Schechter theory [114]. Such a
calculation will be possible once clumps are discovered.

%%%

\subsection{Nonspherical models}

The form of primordial perturbations is typically not
spherically symmetric, and the asymmetry can increase with
time. Therefore, in studying clump formation, it is necessary
to go beyond simple spherical models.

The simplest generalization of the spherical model is the
model of a homogeneous ellipsoid (see its detailed description, e.g., in [128,130]). Although the homogeneous ellipsoid
approximation does not take the density increase toward the
center and other inhomogeneities into account, it is useful
because it allows taking the evolution of nonsphericity into
account in the framework of a simple analytic model, as well
as considering the interaction of external tidal forces with the
quadrupole moment of the perturbation. This interaction
increases the angular momenta of protogalaxies at their
formation stage [130]. The evolution of a self-gravitating
homogeneous ellipsoid preserves the ellipsoidal form and
volume homogeneity of the object, although both the density
and the ratio of different ellipsoid axes can change. The initial
compression velocities along different ellipsoid axes can be
inferred from the peculiar velocities.

The use of the homogeneous ellipsoid model in [131] to
investigate the problem of superdense clump formation at the
radiation-dominated stage (a generalization of the model
described in Section 3.3) allowed finding the growth factor
of perturbation nonsphericity $s(t)$. We can assume that the
boundary of maximum admissible nonsphericity satisfies the
condition $s_f/b_f<1$, and at higher nonsphericities, the clump
disintegrates and is not virialized. If the growth factor is
known, the clump formation condition can be expressed
through the initial data nonsphericity: $s_i/b_i<(b_f/b_i)(s_i/s_f)$.
The shape distribution of Gaussian perturbations can be
calculated using the results in [132] or [89, 133]. The closer
the perturbation shape is to the spherical one, the more rarely
such a perturbation occurs. As a result, for the least dense
clumps forming at $t\leq t_{\rm eq}$, the number of clumps decreases by
an order of magnitude due to the initial perturbation
nonsphericity. For dense clumps that are formed earlier, at
$t\ll t_{\rm eq}$, this decrease can be as high as 4-5 orders of magnitude.

For objects that are formed at the dust-like stage, the
model of ellipsoidal collapse allows precisely determining the
halo formation criterion in the Press--Schechter formalism. In
[133], corrections to the spherical model were obtained that
improved the Press--Schechter mass function, and the modified mass function is widely used in theoretical calculations of
galaxy and galaxy cluster formation.

Another method to go beyond the spherical approximation is to use the Lagrangian method to describe the
perturbation growth. The Zeldovich approximation [134,
135] provides a remarkable example of such an approach. It
allows studying different aspects of three-dimensional compression of dark matter. The Zeldovich approximation was
generalized to the evolution at the RD stage in [129], where
corrections to the linear theory were obtained. Simpler
Lagrangian approaches are also used, in which the dark
matter density growth due to the velocity boost is studied,
and gravitational forces are neglected over most of the time of
evolution. This method is used, for example, to study the
evolution of clumps from strongly nonspherical topological
defects [107, 136].

%%%

\subsection{Clumps around topological defects}

Topological defects (infinite strings, cosmic string loops, and
textures) are considered as clump seeds in [107]. Moving
infinite strings or loops produce a velocity boost for dark
matter particles, which leads to the density increase along the
trajectories of motion of the topological defects.

We focus on the scenario of clump formation around
cosmic string loops, which serve as seeds for entropy
perturbations, as described in Section~2.4. These clumps are
formed at the RD stage, and their density can greatly exceed
$\gg\rho_{eq}$. Linear topological defects, cosmic strings, can be
formed during the early cosmological phase transitions (see
reviews [137, 138]). In addition to the formation of infinite
strings, closed string loops can be formed due to string self-crossing [139, 140]. Only low-velocity loops can form very
dense clumps around them [85]. We require the shift of a
string, starting from its formation time $t_i$ to its destruction
time $t_d$, to be small in comparison with the loop radius $l/(2\pi)$.
For the Maxwell loop velocity distribution, the string
formation probability satisfying this condition is estimated
to be $P_{\rm lv}\sim10^{-7}$ [141]. However, the requirement of a small
shift compared with the loop radius is redundant. Even if it is
violated, a clump can be formed, although with a lower mean
density [136].

Clump formation around loops at the RD stage was
studied in [85] using Eqn~(19). In the approach in [85], the
maximum clump density is restricted by the adiabatic
expansion of clumps that have already formed due to the
evaporation of loops. In [141], this density restriction was
modified in the case where loop decay occurs before clump
virialization. The clump density can then be $\rho_{\rm cl}\gg140\rho_{\rm eq}$.

The loop length distribution is obtained in the form [142]
\begin{equation}
dn_{\rm loop}=\frac{Ndl}{c^{3/2}t^{3/2}l^{5/2}}, \label{loopsdis}
\end{equation}
with $N\sim2$. This distribution is translated into the clump
distribution, which can be used to calculate dark matter
annihilation signals [141] or to study gas ionization [136].

%%%

\section{Internal structure of clumps}

Besides the typical characteristics of clumps, their internal
structure is very important, especially in the central core,
where dark matter particle annihilation can occur most
efficiently due to the high density. The model of galaxies
with a power-law increasing central density profile was
elaborated in [105, 106] and [76] in the framework of the
``secondary accretion model''. This model assumes that dark
matter `accretes' onto an initially existing central object, a
``seed mass'' $M_c$, which can be a compact clump or a black
hole. This central object produces a perturbation $\delta=M_c/M$,
as shown in Section~2.4. In an approximate approach, the
evolution of caustics that arise during gravitational contraction and the crossing of dark matter layers was also discussed
in [76]. Later, the model of a dissipationless gravitational
singularity was proposed in [55-57], which also produces a
power-law density profile, and the calculations in [55-57]
were performed by a powerful method of adiabatic invariants, enabling the study of evolution in the nonlinear regime.
We note the fundamentally important advantage of the model
developed in [55-57] over the secondary accretion models. As
shown in [55-57], under the action of self-gravity, a density
perturbation with an initially smooth profile takes an
approximately power-law form without any seed mass:
rint
\begin{equation}
\rho_{\rm int}(r)=\left\{
\begin{array}{lr}
\rho_c, & r<R_c; \\
\displaystyle{\rho_c\left(\frac{r}{R_c}\right)^{-\beta}}, & R_c<r<R;\\
 0, & r>R,
 \label{rho}
\end{array}
\right.
\end{equation}
where $\rho_c$ is the clump central density, $R_c$ is the radius at which
the density growth stops, $R$ is the virial radius or the radius of
the last layer that was singled out from the cosmological
expansion and stabilized after the nonlinear compression
stage, and $\beta=1.7-1.9$.

Because Gaussian primordial density perturbations yield
smooth initial conditions, the model in [55-57] describes the
physical processes more realistically. However, so far we have
considered the formation of an individual object from an
isolated density perturbation. Coalescences of objects during
hierarchical clustering can appreciably modify the density
profile (see Section~4.9).
The internal structure of the clumps has been studied in
many papers, but this problem remains unsolved in one
important aspect: which physical process stops the density
increase during the clump formation and at what radius? The
region bounded by a circle of this radius is called the clump
core. The core radius is very important for studying the
annihilation signals, and we discuss some processes that are
possibly responsible for the central core formation in
Sections~4.1-4.9.

%%%

\subsection{Physics of violent relaxation and virialization}

Virialization is the mixing of dark matter streams until the
entire halo reaches an equilibrium inside some radius, the so-called virial radius. Outside this radius, the mixing is not
completed, and new layers of dark matter, which decouple
from the cosmological expansion, continue to fall toward the
center. The virialization occurs due to radial oscillations, the
presence of large inhomogeneities, and the nonspherical
motions of matter. According to Lynden-Bell [143], these
processes are called ``violent relaxation''. The virialization
leads to an energy redistribution between different density
clumps and between the kinetic and potential energy.
Inasmuch as dark matter particles scatter mainly on large
clumps and inhomogeneities of matter during `violent
relaxation', a nearly Maxwell velocity (not energy) distribution is established. The virialization also leads to the tidal
destruction of large clumps and their mixing and results in the
formation of a sufficiently smooth density profile. The total
energy $E=T+U$ in some dark matter layer is the sum of the
kinetic and potential energies. At the linear stage,
$|E|\ll |T|\approx|U|$, but after the virialization, $E\approx-T\approx U/2$.
This state is reached when the layer radius decreases by a
factor of two, $R=R_{\rm max}/2$, after the expansion stop at the
time $t_s$ (because the potential energy of the layer $U\approx GM/R$)
and, accordingly, the mean density of the stabilized object is
$\rho=8\rho(t_s)$.
In the approach considered here, we reliably assume that
most of the mass of dark matter is concentrated in the outer
layers, which have just been virialized; therefore, the contribution of the internal mass to the mean density is not very
significant.

Initially, the ``violent relaxation'' concept by Lynden-Bell
was based on radial oscillations of an object, which are
accompanied by rapid variations of the gravitational potential. As the picture of the hierarchical clustering of large-scale
structures emerged, it became clear that nonradial motions
during the aggregation of objects, which represent large
inhomogeneities, also produce a variable gravitational
potential leading to violent relaxation in the forming
halo [144]. Hierarchical clustering and the relaxation related
to it produce, besides virialization, a universal density profile,
for example, an NFW one. According to the results of
calculations in [145], an instability in the phase space, phase
mixing, and a moderate `violent relaxation' should be present
for the relaxation to begin.

Nonradial motions of dark matter particles in the halo
play an important role in the relaxation process because
these motions lead to dark matter mixing inside each dark
matter layer with energy redistribution until the virial
equilibrium is reached. Nonradial motions occur due to
coalescences of different subhalos, the initial inhomogeneities, and tidal forces from the external mass distribution.
Adiabatic capture in the multi-stream regime [55-57] is an
important mechanism of energy redistribution between the
layers. Although this mechanism operates only in the radial
direction, it also leads to the universal density profile
$\rho\propto r^{-1.8}$ in isolated halos.

In a real situation, even in the case of an isolated object
formed from a peak in the perturbation spectrum at some
scale $k$, both the inhomogeneity and nonsphericity appear on
scales $\Delta k\sim k$. At the beginning of the nonlinear stage, the
proto-object consists of several (for example, $N$) large
inhomogeneities with masses $M_s\sim M/N$. The two-body
relaxation time in this `cluster' is
\begin{equation}
t_r\simeq \frac{Nt_{\rm dyn}}{10\ln(N)} \label{trel}
\end{equation}
and does not differ significantly from the dynamical time
$t_{\rm dyn}$. Thus, the object experiences a ``violent relaxation''
accompanied by energy redistribution due to binary scattering of large inhomogeneities, which ultimately leads to
virialization.

The virialization of early formed clumps can occur even
at the RD stage. Does this process differ from the ordinary
violent relaxation at the matter-dominated stage? Disintegration and dissociation of very irregular halos can be a new
feature of the nonlinear evolution of clumps at the RD
stage. Indeed, if the protohalo is sufficiently elongated, the
mass of radiation inside the minimal sphere encircling the
protohalo exceeds that of dark matter, and the object self-gravity cannot hold its parts together. The protohalo decays
and its pieces fly apart due to cosmological expansion. To
find the nonsphericity above which such an evolution is
possible, numerical simulations of clump formation at the
RD stage are required.

%%%

\subsection{Secondary accretion, self-similar solutions,
and ``ultra-compact minihalos''}

In the spirit of secondary accretion models [76,105,106], we
assume that at the radiation-dominated stage, there is a seed
mass $M_c$, and dark matter is distributed homogeneously in its
nearby surroundings prior to the dust-like stage (the effect of
nearby perturbations is to be discussed below). At $t\ll t_{\rm eq}$ for
mass scales $M\gg M_c$, a dark matter perturbation $\delta_i=M_c/M$
does not evolve. Indeed, according to the Meszaros solution,
$\delta=\delta_i(1+3x/2)$, where $x=a/a_{\rm eq}$ [128]. This solution can
easily be found from Eqn~(19) in the linear approximation.
Later, at $t>t_{\rm eq}$, we can use solution (15) with the substitution
$5\delta_i/3\to \delta_i$, as noted in Section~3.1. The object formation
threshold in the spherical model in this case is $\delta(t)=\tilde\delta_c=(3\pi/2)^{2/3}\approx2.81$. Then the virialized object mass as a
function of the redshift takes the form
\begin{eqnarray}
M(z)&=&\frac{3}{2}\left(\frac{2}{3\pi}\right)^{2/3}
 \frac{1+z_{\rm eq}}{1+z}M_c \label{rcol2}
\\
&\approx&1.7\times10^3\left(\frac{M_c}{10^2M_{\odot}}\right)
\left(\frac{1+z}{100}\right)^{-1}M_\odot, \nonumber
\end{eqnarray}
with $(3/2)(2/(3\pi))^{2/3}\approx0.53$, and the radius $r_c=r_s/2$ of the
layer that was virialized is
\begin{eqnarray}
r_c&=&\frac{1}{3}\left(\frac{3}{4\pi} \right)^{1/3} \frac{M^{4/3}}{\rho_{\rm eq}^{1/3}M_c}
\label{rcol}
\\
&\simeq& 3.2\left(\frac{M_c}{10^2M_{\odot}}\right)^{1/3} \left(\frac{1 +
z}{100}\right)^{-4/3}\mbox{~pc}\nonumber
\end{eqnarray}
Using halo's mass-radius relation (35), we find the density
profile
\begin{eqnarray}
\rho(r)&=&\frac{1}{4\pi r_c^2}\left.\frac{dM(r_c)}{dr_c}\right|_{r_c=r} \label{rhoih}
\\
&\simeq &3\times10^{-21}\left(\frac{r}{1\mbox{~pc}}\right)^{-9/4}
\left(\frac{M_c}{10^2M_\odot}\right)^{3/4}\mbox{g~cm$^{-3}$}. \nonumber
\end{eqnarray}
In the case of a noncompact central mass, for example, an
extended PBH cluster, the density profile differs from
$\rho\propto r^{-9/4}$. Adding new layers at the periphery does not
strongly affect the already formed internal density profile
due to a relatively high central density, which is confirmed by
analytic estimates and numerical simulations [145]. The
constant flux of a new mass onto the system shifts its state
by 10\% from virial equilibriums, i.e., the constant in the
energy relation $2T/|U|=const$ differs from unity [145].

The total mass of the induced halo increases with time, as
more and more remote layers decouple from the cosmological
expansion and are virialized around the seed mass. The
induced halo growth stops when ordinary dark matter
fluctuations (those originating from inflation perturbations)
with mass M equal to the induced halo mass enter the
nonlinear stage. The growth law $\propto t^{2/3}$ is the same for usual
perturbations and those induced by a seed mass; therefore,
the halo growth stopping condition is simply [145,147]
\begin{equation}
\label{Deltaeq} \nu\sigma_{\rm eq}(M)=\frac{9}{10}\frac{M_c}{M},
\end{equation}
where $\nu$ is the density perturbation amplitude in units of the
rms fluctuation value. The right-hand side of (37) is the
fluctuation amplitude due to the mass $M_c$ taking the two
correction factors described above into account. The left-hand side describes usual Gaussian fluctuations taken at the
time $t_{\rm eq}$. The numerical solution of Eqn~(37) yields the
ultimate mass of the induced halo, which for $M_c\sim1-10^4M_\odot$ is $M\sim10^2M_c$. In [103,148-152], it was
assumed that the growth of clumps stops at $z\sim10-30$,
when large-scale dark matter structures start forming intensively. A more accurate quantitative criterion of the cessation
of growth is given by Eqn~(36).

Notably, clumps can be formed by this mechanism
around PBHs [146-148,150], and the idea that PBHs can be
seeds of more extended objects was first put forward in [153].
Dark matter annihilation in mini-peaks around black holes
(of stellar origin or resulting from collapses of gas clouds in
the first halos) was considered in [154]. Of special interest is
the case where dark matter particles annihilate to produce a
narrow line. This line is broadened due to relativistic effects
near the black hole horizon, which offers an observational
signature of dark matter annihilation around black holes
[155]. The character of line broadening must depend on the
dark matter density distribution at distances up to several
gravitational radii from the black hole; however, no detailed
calculations of this dependence have yet been performed.

In [76,156], self-similar accretion regimes were found in
the framework of the secondary accretion model, in which
collapses of new spherical layers maintain a power-law
density profile. However, for this to be possible, the initially
singular density profile $\delta\rho_i\propto r^{-\varepsilon}$ with an infinite central
density in the protohalo is required. The rate of the virial
radius $r_s$ increase then depends on the parameter $\varepsilon$, and
characteristics of all halos in the self-similar solutions are
expressed in terms of the variable $X=r/r_s$. Initially, the
density perturbation is not singular; therefore, the central
density can be lower than in self-similar solutions, and the
completion of the accretion of fresh outer layers can violate
the self-similar regime and flatten the density profile $\rho\propto
r^{-3}$
in the halo periphery [145]. The formation of singular profiles
from the initially smooth perturbations cannot be explained
by the secondary accretion model and self-similar solutions.
This problem was solved by Gurevich and Zybin [55-57].

Recently, some papers [103,148-152,154,157-163] have
discussed so-called ultra-compact mini-halos?clumps
formed at the very early stages (at the beginning of the dust-like stage) from large density perturbations of different
natures: spectral peaks or perturbations enhanced in phase
transitions in the early Universe. The important idea formulated in [158, 159] is that the annihilation limits, in principle,
allow constraining the primordial perturbation spectrum from
which clumps are formed. A sufficiently strong annihilation
signal would violate the gamma-ray background limits;
hydrogen recombination retardation or a delay in the
reionization in the Universe may also have taken place,
which would leave imprints in the CMB spectrum. However,
to obtain reliable constraints, it is necessary to know, in
addition to dark matter particle properties, the internal
structure of dark matter clumps, especially in the core.
If the clump was formed around a superdense clump,
which in turn originated at the RD stage, a modification of
the object density at the center would be required. To
calculate the mean density of the central superdense clump,
the formalism in [85, 88, 164], described in Section 3.3, is
required. In superdense clumps, due to their individuality, the
Gurevich--Zybin profile is likely to arise. The issue of
superdense clump cores also remains open.

%%%

\subsection{Nondissipative gravitational singularity (Gurevich--Zybin theory)}

The dissipationless gravitational singularity is reviewed in
detail in [57], and therefore this model, in spite of its
importance, is discussed here only very briefly.
We consider some individual perturbation with a smooth
density profile. Given that adiabatic perturbations are
characterized by one scalar function whose derivatives give
peculiar velocities (perturbations relative to the homogeneous
Hubble flow), the problem was reduced in [55-57] to the
collapse of a density perturbation at rest by introducing some
effective density. By assuming a small deviation from
spherical symmetry, a nonlinear growth law for density
perturbations was found. The central density of the perturbation increases and becomes formally infinite at some instant,
after which dark matter flow crossing emerges.

The novelty of the approach in [55-57] and the key factor
of its success are the use of the adiabatic invariant method in
the multi-stream evolution regime. The conservation of
adiabatic invariants allowed the authors to find the dark
matter particle distribution function in forming halos. It was
shown, as a result, that in the nonlinear regime with $\delta\ge1$, a
multi-stream instability arises, and power-law density profile
(32) is formed in the clump. After the cessation of cosmological expansion, the forming clump starts compressing to the
radius $R=\lambda R_{\mathrm{max}}$, where $\lambda$ is a nonlinear compression factor.
It is usually assumed [165] that after the cosmological
expansion stops, the clump is virialized by compressing by a
factor of two in radius, i.e., $\lambda=0.5$. But according to the
gravitational instability theory, $\lambda\simeq0.3$ in the multi-stream
region [166]. In [55-57], the clump core radius is estimated to
$R_c/R\simeq\delta^3_{\rm eq}\ll1$ from the analysis of the velocity field in the
decaying perturbation mode.

As mentioned in Section~4.2, an important advantage of
the Gurevich--Zybin theory is that it explains the power-law
density halo formation from initially smooth perturbations
without the central seed mass that is a prerequisite in the
secondary accretion models. What is not fully accounted for
by this theory so far is the effect of hierarchical clustering of
halos with different scales, which is now considered to be
the main factor for the universal density profile formation.
As a whole, the theory in [55-57] is in good agreement with
observations and numerical results at intermediate scales.
However, in the central parts of galactic halos, the power-law exponent $\beta$ in profile (32), apparently cannot take values
$\beta=1.7-1.9$, as pointed out in [55-57], because microlensing
observations and measurements of star dynamics suggest
that $\beta\le1.5$ near the galactic center [167]. By contrast, in the
cores of minimal-mass clumps and cores formed in spectral
peak models, the values $\beta=1.7-1.9$ are quite possible,
because these clumps are formed from decoupled smooth
density perturbations.

%%%

\subsection{Constraints on the core radius}

from the Liouville theorem
The constraint for the maximum density of an object,
following from the constraint for the phase density
$F_L(p,q,t)$, which is conserved according to the Liouville
theorem, was considered in [168] in application to hypothetical heavy leptons. For the Liouville theorem to be applied,
only the Hamiltonian character of the system is required. The
Liouville theorem remains valid when multi-stream flows
emerge (the formation of caustics), because the distribution
function remains single-valued in the phase space.

If particle annihilation and other effects violating the
Liouville theorem conditions are unimportant, $F_L(p,q,t)=const$ along the particle trajectories. Unfortunately, $R_c$ can
only be restricted but not obtained from the Liouville theorem
applied only to the initial and final clump states. This is related
to entropy generation in the intermediate processes, which is
taken into account in the entropy theory [79]. In other words,
the phase volume is rendered very dispersive by the process of
nonlinear dark matter mixing (see Fig.~8.3 in [2]).

For a comparison with observations in astrophysics, it is
convenient to use the quantity $Q=\rho/\sigma^{3/2}$ as a distribution
function, where $\sigma=\langle v_{||}^2\rangle=\langle
v^2\rangle/3$ is the one-dimensional
velocity dispersion. Numerical modeling shows that over
several orders of magnitude in radius, $Q\propto r^{-1.875}$ [169]. For
theoretical estimates, we can use $f_c=\rho_c/v^3$ as the distribution function in the clump core, where $\rho_c$ is the core density
and $v$ is the characteristic core particle velocity. For nearly
isothermal profile $\rho(r)\propto r^{-2}$, the velocity $v$ is of the order of
the virial velocity in the entire clump.

We find the constraint on $R_c$ from the Liouville theorem
applied to the initial and final states. There are two sources
of the initial entropy or the initial velocity dispersion $\sigma$:
thermal velocities of dark matter particles at the decoupling
and peculiar velocities in the case of adiabatic perturbations.

The thermal part can be found from the distribution
function at the kinetic decoupling instant $t_d$ [2, Section~8.3.2].
The neutralino gas is nonrelativistic and nondegenerate at the
decoupling time $t_d$, and hence the Maxwell distribution is a
good approximation at that time:
\begin{equation}
f_p(p)d^3rd^3p=\frac{\rho_m}{m(2\pi mkT)^{3/2}}e^{-\frac{p^2}{2mkT}}d^3rd^3p, \label{maxdistr}
\end{equation}
Here, m is the dark matter particle mass and $\rho_m$ is the dark
matter density, which can be expressed through the temperature in the Universe at any time from the local entropy
conservation law $g_*T^3a^3=const$, where $g_*$ is the effective
number of the degrees of freedom at the temperature $T$. The
core distribution function is smaller than the initial distribution function that has a maximum at $p=0$. Hence,
$f_c<f_p(p=0)$. For the isothermal density profile in the
clump, this condition yields a constraint on the relative core
radius:
\begin{equation}
\frac{R_c}{R}>\frac{2\pi^{1/2}\bar\rho^{1/4}T_d^{3/4}}{3^{1/4}G^{3/4}M^{1/2}m^{3/4}\rho_m^{1/2}(t_d)}.
\label{rcliwtherm}
\end{equation}
In this calculation, we have used the fact that for the
isothermal density profile $\rho(r)=\rho_c
(r/R_c)^{-2}$, the relative
core radius $x_c=R_c/R$ is expressed as $x_c=(\bar\rho/3\rho_c)^{1/2}$, where
$\bar\rho$ and $\rho_c$ are the mean and central clump density.

In the constraints found from the Liouville theorem,
peculiar velocities that are generated by the gravitational
instability can play a role similar to that of thermal velocities.
The velocity of the increasing perturbation mode is smaller
than the peculiar velocity in (14) by a factor of $2/5$ [128]. After
the time $t_{\rm eq}$, the decreasing mode rapidly decays, and peculiar
velocities are formed that later increase with time as $\propto t^{1/3}$, darck matter density $\propto
t^{-2}$, and hence the distribution function in the phase volume
decreases. We consider a time instant close to $t\sim t_{\rm eq}$. By
calculations similar to those given above with the substitution
of thermal velocities by peculiar ones, we find the core radius
as $R_c/R\simeq 0.01\delta_{\rm eq}^{9/2}$. The results of calculations of $R_c/R$ for
three example cases are listed in the Table (for the neutralino
mass $m=100$~GeV and the decoupling temperature
$T_d\simeq25$~MeV), which shows that the peculiar velocities are
unimportant for standard clumps, but can become decisive
for superdense clumps.

\begin{widetext}
\begin{center}
\begin{table}[t]
\centering \caption{\label{tabrc}Three examples of clump parameters and the relative core radii caused by different effects discussed in the text. * --- bounds on $x_c=R_c/R$ from the Liouville theorem. ** --- bounds on $x_c$ from peculiar velocities. *** --- bounds on $x_c$ from annihilation.}
%\begin{ruledtabular}
\begin{tabular}{|c|c|c|c|c|c|c|}

\hline

Example & $M/M_\odot$  & $\bar\rho$, g~cm$^{-3}$   &  $\delta$ &  $x_c$, thermal velocities$^*$
 & $x_c$, peculiar velocities$^{**}$  & $x_c$, annihilation$^{***}$  \\

\hline 1 & $10^{-6}$ & $3\times10^{-23}$   & $\delta_{\rm eq}=0.009$  &  $4\times10^{-3}$  &
$6\times10^{-12}$  & $2.6\times10^{-5}$  \\

 \hline

2 &  $10^{-6}$  & $4.2\times10^{-16}$   &  $\delta_H=0.05$ & 0.24  &  0.1 & 0.1  \\

\hline
3 &  $0.1$  &  $2.5\times10^{-17}$  & $\delta_{\rm eq}\simeq 1$  &  $4\times10^{-4}$  &
0.01   & $2.5\times10^{-2}$  \\

 \hline

 \end{tabular}
%\end{ruledtabular}
\end{table}
\end{center}

\end{widetext}

%%%

\subsection{Entropy theory}

To date, theoretical calculations and numerical $N$-body
simulations of the halo density profiles on galactic scales
have not been able to fully reproduce the results of
observations. Some galaxy types (low surface brightness
(LSB) galaxies and dwarf spheroids) demonstrate a sufficiently large core with the density that is constant or slowly
increases toward the center. The density profile outside the
central region is also better described by the Burkert profile
than by analytic approximations of numerical simulations
(see Section 4.9 for more details). In this connection, new
factors should be sought to describe this discrepancy.

In [79], effects were found that can be responsible for the
observed density profiles, including the initial entropy due to
the peculiar velocities of small perturbations on the halo
background and entropy generation during nonlinear clustering. In [79], the initial entropy was calculated analytically,
and the generated entropy was taken from the results of
N-body calculations. Presently, numerical modelling is the
only tool capable of tracking the mean distribution function.
In [79], by uniting the initial and generated entropies, the
entropy function was found and the dark matter halo density
profiles on galactic scales were calculated. It turns out that
taking these effects into account enables significantly improving the agreement of the calculated density profile with
observations. The additional entropy provides more effective core formation (in [79], the profiles with $\beta\leq1$ are
referred to as the core). An important element of the entropy
theory is also that the initial Gaussian density distribution
produces a sufficiently broad distribution of the galactic
central core properties.

The entropy theory enabled explaining density profiles of
sufficiently massive galaxies.However, for low-mass galaxies,
the entropy corrections are small due to the low initial
entropy, which is caused by small peculiar velocities in low-mass halos [79]. Therefore, the applicability of the entropy
theory to small-scale clumps considered here requires additional inspection. As shown in Section 4.4, the initial entropy
can be important for superdense clumps.
Calculations using entropy and the Liouville theorem are
useful because they give some general constraints and
predictions without the need for detailed study of complicated gravitation dynamics. Another version of the entropy
function was also used in [170] to study phase mixing
processes and violent relaxation.

%%%

\subsection{Tidal effects on the density profile}

Modification of the density profile at the formation stage is
possible due to tidal forces from external perturbations or
internal inhomogeneities of the medium. These factors cause
particles in a forming clump to acquire angular momentum,
which leads to the central density profile flattening and,
possibly, to the formation of a core [39, 119, 171].
In the framework of the simple spherical model considered in Section 3.1, with the correction due to tidal forces
taken into account, we find the characteristic scale at which
the density profile can have a kink [39]:
\begin{equation}
 x_c=\frac{R_c}{R}\simeq
 0.3\nu^{-2}f^2(\delta_{\rm eq}),
 \label{coreit}
\end{equation}
where $\nu$ is the peak amplitude expressed in units of the rms
perturbations introduced in Section~3.1, and the function
$f\sim1$. For perturbations with $\nu\sim 0.5-0.6$, $x_c \sim 1$, i.e., the
corresponding clumps are destroyed by tidal forces already at
the formation stage. In [119], it was found that the angular
momentum of dark matter particles modifies the density
profile to produce a kink similar to that in the NFW profile
and to form an effective core in the dark matter halo center.
Similar results were also obtained in [171-174]. In [159], the
core radius $R_c/R\sim3\times10^{-7}$ was estimated from an examination of nonradial velocities.

The tidal forces partially prevent the appearance of the
central singularity of the density (or decrease the density
divergence) during evolution of the clump, but if this
singularity (a very high density in the clump center) has
already been formed in some way, then the tidal interactions
cannot disintegrate it any more due to the adiabatic invariant
conservation. Indeed, core particles in the clump oscillate
along orbits with a very high frequency, and therefore slowly
varying tidal forces cannot affect their motion. Clumps, and
especially their cores, can be destroyed only at later stages of
evolution due to tidal interaction with stars and galactic discs.

%%%

\subsection{Annihilation limit of the maximum density}

In [175], the maximum central density in the clump was
estimated using the annihilation rate and the time elapsed
after clump formation,
\begin{equation}
\rho(r_{\rm min})\simeq \frac{m}{\langle\sigma v\rangle(t_0-t_f)} \label{annwr}
\end{equation}
where $t_0$ is the present time and $t_f$ is the clump formation time.
According to this estimate, the core radius increases with time
due to annihilation losses of particles in orbits passing near
the clump center. For the isothermal density profile $\rho(r)\propto r^{-2}$, the core radius
\begin{equation}
\frac{R_c}{R}\simeq\left(\frac{\langle\sigma v\rangle t_0\bar\rho}{3m}\right)^{1/2},
\label{rcann1}
\end{equation}
is listed in the Table for the annihilation cross section
$\langle\sigma
v\rangle\simeq3\times10^{-26}$~cm$^3$~s$^{-1}$
corresponding to thermal particle creation
and $m=100$~GeV. This approach assumes that particle orbits
near the clump center are not refilled after annihilation of
their particles. The value in (41) was used to calculate the
annihilation signal from ultra-compact mini-halos in [103,149-152,158,159].

The opposite case with compensation of losses was
considered in [176, 177]. The authors of [176] inferred the
core radius from the annihilation limit at the dark matter halo
formation stage. The minimal radius was obtained from the
condition that the characteristic annihilation time is of the
order of the Jeans instability time, $t_h \sim
(G\bar\rho)^{-1/2}$, because
this time gives the characteristic halo formation time scale.
The characteristic core radius in that case is
\begin{equation}
x_c^2\simeq\frac{\langle\sigma_{\rm ann} v\rangle\rho^{1/2}}{G^{1/2}m}.
\end{equation}
In [177], the core radius of the already formed clump was
found by assuming a permanent hydrodynamic flow of dark
matter onto the clump center. The free-fall time of a particle
onto the clump center was assumed to be equal to the
characteristic annihilation time in the core, and the flow was
assumed to be permanently refilled by new particles. This
approach yields the most conservative estimate of the core
radius.

In a real clump, external gravitational perturbations must
lead to the complete or partial regeneration of orbits with low
angular momenta passing through the clump core. The orbit
refilling leads to a higher central density than the one in (41).

%%%

\subsection{Gravothermal catastrophe for superheavy particles}

Superheavy particles with masses $m\geq10^{11}$~GeV can be born
immediately after the inflation stage due to a nonstationary
gravitational field or other mechanisms; they can also be dark
matter particles [20-22,178-180]. Several constraints on the
possible properties of super-heavy particles are discussed in
review [113].

If a superdense clump consists of superheavy particles, an
interesting effect, called the ``gravothermal catastrophe'' in
stellar dynamics, can occur [164, 181]. In globular stellar
clusters, the instability and the gravothermal catastrophe are
developed due to binary gravitational scatterings of stars. A
similar process can become the dominant evolutionary factor
of superdense clumps that formed early at the RD stage and
consist of superheavy dark matter particles, which scatter on
each other like stars.

In the gravothermal catastrophe regime, particles are
evaporated (ejected) from the clump cores, which are
decreasing in radius. The time when the gravothermal
catastrophe occurs is smaller than the Hubble time only for
extremely dense clumps (see [164, 181] for more details). The
singular density profile $\rho\propto r^{-2}$ resulting from the gravothermal catastrophe can be formally continued to very small radii
$R_c$. There are physical effects that restrict the density:
electroweak particle scattering, particle annihilation (considered in Section 4.7), and degenerate Fermi-gas pressure.
January 2014 Small-scale clumps of dark matter 15
The results of calculations in [181] suggest that for superheavy
fermions in superdense clumps, the Fermi degeneration is the
crucial effect. The maximum core density and, accordingly,
the core radius can be found by equating the momentum of a
degenerate Fermi-gas particle to its virial momentum at the
core boundary, $p_F=(3\pi^2)^{1/3}(\rho_c/m)^{1/3}=m V_c $, where $V_c=
\sqrt{GM_c/r_c}$ is the particle velocity at the core boundary, $M_c=(4\pi/3)\rho_c r_c^3$ is the core mass.
Expressing the core radius as $x_c=(\bar\rho/3\rho_c)^{1/2}$, we obtain
\begin{equation}
x_c^2=\pi^2 \frac{\bar\rho}{m^4}\left (\frac{GM}{R}\right )^{-3/2}. \label{x_c}
\end{equation}
For example, for superdense clumps with the mass $M \sim 1\times 10^5$~g, the central density
$\bar\rho \sim 3\times 10^3$~g/cm$^3$, and
$R \sim 3$~cm, the core radius is $x_c
\sim10^{-11}$. We also note that
superdense dark matter objects that are kept in equilibrium
by dark matter Fermi-gas pressure were considered in [182].

%%%

\subsection{Numerical $N$-body simulations}

Presently, numerical $N$-body simulations already use arrays
of $N\sim10^9$ points, which allows achieving a mass resolution
of three orders of magnitude. The galaxy formation
modeling has shown that halos with simple scaling density
profiles are formed. For example, the NFW profile [120] is
expressed as
\begin{equation}
 \rho_{\rm H}(r)=
 \frac{\rho_{0}}{\left(r/R_s\right)\left(1+r/R_s\right)^2}.
 \label{halonfw}
\end{equation}
For the Galactic halo, $R_s=20$~kpc, the halo size (its virial
radius) is $R_h=200$\,kpc, and the mean halo density at the
distance $r=r_{\odot}=8.5$~kpc from the Galactic center (the
distance of the Sun from the center) is 
$\rho_{\rm h}(r_\odot)=0.3$\,GeV/cm$^3$. We also note that at distances from the center
up to $r=r_{\odot}$, baryonic matter dominates in the gravitational
potential. An alternative density profile was obtained in [183,
184]. This profile differs from the NFW one mainly in the
central part. In the Moore profile, $\rho_{\rm H}(r)\propto r^{-1.5}$.
Another popular representation of the dark matter halo
numerical simulations is the Einasto profile [185],
\begin{equation}
 \rho_{\rm H}(r)=
\rho_{0}\exp\left[-\frac{2}{\alpha}\left(\left(\frac{r}{r_s}\right)^\alpha-1\right)\right],
 \label{haloein}
\end{equation}
where $\alpha=0.16-0.3$ and $r_s\simeq20$~kpc, and the best phenomenological description of observational data is provided by
the Burkert profile [186] (see [79] for a detailed discussion).
The halo boundary is usually associated with the sphere of a
radius $R_{200}$ inside which the mean dark matter density
increases by a factor of 200 relative to the mean cosmological
density.

There have been attempts to explain the nature of the
universal density profile obtained in numerical simulations.
Some of them were mentioned above. In [187], the universal
density profile appearance in hierarchical clustering models
was explained by the competition of the dynamical friction
process and tidal stripping of smaller halos assembled in
larger ones. The dynamical friction tends to shift small dense
halos toward the center of a large halo, such that its central
density increases, and the tidal stripping destroys small halos
as they move toward the center, which increases the mass at
the periphery of a large halo.

In observations, the structure of the central parts of dark
matter halos has not yet been finally established. It remains
unknown how close to the center the density continues to
increase. There are some galaxy classes (LSB galaxies and
some dwarfs) that, according to dynamical models, demonstrate not a central density peak but a core with a slowly
increasing or even constant density. This discrepancy with
numerical modeling results can be successfully explained by
the entropy theory discussed in Section~4.5.

Numerical simulations, for example, Aquarius [188]
(extensive numerical simulations are sometimes given proper
names, such as Aquarius, Millenium, or Aquila) have revealed
the presence of substructures (clumps) on mass scales down
to $\sim10^6M_\odot$, which is the dynamical mass resolution of the
calculations. The obtained clump mass function is $\propto
M^{-1.9}$, and the fraction of clumped dark matter mass decreases from
the periphery to the center. Apparently, this is due to a more
effective tidal disruption of clumps in the central dense parts.
The annihilation or decay of dark matter particles in such
clumps can produce an inhomogeneous high-energy photon
sky background, but this effect has not yet been discovered
[63-65]. There is a problem of large-clump overproduction
here: according to numerical simulations, too many clumps
with dwarf-galaxy masses are formed, but astronomical
observations of the Local Group of galaxies have not found
the predicted number of dwarf galaxies. Presumably, the issue
can be resolved by assuming that the ``extra'' large clumps form
dark galaxies consisting mostly of dark matter, almost
without stars. Stars may not have been formed there due to
the absence of effective cooling mechanisms of low-mass
baryonic halos.

Many numerical simulations of large halos in galaxies and
galaxy clusters have been performed, but only a few studies
have dealt with low-mass clump modelling [44,51,77,82].
Numerical modelling of small-scale clumps usually stops
before the galaxy formation epoch. For example, in [51], the
calculations are restricted to the redshift interval $z\sim500-30$.
The point is that the modelling space region itself enters the
nonlinear regime at $z\sim30$ (this occurs quite rapidly, because
the perturbation spectrum at small scales is almost flat), after
which it is already difficult to follow the evolution of small-scale clumps inside this region. Numerical simulations have
revealed that the mass function of substellar-mass clumps is
close to the power-law form $\propto M^{-1.9}$, as obtained in
modelling large substructures in galaxies, and these mass
functions match quite well when extrapolating to intermediate scales.

The internal density profile in a single clump with a
minimal mass was numerically calculated in [77], which is an
important result. The obtained density profile can be
approximated by a power law $\propto r^{-\beta}$ with $\beta=1.5-2.0$,
which is in good agreement with the theoretical prediction
$\beta=1.7-1.8$ in [55-57]. In [82], the value $\beta=1.5$ was
obtained. The tendency of $\beta$ to increase in passing from
clumps formed via hierarchical clustering of small-scale
objects to clumps originated from isolated density perturbations was revealed in [51]. It was shown that in hierarchical
clustering, an NFW profile with $\beta=1$ at the center and a
density profile with $\beta\simeq1.4$ at the center are produced from
isolated perturbations. Clumps with the minimal mass $M_{\rm min}$
are also related to clumps formed from isolated perturbations
because the perturbation spectrum has a cut-off at smaller
masses (more precisely, the Gaussian distribution of perturbations should be taken into account, and therefore even
clumps with equal masses $M_{\rm min}$ can show different properties). In [51], from the standpoint of density profile formation,
the boundary between hierarchical and nonhierarchical
regimes was found to be at $\sim10^2M_{\rm min}$.

Numerical modelling of clumps on galaxy scales revealed
the dependence of the concentration parameter $C_{\rm NFW, 200}=R_{200}/r_s$ in Navarro-Frenk-White formula (45) on the halo
mass: $C_{\rm NFW, 200}=8,45\times(M/10^{12}M_\odot)^{-0.11}$. However, as
pointed out in [51], this dependence cannot be applied to
low-mass clumps. A more complicated non-power-law
dependence found in [189] can be used. If we ignore this fact
and use the mass function $\propto M^{-0.11}$, as has been assumed
in some papers, then the annihilation signal enhancement
by clumps can be overestimated by 2 to 3 orders of
magnitude [51].

In the calculations in [51], caustics similar to those in the
Bertschinger [76] and Gurevich--Zybin [55-57] models were
found in the formed clumps. According to the estimates
in [51], the presence of these caustics can increase the
annihilation luminosity of the clump immediately after
formation by a factor of more than 1.5 in comparison with
that of a clump with a smooth density profile. However, the
caustic enhancement of the annihilation signal decreases with
time, as the caustics are destroyed by dark matter stream
mixing inside the clumps.

We emphasize that the problem of the maximum central
clump density, or the problem of the core radius in numerical
modelling, has not been solved so far, apparently due to the
insufficient mass resolution. Simulations in [77] found a trace
of a core with $x_c\simeq10^{-2}$ (see Fig.~2 in [77]); however, in the
calculations in [51,82], the core was not resolved for $x_c$ as
small as $\simeq10^{-3}$.

%%%

\section{Clumps with minimal mass}

Although the spectrum of perturbations generated at the
inflation stage can be extended down to microscopic scales,
there are some effects that cut off (suppress) the low-scale part
of the spectrum at later epochs of cosmological evolution,
preventing low-mass clump formation. Many papers have
been devoted to the calculation of the minimum possible mass
$M_{\rm min}$, with results that differ by several orders of magnitude,
even using similar assumptions about the nature of dark
matter particles. However, to date, the problem of $M_{\rm min}$ has
largely been solved and uncertainties are not too large.

We illustrate the spectral cutoff using the ``free streaming''
effect. Cold dark matter particles at high temperatures
$T>T_f \sim 0,05 m_{\chi}$ are found in chemical equilibrium with
cosmic plasma, when the particle number density is determined solely by the temperature. After freezing at $t>t_f$ and
$T<T_f$, the dark matter particles stay in kinetic equilibrium
with the plasma for some time, as the particle gas temperature
$T_{\chi}$ follows the plasma temperature $T$, but their number in a
fixed volume remains constant. However, at this stage, the
dark matter particles are not fully connected with the plasma
any more. The exchange of momentum between the dark
matter particles and radiation leads to spatial diffusion of the
dark matter particles. They escape small-scale perturbations
via diffusion, and therefore the flattening (decay) of perturbations occurs at some mass scale $M_D$.

When the energy relaxation time $\tau_{\rm rel}$ of dark matter
particles decreases below the Hubble time $H^{-1}(t)$, the
particles lose equilibrium with plasma. These conditions
determine the kinetic decoupling time $t_d$. The kinetic decoupling time was estimated, for example, in [61]. At $t
\geq t_d$, the dark matter particles move in the free steaming regime, and
all perturbations are flattened on the scale
\begin{equation}
\lambda_{\rm fs}=a(t_0)\int_{t_d}^{t_0}\frac{v(t')dt'}{a(t')} \label{fs}
\end{equation}
(where $v(t)$ is the velocity of a dark matter particle) and on
smaller scales. For a nonrelativistic particle, $v(t)\propto1/a(t)$.
The corresponding minimal mass of a dark matter clump at
the time $t_0$ is
\begin{equation}
M_{\rm fs}=\frac{4\pi}{3}\rho_{\chi}(t_0)\lambda_{\rm fs}^3, \label{Mfs}
\end{equation}
which is much larger than $M_D$.

The choice of a mathematical formalism needed to
calculate the minimal mass depends on how close the
considered scales are to the cosmological horizon and
whether macroscopic motions of radiation (including deviations from the perfect fluid model) are important. On scales
much smaller than the horizon size, the nonrelativistic
Boltzmann equation is sufficient to describe the physics of
the phenomena if there are no such macroscopic motions. In
Sections~5.1-5.6, we perform nonrelativistic calculations, and
then show in which situations relativistic effects become
important and describe the results of relativistic calculations.

%%%

\subsection{Neutralino-lepton scattering cross-section}

To calculate theminimal mass $M_{\rm min}$ of a clump, the scattering
cross section of dark matter particles on cosmic plasma
particles is of fundamental importance. This cross section is
strongly model dependent. We consider the ``standard''
neutralino dark matter model in detail. This model is now
believed to be the most plausible one; however, other models
cannot be excluded. Eventually, the true model can be
established only after experimental registration of dark
matter by direct or indirect means or by producing dark
matter particles in accelerator experiments. We consider the
neutralino to be a pure bino ($\chi=\tilde B$).

The scattering cross section of left fermions on a
neutralino $f_L+\chi\to f_L+\chi$ by angles $\theta_{12}$ in the neutralino
rest frame is given by [39]
\begin{equation}
\left(\frac{d\sigma_{\rm el}}{d\Omega}\right)_{f_L\chi}= \frac{\alpha_{\rm
e.m.}^2}{8\cos^4\theta_{\rm W}} \frac{\omega^2(1+\cos\theta_{12})}{(m^2-\tilde m_L^2)^2},
\label{crosss1}
\end{equation}
where $\omega\gg m_f$ is the fermion $f_L$ energy in the neutralino rest
frame, $m$ is the neutralino mass, $\tilde m_L$ is the left fermion mass,
and $\alpha_{\rm e.m.}$ is the electromagnetic coupling constant. For the
scattering $f_R+\chi\to f_R+\chi$, we have
\begin{equation}
\left(\frac{d\sigma_{\rm el}}{d\Omega}\right)_{f_R\chi}=16 \left(\frac{d\sigma_{\rm
el}}{d\Omega}\right)_{f_L\chi}, \label{crosss2}
\end{equation}
under the condition that $m_L=\tilde m_R$, where $\tilde m_L$, and $\tilde m_R$ is the right
sfermion mass. We are interested in the scattering processes
$\nu+\chi\to\nu+\chi$ and $e+\chi\to e+\chi$. In the first case, the cross
section is given by (49), and in the second case, it is determined
by the sum of scatterings $f_L+\chi\to f_L+\chi$ and
$f_R+\chi\to f_R+\chi$, i.e., is larger than (49) by a factor of 17.
We use the notation $\tilde{m}$ for both left and right selectrons and
sfermions and $\tilde{M}^2=\tilde{m}^2-m_{\chi}^2$.

We note that for order-of-magnitude estimates, the
scattering cross section of neutralinos on leptons can be
written in the simple form $\sigma\approx T^2/M_{\sigma}^4$
s [62], where T is the
cosmic plasma temperature and $M_{\sigma}$ is of the order of the
electroweak interaction scale ($\sim100$~GeV).

%%%%%%%%%%%%%%%%%%%%%%%%%%%%%%%%%%%%%%%%

\subsection{Kinetic decoupling}

We use the kinetic equation formalism to study the process of
dark matter particles losing kinetic equilibrium with the
cosmic plasma. Following [128], we introduce the distribution function $f(x,p,t)$ in comoving coordinates $\vec x$ and
momenta $\vec p=ma^2\dot{\vec x}$ (the momentum of a freely moving
particle defined in this way is constant). The dark matter
particle density is
\begin{equation}
\rho(x,t)=\frac{m}{a^3}\int d^3pf(x,p,t)=\bar\rho_{\chi}(t)(1+\delta(x,t)) . \label{rhobb}
\end{equation}
The kinetic equation with the collision integral in the Fokker-Planck form [190] can be written as
\begin{equation}
\frac{\partial f}{\partial t} \!+\! \frac{p_i}{ma^2}\frac{\partial f}{\partial x_i} \!-\!
m\frac{\partial \phi}{\partial x_i}\frac{\partial f}{\partial p_i} \!=\! D_p(t)\frac{\partial
}{\partial p_i}\!\left(\frac{p_i}{mTa^2}f \!+\! \frac{\partial f}{\partial p_i}\right)\!,
\label{mainbb}
\end{equation}
where $\phi$ is the gravitational potential, which can be neglected
in the epochs $t\leq t_{\rm eq}$, $T(t)$ is the temperature of the surrounding plasma,
\begin{equation}
t=\frac{2,42}{\sqrt{g_*}}\left(\frac{T} {1\mbox{ MeV}}\right)^{-2}\mbox{
 s}, \label{ttime}
\end{equation}
where $g_*$ is the effective number of degrees of freedom, and
$D_p(t)$ is the diffusion coefficient in the momentum space. We
let
\begin{equation}
n_0(\omega)=\frac{1}{2\pi^2}\frac{\omega^2}{e^{\omega/T}\pm1},
\end{equation}
denote the number density of background relativistic
fermions or bosons with one polarization and with the
energy $\omega$. According to [190],
\begin{equation}
D_p(t)=\frac{g_f}{3} \int d\Omega\int d\omega\, n_0(\omega) \left(
\frac{d\sigma_{{el}}}{d\Omega} \right)_{f_L\chi} (\delta p)^2, \label{bebb}
\end{equation}
where the factor $g_f=40$ is obtained from counting the
degrees of freedom in neutralino-fermion collisions: three
neutrinos and antineutrinos (or $\nu_L^c$
in the case of a Majorana
neutrino) yield six degrees of freedom, $e_L$ and $e_L^c$
(which have
two degrees of freedom and two right (singlet) states for
electrons and positrons) yield 34 degrees of freedom because
their cross section is larger by a factor of 17. Equation (52)
with diffusion coefficient (55) coincides with Eqn (16) in [191],
up to a numerical factor of the order of unity in the expression
for $D_p$.

We consider the exit (decoupling) of neutralinos from
kinetic equilibrium in a homogeneous universe, when the
term $\partial/\partial x_i$ in (52) can be neglected. The temperature $T_{\chi}$ of the
neutralino gas is determined as
\begin{equation}
\int p_ip_jfd^3p=\bar\rho_{\chi}a^5T_{\chi}(t)\delta_{ij}.
\end{equation}
Multiplying both sides of (52) by $p_ip_j$ and integrating over
$d^3p$, we obtain
\begin{equation}
\frac{dT_{\chi}}{dt}+2\frac{\dot a}{a} T_{\chi} -\frac{2D_p(t)}{ma^2}\left(
1-\frac{T_{\chi}(t)}{T(t)} \right)=0, \label{ttbb}
\end{equation}
The initial condition $T_{\chi}(t_i)=T(t_i)$ for Eqn~(57) can be chosen
at the freezing time $t=t_f$, as in [191], or at any time $t_i$ from the
interval $t_f< t_i\ll t_d$, which is more convenient. The solution
of Eqn~(57) has the form
\begin{equation}
 \frac{T_{\chi}(t)}{T_d}\!=\!\frac{1}{\tau}\!\!\left(\!
 \tau_i^{-1/2}e^{1/4\tau^2-1/4\tau_i^2}\!+\!
 \frac{1}{2}e^{1/4\tau^2}\!\!\int\limits_{\tau_i}^{\tau}
 \!\!d^3xx^{-5/2}e^{1/4x^2} \!\right)\!, \label{soltau}
\end{equation}
where we introduce the dimensionless variable $\tau=t/t_d$ and
the notation
\begin{equation}
t_d\simeq10^{-3}\!\left(\frac{m}{100\mbox{GeV}}\right)^{-1/2} \!\left(\frac{\tilde
M}{0.2\mbox{~TeV}}\right)^{-2} \!\!\left(\frac{g_*}{10}\right)^{-3/4}\mbox{s}, \label{tsmd}
\end{equation}
\begin{equation}
T_d=30\left(\frac{m}{100\mbox{GeV}}\right)^{1/4} \left(\frac{\tilde M}{0.2\mbox{~TeV}}\right)
\left(\frac{g_*}{10}\right)^{1/8}\mbox{MeV}. \label{tbigd}
\end{equation}

The asymptotic form of solution (58) is $T_{\chi}/T_d=\tau^{-1/2}$ at
$\tau\ll1$ and $T_{\chi}/T_d=\tau^{-1}\Gamma(3/4)/2^{1/2}$ (where $\Gamma$ is the
gamma function) at at $\tau\gg1$ , as it must be. From solution (58),
we can see that the transition from the kinetic equilibrium of
neutralinos with relativistic fermions to the nonequilibrium
regime occurs very rapidly. Therefore, the treatment of
diffusion separately from free streaming seems to be justified.

The time $t_d$ and the temperature $T_d$ of the kinetic
decoupling of the neutralino can also be derived from the
simple condition
\begin{equation}
\frac{1}{\tau_{\rm rel}}\simeq H(t), \label{tauH}
\end{equation}
where $H(t)=1/(2t)$ is the Hubble parameter and $\tau_{\rm rel}(T)$ is
the energy relaxation time at the electron-neutrino gas
temperature $T$. The relaxation time $\tau_{\rm rel}$ is determined by the
scattering of neutralinos on the fermions $\nu_L,~~e_L$ and $e_R$. The
neutralino can be considered to be at rest because its rest-mass
frame coincides with the center-of-mass frame up to
$\sqrt{T/m_{\chi}}$.

Let $\delta p$ be the momentum acquired by the neutralino in one
scattering: $(\delta p)^2=2\omega^2(1-\cos\theta)$, where $\omega$ and $\theta$ are the
energy and the scattering angle of the fermion. Then the
relaxation time $\tau_{\rm rel}$ can be written as
\begin{equation}
\frac{1}{\tau_{\rm rel}}\!=\!\frac{1}{E_k}\frac{d E_k}{dt}\!= \!\frac{g_f}{2E_km}\! \int
\!\!d\Omega\int\!\! d\omega\, n_0(\omega) \!\! \left( \frac{d\sigma_{\rm el}}{d\Omega}
\right)_{f_L\chi}\!\!\!\! (\delta p)^2, \label{tau}
\end{equation}
where $E_k\simeq(3/2)T$ is the mean kinetic energy of the
neutralino, and $(d\sigma_{\rm
el}/d\Omega)_{f_L\chi}$ is given by (49). After integrating in (62), we arrive at
\begin{equation}
\frac{1}{\tau_{\rm rel}}=\frac{40\Gamma(7)\alpha_{\rm e.m.}^2} {9\pi\cos^4\theta_{\rm W}}
\frac{T^6}{\tilde M^4m}. \label{taurel}
\end{equation}
With (53), Eqn~(61) implies expressions (59) and (60).

%%%%%%%%%

  \subsection{Diffusion cut-off of the perturbation spectrum}

We consider Eqn (52) prior to the kinetic decoupling epoch,
i.e., for $t\ll t_d$. It is possible to find the first two moments by
integrating (52), first over $d^3p$ and then over $p_id^3p$. After
substituting the first of the obtained expressions into the
second one, we find an equation for the Fourier components:
\begin{equation}
\frac{\partial^2 \delta}{\partial^2 t}+ 2\frac{\dot a}{a}\frac{\partial \delta}{\partial t}
+D_p(t)\frac{1}{mTa^2}\frac{\partial \delta}{\partial t}= \frac{k_ik_j}{\bar\rho_{\chi}a^7m}
\int p_ip_jfd^3p. \label{main1}
\end{equation}
In the limit $\tau\ll 1$, we can neglect the first and second terms in
(64) to obtain the equation
\begin{equation}
\frac{\partial\delta(\vec x,t)}{\partial t}= \frac{D(t)}{a^2(t)}\Delta_{\vec x}\delta(\vec
x,t), \label{diff}
\end{equation}
with the diffusion coefficient
\begin{equation}
D=\frac{3\pi\cos^4\theta_{\rm W}\tilde M^4}{40\Gamma(6) \alpha_{\rm e.m.}^2T^5}. \label{ddd}
\end{equation}
The diffusion coefficient $D(t)$ depends on time via $T(t)$.

We find the minimal mass in the perturbation spectrum
caused by neutralino diffusion from the perturbation region.
The solution of (65) for the Fourier components has the form
\begin{equation}
\delta_{\vec k}(t)= \delta_{\vec k}(t_f)\exp\left\{ -k^2Cg_*^{5/4}\tilde M^4
\left(t^{5/2}-t_f^{5/2}\right) \right\}, \label{dkdif}
\end{equation}
where $C=const$. The factor $Cg_*^{5/4}\tilde{M}^4(t^{5/2}-t_f^{5/2})$ in front of
$k^2$ in (67) is the diffusion length square $\lambda_{\rm D}^2(t)/a^2(t)$  in
comoving coordinates. Then the minimal mass due to the
neutralino diffusion from the fluctuation region is
\begin{eqnarray}
M_{\rm D}\!&=&\!\frac{4\pi}{3}\rho_{\chi}(t_d)\lambda_{\rm D}^3(t_d) =5\times\!10^{-12}
\!\left( \frac{m}{100\mbox{ GeV}}\right)^{-15/8} \nonumber
\\
&\times&\!\! \!\!\left(\frac{\tilde M}{0.2\mbox{~TeV}}\right)^{-3/2}
\left(\frac{g_*}{10}\right)^{-15/16} M_{\odot}. \label{mdif}
\end{eqnarray}
The expression for functional dependence (67), obtained in
the diffusion approximation, coincides with the correspond-
ing expression obtained by other means in [191].

%%%%%%%%

  \subsection{Free streaming}

In the limit case $\tau\gg1$, i.e., after the kinetic decoupling,
Eqn~(52) takes the simple form
\begin{equation}
\frac{\partial f}{\partial t}+ \frac{p_i}{ma^2}\frac{\partial f}{\partial x_i}=0, \label{main2}
\end{equation}
Its solution in the Fourier space is
\begin{equation}
f\propto \exp\left[\frac{ik_jp_j}{ma_d}g(t)\right], \label{bnv}
\end{equation}
where
\begin{equation}
g(t)=a(t_d)\int \limits_{t_d}^{t}\frac{dt'}{a^2(t')}. \label{defg}
\end{equation}
Solution (70) is also valid with good accuracy for $\tau\geq1$,
because, according to (58), the kinetic decoupling proceeds
very rapidly. The momentum distribution at the decoupling
time is given by Maxwell distribution (38). Integrating the
product of (38) and (70) over $d^3p$, we obtain
\begin{equation}
n_{\vec k}(t) = n_{\vec k}(t_d) e^{-(1/2)k^2g^2(t)T_d/m}, \label{promnk}
\end{equation}
i.e., until the instant $t$, all perturbations are flattened due to
free steaming on the physical scale
\begin{equation}
\lambda_{\rm fs}(t)=a(t)g(t)\left(\frac{T_d}{m_{\chi}}\right)^{1/2}. \label{lambdafs}
\end{equation}
This scale corresponds to clumps with the minimal mass
\begin{equation}
M_{\rm fs}(t)=\frac{4\pi}{3} \rho_m(t)\lambda_{\rm fs}^3(t), \label{mfs}
\end{equation}
where $\rho_m(t)=\rho_{\rm eq}a_{\rm eq}^3/a^3(t)$. In the RD epoch, $M_{\rm fs}(t)$ increases logarithmically with time. This growth saturates at
the matter-dominated stage. The resultant mass $M_{\rm min}$ at the
instant $t_0$ can easily be found from the Friedmann equation:
\begin{equation}
M_{\rm min}\!=\!\frac{\pi^{1/4}}{2^{19/4}3^{1/4}} \frac{\rho_{\rm eq}^{1/4}t_d^{3/2}}{G^{3/4}}
\left(\frac{T_d}{m}\right)^{3/2}\!\!\!\ln^3\left\{\frac{24}{\pi G\rho_{\rm eq}t_d^2} \right\}.
\label{mminbukv}
\end{equation}
Using (59) and (60), we obtain
\begin{eqnarray}
M_{\rm min}&\simeq&2\times10^{-7}\left(\frac{m}{100\mbox{GeV}} \right)^{-15/8} \left(\frac{\tilde
M}{0.2 \mbox{~TeV}}\right)^{-3/2}
\nonumber\\
&&\times\left(\frac{g_*}{10}\right)^{-15/16} \left(\frac{\Lambda^*}{83}\right)^3 M_{\odot},
\label{mminnum}
\end{eqnarray}
where $\Lambda^*$ is the logarithm from Eqn~(75).

Thus, there are two processes of cosmological perturbation flattening in the neutralino gas. The first is neutralino
diffusion due to neutralino scattering on neutrinos, electrons,
and positrons. This process is effective as long as neutralinos
remain in kinetic equilibrium with the cosmic plasma. Before
the decoupling instant $t_d$, all perturbations with masses
$M<M_{\rm D}\simeq 10^{-13}-10^{-12}M_{\odot}$ are smoothened. The second
process is the free streaming of neutralinos. It starts at later
epochs at $t>t_d$, flattens larger perturbations with $M \leq M_{\rm fs}$,
and ultimately determines the minimum possible mass $M_{\rm min}$
[formula (76)] in the present-day mass function of clumps. We
note that the supersymmetry parameters are usually chosen in
the literature such that $M_{\rm min}$ is of the order of Earth's mass
($\sim10^{-6}M_{\odot}$).
Accurate calculations of the perturbation spectrum
transformation due to diffusion and free streaming, taking
GR corrections to the evolution equations into account, are
presented in [41,90].

%%%

\subsection{Cosmological horizon and acoustic oscillation effects}

In Sections 5.2-5.4, we assumed that all the considered scales,
including the free streaming scale, are much smaller than the
horizon size. This is indeed the case for a wide range of
supersymmetry parameters. However, the supersymmmetry
parameters that yield a low value of $T_d$ are often considered.
In this case, horizon effects, such as acoustic oscillation and
an effect similar to Silk decay, become important. These
effects cut off the perturbation spectrum at high masses and
thus become decisive for $M_{\rm min}$ calculations. Moreover,
according to the calculations in [46], the mass spectrum
cutoff may not be exponential, and the mass function can
increase in accordance with a power law, $\propto M^{-1/3}$, toward
low-mass clumps, although the integral contribution of
clumps described by this asymptotic behavior to the total
mass of clumps is small.

Qualitatively, the role of the cosmological horizon in
cutting off the perturbation spectrum can be represented as
follows. The evolution of perturbations with masses $M\ll M_d$
and $M\gg M_d$ differ greatly after the horizon crossing [62].
Perturbations $M\ll M_d$ run as sound waves. Such fluctuations do not produce a boost in the dark matter
peculiar velocity, and therefore perturbations in dark matter
do not grow logarithmically. After the kinetic decoupling,
their amplitude is ``frozen'' until the matter-dominating epoch
begins, and their evolution is described by the Meszaros
solution. Thus, at masses close to $M\sim M_d$, the perturbation
spectrum demonstrates a cutoff and is flattened at smaller
masses (or, possibly, has the form $\propto M^{-1/3}$ [46]).

In the opposite case $M\gg M_d$, the peculiar velocities take
the form vph $v_{pH}\simeq\delta_Hc/3$ immediately after the horizon crossing
[129]. Unlike thermal velocities, these peculiar velocities are
regular and directed toward the perturbation center. Adiabatic perturbations grow according to the law d$\delta\propto \ln(t)+const$ due to the peculiar velocity evolution, $v_{p}(t)\simeq
v_{pH}a(t_H)/a(t)$.

The effect of acoustic oscillations on dark matter density
perturbations near the horizon crossing time was studied in
[62]. As in the case of baryon acoustic oscillations, oscillations
near the horizon scale appear in the dark matter perturbation
spectrum.

The most detailed calculations of quasi-free streaming
with friction for neutralinos were performed in [46],
complementing the calculations in [62]. In [46], expressions
for the decoupling temperature and minimal mass were
obtained as
\begin{equation}
 \label{kindec}
  T_d=7.65\,C^{-1/4}g_{*}^{1/8}\left(\frac{m}{\hbox{100
    GeV}}\right)^{5/4}\ \hbox{MeV},
\end{equation}
\begin{equation}
 \label{mminbert}
 M_{\rm min}=
 7.59\times10^{-3}\,C^{3/4}\!\left(\frac{m\sqrt{g_{*}}}
    {\hbox{100 GeV}}\right)^{-15/4}\,M_\odot,
\end{equation}
with the dimensionless constant
\begin{equation}
 \label{Cneut}
  C=256\,(G_Fm_W^2)^2 \left(\frac{\tilde m^2}{m^2}
  -1\right)^{-2} \sum_{L}(b_L^4+c_L^4),
\end{equation}
where $G_{\rm F}$ is the Fermi coupling constant, $G_Fm_W^2=0.0754$,
$m_W$, $\tilde m$ and $m$ are the respective masses of the $W$-boson,
slepton, and neutralino, and $b_L$ and $c_L$ are the respective left
and right chiral coefficients; the number of degrees of
freedom at the decoupling time is $g_*=43/4$. The value of $C$
calculated in [39], which is related to the squared matrix
element for $l+\chi\to l+\chi$ scattering, differs from that in
Eqn~(78) by a factor of 1.6. According to (78), the kinetic
decoupling under typical supersymmetry parameters, as
assumed in [46], occurs at the temperature $T_d=22.6$~MeV,
i.\,e., after the muon-antimuon annihilation but before the
$e^+e^-$ annihilation epoch, and the $e^+e^-$ annihilation slightly
modifies the perturbation spectrum [46]. The mass in (78), as
in [62], is close to the mass inside the horizon at the decoupling
time by the order of magnitude. Modelling of the cosmic
plasma by a nonperfect fluid [61] gives another way to
calculate the minimum mass.

We note that in calculations like those in [46,62],
methods of the relativistic theory of density perturbation
evolution, taking the gravitational potential of radiation into
account as long as effects near the cosmological horizon are
considered.

%%%

\subsection{The $M_{\rm min}$ mass for superheavy neutralinos}

The assumption that dark matter was in chemical and thermal
equilibrium with radiation is not necessary and is invalid for
sufficiently heavy particles. In this case, the constraints on the
minimal halo mass considered in Sections 5.2-5.5 are absent.
If the equilibrium did occur, $M_{\rm min}$ would be very small for
superheavy particles due to the early kinematic decoupling of
clumps from the cosmic plasma [102].
Neutralinos in the superheavy supersymmetry proposed
in [22] were considered to be superheavy dark matter particle
candidates that interacted with radiation at early epochs.
Superheavy supersymmetry is a unique renormalizable model
that respects unitarity despite the particle masses that are
much larger than the electroweak scale. For example, a
neutralino with the mass $m=10^{11}$\,GeV$=1.78\times10^{-13}$~g
can be created gravitationally at the end of the inflation
stage to provide the value $\Omega_\chi h^2\approx 0.1$ derived from WMAP
observations.

In the conditions of decoupling, $\tau_{\rm rel}^{-1}\simeq H$, the running
coupling constant and mixing parameters at a temperature $T$
are used as obtained from the Standard Model,
$\sin^2\theta_W(T)=1/6+5\alpha(T)/[9\alpha_s(T))]$. For $M_{\rm SUSY}=10^{12}\,$GeV,
in the case of a bino and a Higgsino, we respectively find
$T_d \simeq2\times10^{11}$~GeV and $T_d \simeq2$~GeV. The dark matter mass
inside the horizon at these temperatures is respectively $M_d \simeq6\times10^{-12}$~g and $M_d
\simeq6\times10^{21}$~g. That is, in the case of the
bino, the mass $M_d$ is higher than the particle mass
$m\sim10^{11}$~GeV$=1.78\times10^{-13}$~g by only a factor of 34.

The free-streaming scale and mass $\lambda_{\rm fs}$ for superheavy dark
matter particles are very small. In the case of the bino, the
decoupling time is $t_d=7\times10^{-30}$~s and $ M_{\rm fs}\simeq
4.6\times10^{-11}\mbox{~g}$.
The latter quantity is larger than the particle mass by only a
factor of 260, and all masses of clumps starting from
$M \sim 260 m$ are possible. In the case of the Higgsino, $M_{\rm
fs}\ll m$, and free streaming plays no role in the perturbation evolution.
Hence, two mass scales, $M_d$ and $M_{\rm fs}$, could play the role of the
minimal clump mass $M_{\rm min}$. In the case of the bino, $M_{\rm fs}
> M_d$,
and the cutoff in the mass function starts at $M_{\rm min} \sim M_{\rm fs}$. In the
case of the Higgsino, $M_{\rm fs}$ is very small and $M_{\rm min} \sim M_d$.

%%%

\section{Formation of the clump mass function in early hierarchical clustering processes}

Dark matter objects with the minimal mass $M_{\rm min}$ were the
first gravitationally bound objects in the Universe. Clumps
with larger masses, which in the case of the usual power-law
spectra were formed at later epochs, consist of less massive
clumps and are themselves captured by larger clumps. Most
of the small-scale clumps that are present in larger and
typically more massive clumps (which we call `host' clumps)
are destroyed by tidal forces. Hierarchical clustering on small
scales is a rapid nonlinear process. The formation of new
clumps and their capture by more massive objects occur
almost simultaneously. Indeed, it follows from the form of
the transfer function $T(k)$ that the power-law exponent of the
perturbation spectrum on small scales is $n\approx n_s-4\approx-3$,
and the perturbation $\sigma(M)\propto M^{-(n+3)/6}$ is almost independent of $M$. This leads to a complication in $N$-body simulations [192]. The clumps are not fully virialized by the time they
are captured by the host clumps; therefore, the adiabatic
invariants do not prevent clumps from being destroyed at this
stage, since the singular density profile in the clumps had not
been formed by that time. The internal dynamical time in the
clump is of the same order of magnitude as the time of its
capture by the host clump.

The described picture of hierarchical clustering is valid
only when the perturbation spectrum has no sharp peak, for
example, is close to a power law. In peaked spectra, there are
long periods without clustering: clumps formed from the peak
region can be the only clump population for a long time, and
only after a long time delay does the clump formation from
spectral parts with larger scales begin and the clumps are
captured by these new structures. In Sections 6.1-6.3, we
consider the standard power-law primordial perturbation
spectrum and discuss the hierarchical clustering of clumps in
a wide mass range.

%%%

\subsection{Press--Schechter formalism}

The Press--Schechter theory is based on the spherical model
considered in Section 3.1. The spherical model yields the
clump formation threshold $\delta(t)\geq\delta_c$ (a more precise criterion
is given by the ellipsoidal collapse model), which allows
calculating the statistical characteristics of dark matter halos
being formed using perturbation properties at the linear
stage. The probability that a dark matter particle is in the
region with $\delta(t)\geq\delta_c$ (with flattening on the scale $M$) is
expressed as
\begin{eqnarray}
P(M)=\frac{1}{\sqrt{2\pi}\sigma(M)}\int\limits_{\delta_c}^\infty
d\delta'\exp\left(-\frac{\delta'^2}{2\sigma^2(M)}\right).
\end{eqnarray}
Then the differential number density of noncaptured (free)
clumps, i.e., those that are not included in large host clumps,
is given by the Press--Schechter formula [34]
\begin{eqnarray}
dn(t,M)&=&-2\frac{\bar\rho_0}{M}\frac{dP(M)}{dM}dM  \label{ps1}   \\
 &=&-\left(\frac{2}{\pi}\right)^{1/2}\frac{\bar\rho_0}{M\sigma(M)}
 \frac{d\sigma(M)}{dM}\nu e^{-\nu^2/2}dM\nonumber
\end{eqnarray}
where $\nu=\delta_c/\sigma(M)$. In the generalized Press--Schechter
theories [88] (see [33] for a very clear presentation), it was
possible to construct a powerful mathematical formalism and
to make many statistical predictions; in particular, the factor
2 in formula (81) was explained, which was not present in the
original theory [91].

From the standpoint of searching for annihilation
signatures, it is not only individual clumps with mass
function (81) that are of interest; so are clumps captured
inside other objects, in particular, inside the dark matter halo
of our Galaxy. In the first approximation, the mass function
of clumps in the halo at the halo formation time is given by
formula (81), with the clump number density increase in
proportion to the halo density growth taken into account.
This statement is valid in the case where a big difference exists
between the clump and halo masses, when the biasing\footnote{Biasing in astrophysics means the dependence of the properties of
galaxies being formed on the mean density of matter on scales greatly
exceeding the galaxy scales.} effects are small.

The initial mass function of clumps entering a large-scale
halo at the instant of formation was calculated more precisely
in [92] using the generalized Press--Schechter theory. Another
original approach to the calculation of this mass function was
used in [193], where the host halo was considered a part of a
closed universe and the Press--Schechter mass function was
calculated in this background. The Press--Schechter formalism was used in [194,195] to study substellar mass function
formation, and the authors of [195] concluded that effective
clump formation begins only for masses $M\gg M_{\rm
min}$. Those
approaches ignore small-scale clump destruction by tidal
forces, which modifies the initial mass function. To calculate
the mass function of clumps, including clumps inside other
structures, using only the statistical Press--Schechter theory is
insufficient, and the dynamical clump destruction processes
should be taken into account. The dynamical effects (tidal
stripping and dynamical friction) have been studied only for
large subhalo masses, $M\ge10^6M_{\odot}$ (see, e.g., [196]). For the
small-scale clumps considered here, a quite different treatment is required, because the hierarchical clustering of such
clumps proceeds very rapidly and on the same time scale that
their internal density profile is established.

%%%

\subsection{Tidal processes}

The tidal destruction of clumps is a complicated process that
depends on many factors: the clump formation history, the
density profile in the host clump, the presence of other
structures in the host clump, the orbital parameters of
individual clumps in the host clump, and so on. These factors
can be sufficiently fully taken into account in numerical
modeling only; nevertheless, it is possible to make simple
analytic estimates.

Let us consider a host clump with mass $M_h$, radius $R_h$,
and some distribution of smaller clumps inside it. These small
clumps move in the common gravitational potential with the
velocity dispersion$\sim V_h\simeq (GM_h/R_h)^{1/2}$. By tidally
interacting with the environment, a small clump is `shaken'
and its internal energy (the kinetic energy of dark matter
particles) increases.
We first consider the interaction of the clump with
another, target clump when the first flies with an impact
parameter $l$ by the target. The target clump is characterized by
a mass $M'$, a radius $R'$, the core radius $R'_c=x_cR$, and some
internal density distribution.

The internal energy increase of a clump with mass $M$
during one collision in the momentum approximation [197] is
expressed as
\begin{equation}
\Delta E=\frac{1}{2}\int d^3r\,\rho_{\rm int}(r)(v_x-\tilde v_x)^2, \label{tiddeier}
\end{equation}
where $v_x$ is the dark matter particle velocity increase along the
$x$ axis and $\tilde v_x$ is the same quantity for the center of mass of the
clumps. It is easy to obtain
\begin{equation}
v_x=\frac{2GM'}{v_{\rm rel}R'}g(y),\label{vx2GM}
\end{equation}
where $y=l/R'$, and the function $g(y)$ depends on their internal structure [39]. The
rate of the internal energy increase due to collisions of the
selected clump with all other clumps is determined as
\begin{equation}
\dot E=\int 2\pi l v_{\rm rel}dl \int dM'\psi(M',t)\Delta E, \label{dee1ier}
\end{equation}
assuming that the mass function $\psi(M',t)$ inside the host
clump is known.

As another process that can be responsible for tidal
destruction, we consider clump interaction with the common
gravitational potential of the host clump. The energy increase
per unit mass at a distance $r$ from the barycenter of the
selected small clump over the time of one periastron passage
(if the orbit can be treated as elliptical) is [197]
\begin{equation}
\langle E_p\rangle=\frac{GM_h}{R_h^3}r^2 \left(\frac{R_h}{R_p}\right)^{\beta} \chi_{\rm
ecc}(e)A(\omega\tau),
\end{equation}
where $e$ is the orbital eccentricity, $R_p$ is the periastron
distance, the function $\chi_{\rm ecc}$ is presented in [197], and the
adiabatic correction is $A(x)=(1+x^2)^{-\gamma}$, $\gamma\simeq2.5 - 3$. The
clump energy increase over one period $T_{\rm orb}$ has the form
$\Delta E=\int\langle E_p\rangle\rho_{\rm int}(r)d^3r$, and the energy rate increase is
\begin{equation}
\dot E=\frac{2\Delta E}{T_{\rm orb}}. \label{deorbier}
\end{equation}

As shown in Section~6.3, clumps are tidally destroyed with
a high probability, and the survival probability of each clump
is $\xi\ll 1$. During the rest lifetime, the surviving clumps are
surrounded by other clumps with the mass function $\psi(M',t)$.
When the host clump is destroyed, the selected small-scale
clump under consideration turns out to be inside a clump with
a larger mass and with a small-scale clump distribution
$\psi(M',t)$, but with a larger time $t$. The characteristic
formation time of the minimal host clump is close to $t_f$, and
the destruction time can be much longer than $t_f$. The energy
increase rate due to both mentioned processes is given by the
sum of (84) and (86).

Above, we sketched the ideas of more precise calculations,
which were carried out by extracting individual processes
from the general complex picture of tidal destructions.
However, real processes are so chaotic that it is worth using
a simplified approach by assuming that for a clump with mass
$M$ and radius $R$, after the typical gravitational impact, the
internal energy increases by the value
\begin{equation}
 \Delta E\sim\frac{4\pi}{3}G\rho_hMR^2,
 \label{dele}
\end{equation}
where $\rho_h$ is the mean density of the host clump.

%%%

\subsection{Hierarchical clustering taking destructions
into account}

The process of transition of a clump from one host clump to
another occurs almost continuously in time until the last host
clump is formed, in which tidal destructions become ineffective. The survival probability of the clump is defined as the
fraction of clumps with mass $M$ that survived tidal destructions during the hierarchical clustering. The first host clump
mostly contributes to the tidal destruction of the selected
clump, especially if their densities are similar. With the
dynamical effects in the Press--Schechter theory taken into
account, the distribution function of clumps that avoided
destruction was calculated in [83] as
\begin{equation}
 \xi\,\frac{dM}{M}\,d\nu\simeq
 \frac{\nu\,d\nu}{\sqrt{2\pi}}\,e^{-\nu^2/2}
 f_1\frac{d\log\sigma_{\rm eq}(M)}{dM}\,dM,
 \label{psiitog}
\end{equation}
where $f_1\simeq0.2-0.3$.
Qualitatively, the first factor $\nu$ in (88) corresponds to the
fact that the clumps that were formed from high peaks (with
higher values of $\nu$) are more resistive to the destruction
processes than those formed from low peaks (with smaller
$\nu$). After integrating (88) over $\nu$, we obtain
\begin{equation}
 \xi_{\rm int}\frac{dM}{M}\simeq0,02(n+3)\,\frac{dM}{M}.
 \label{xitot}
\end{equation}
The effective power-law exponent in (89) is $n=-3(1+2\partial\ln\sigma_{\rm eq}(M)/\partial\ln M)$, which very weakly depends
on $M$. Equation (89) assumes that for the typical $n$, only a
small fraction of clumps (about $0.1-0.5$~\%) survive the
hierarchical tidal destruction in each logarithmic mass
interval $\Delta\ln M\sim1$. It should be noted that the physical
meaning of the survived clump mass function $\xi\,dM/M$ is
different from that of free clumps described by the Press--Schechter formula $\partial F/\partial M$.

Substellar-mass clump formation was studied in numerical simulations in [44, 77]. The form of the differential clump
number density $n(M)\,dM \propto dM/M^2$ turns out to be close to
that obtained for large clumps with masses $M\geq10^6M_{\odot}$. The
simple law, $M^{-1}$, of mass function (89) is in good agreement
with the results of numerical simulation in [77], up to a
normalization coefficient of the order of unity. Good
agreement when extrapolating our calculations to the data
of numerical modeling of large clumps with $M\ge10^6M_\odot$ [43]
must also be noted.

At later times, mass function (89) is transformed due to
tidal interactions with stars of the Galaxy; these processes are
considered in Section 7.

%%%%

\section{Destruction of clumps in the Galaxy}

The problem of the mechanisms and efficiency of destruction
of substructures in galaxies was previously discussed in [198],
and for large (galactic scale) dark matter subhalos, it was
addressed later in many analytic and numerical studies.
Similar studies of small-scale clumps started only recently.
Clumps with distribution function (88) joined the Galaxy at
the time of its formation, and later these clumps lost mass and
were partially or fully destroyed due to tidal interactions with
stars or with the collective field of the disc.

In many papers (see, e.g. [43,48]), a simplified criterion of
the clump tidal destruction was used. Namely, a clump is
Passumed to be destroyed if its total internal energy increase $\sum(\Delta E)_j$ after several disc crossings (or tidal interactions with
stars) becomes of the same order as the initial clump binding
energy $|E|$, i.e.,
\begin{equation}
 \sum\limits_j(\Delta E)_j\sim|E|,
 \label{prevcrit}
\end{equation}
where the summation is made over successive tidal interaction
events. This criterion is valid at the early stages of hierarchical
clustering in the clump formation epoch, because the clump
density profile formation and tidal impacts occur in one epoch,
and the clumps in which the density profiles have not yet been
formed and which have no cores are mostly being destroyed. In
the case of clump destruction in the Galaxy in later epochs, the
gradual mass loss in each of the tidal interactions [199-201]
should be taken into account; in particular, this was noted for
small-scale clumps in [40, 202, 203].

Condition (40) implies that the clump has lost most of its
mass. This simple criterion is useful in calculations if only the
main mass loss is of interest. However, to calculate dark
matter particle annihilation, it is important to know what the
clump remnant is, because the surviving clump cores produce
approximately the same annihilation signal as the original
clumps. The main difficulty in exactly solving the problem is
taking the mass loss in the inner regions of clumps into
account during their complicated dynamical restructuring
immediately after the tidal impacts.

%%%

\subsection{Clump destruction by the disc field}

We consider clumps that move in orbits in the galactic halo
and cross its stellar disc. The Galaxy disc has the surface
density
\begin{equation}
 \sigma_s(r)=\frac{M_d}{2\pi r_0^2}\,e^{-r/r_0},
 \label{diskmass}
\end{equation}
where $M_d=8\times10^{10}M_\odot$ and $r_0=4.5$~kpc. The increase in
the dark matter particle energy relative to the clump center
energy after a single disc crossing is expressed as [204]
\begin{equation}
 \delta E=\frac{4g_m^2(\Delta z)^2m}{v_{z,c}^2}A(a),
 \label{egain}
\end{equation}
where $\Delta z$ is the vertical distance (across the disc plane) from
the dark matter particle to the clump center, $v_{z,c}$ is the vertical
clump velocity relative to the disc at the disc crossing time,
$A(a)$ is the adiabatic correction, and the gravitational
acceleration above the disc is $g_m(r)=2\pi G\sigma_s(r)$. The factor
$A(a)$ in (92) describes the effect of adiabatic protection from
slowly varying tidal forces [205]. This correction is given by
the extra factor $A(a)$ in the energy increase calculated in
the momentum approximation. In [199], the formula
$A(a)=(1+a^2)^{-3/2}$ was proposed. Here, the adiabatic
parameter is determined as $a=\omega\tau_d$, where $\omega$ is the orbital
frequency of the dark matter particle on the clump and
$\tau_d\simeq H_d/v_{z,c}$ is the effective duration (fly-by time) of the
gravitational impact produced by a disc with the half-thickness $H_d$. For the tidal interaction of the clump with
bulge and halo stars, the gravitational collision duration is
estimated as $\tau_s\sim l/v_{\rm rel}$, where $l$ is the impact parameter and
$v_{\rm rel}$ is the relative velocity of the clump and the star.

The internal density profile $\rho_{\rm int}(r)$ and the particle energy
distribution function in the clump $f_{\rm cl}(\varepsilon)$ (per unit mass) in the
model with an isotropic velocity dispersion are related by the
integral formula [206]
\begin{equation}
 \rho_{\rm int}(r)=2^{5/2}\pi\int\limits_{\psi(r)}^{0}
 \sqrt{\varepsilon-\psi(r)}\,f_{\rm
 cl}(\varepsilon)\,d\varepsilon,
 \label{ferho}
\end{equation}
The simplest models assume the isothermal distribution
function $f_{\rm
cl}(\varepsilon)\propto\exp(-2\varepsilon)$.

The energy addition $\delta\varepsilon$ from the tidal interaction leads to
the particle stripping off with energies in the range
$-\delta\varepsilon<\varepsilon<0$, and the clump density change at a radius $r$ is
expressed as [83]
\begin{equation}
\delta \rho(r)=2^{5/2}\pi\int\limits_{-\delta\varepsilon}^{0}
 \sqrt{\varepsilon-\psi(r)}\,f_{\rm cl}(\varepsilon)\,d\varepsilon.
 \label{ferho2}
\end{equation}
The adiabatic correction enters this expression via $\delta\varepsilon$ in the
lower integration limit, according to formula (92). The clump
mass loss in a single disc crossing event is
\begin{equation}
\delta M=-4\pi\int_0^R r^2\delta\rho(r)\,dr.
\end{equation}

We use NFW profile (45) in the Galaxy halo. The relation
between the density profile $\tilde\rho_H(x)$ and the distribution function
is given by the same formula (93) with the substitution
$f_{\rm cl}\Rightarrow F(\varepsilon)$. The distribution function $F(\varepsilon)$ for halo profile
(45) is presented in [207], and the clump orbit distribution in
the halo can be expressed in terms of it [83]. Further, by
choosing the time interval $\Delta T$ much longer than the clump
orbital period $T_t$ but shorter than the Galaxy age $t_0$,
$T_t\ll\Delta T\ll t_0$, it is possible to determine the mean mass loss
by the selected clump due to tidal impacts in successive disc
crossings:
\begin{equation}
\frac{1}{M}\left(\frac{dM}{dt}\right)_d\simeq\frac{1}{\Delta T}\sum \left(\frac{\delta
M}{M}\right)_d,
 \label{deriv2}
\end{equation}
where the summation ranges all successive disc crossing
points.

The destruction of large subhalos with masses
$M\geq10^7M_\odot$ in their disc crossings was studied in numerical
simulations in [208]. Large clumps, unlike the small-scale
clumps considered here, can significantly affect the stellar
kinematics in the disc.

%%%%%%%%%%%%

\subsection{Clump destruction by stars}

The calculation of the clump internal energy increase during a
single stellar encounter with an impact parameter $l$ is similar
to the one carried out in Section~6.2 (the star plays the role of
the host clump with the mass $M'=m_*$). By integrating (82)
over the clump volume with the density profile $\rho_{\rm
int}(r)$ from (32), for $l>R$, we obtain the internal energy increase in the
form
\begin{equation}
 \Delta E= \frac{2(3-\beta)}{3(5-\beta)}
 \frac{G^2MR^2m_*^2}{v^2_{\rm rel}l^4}.
 \label{destar5}
\end{equation}
The opposite case $l<R$ is considered, for example, in [39]. It
is easy to verify that the minimal increase in the internal
energy occurs during a tangential stellar fly-by at the distance
$l\simeq R$ from the clump center.

Depending on the impact parameter value, for different
stellar collisions of the clump, there can be two modes of
clump destruction [43]: (1) in one stellar encounter or (2) after
many tidal interactions with stars. The clump internal energy
increase rate is
\begin{equation}
 \dot E=2\pi
 \int \Delta E(l)\,n_*v_{\rm rel}\,l\,dl,
 \label{dee1}
\end{equation}
where $n_*$ is the stellar number density.

\begin{figure}[t]
\begin{center}
\includegraphics[angle=0,width=0.45\textwidth]{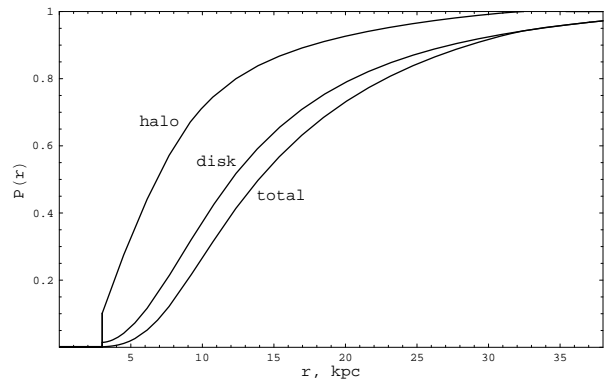}
\end{center}
\caption{Fraction of clumps with the mass $M=10^{-6}M_{\odot}$ and the peak
value $\nu=2$ that were not tidally destroyed by the galactic disc, $P_d$, by
galactic halo stars, $P_h$, and the total fraction, $P_{\rm tot}=P_h P_d$, as
functions of the distance to the galactic center. The rough clump tidal
destruction criterion (90) was used. The cut-off at $r<3$~kpc is due to
complete clump disintegration in the galactic bulge.}
 \label{figdtreal2}
\end{figure}

Using the results in [209], we approximate the radial
stellar number density in the bulge at distances $r=1-3$~kpc as
\begin{equation}
 n_{b,*}(r)=(\rho_b/m_*)\exp\left[ -(r/r_b)^{1,6}\right],
 \label{rhoe}
\end{equation}
where $\rho_b=8M_\odot/$pc$^3$, $m_*=0.4M_\odot$ and $r_b=1$~kpc. The
corresponding stellar number density in the halo at $r>3$~kpc
outside the galactic disc can be approximated as [210]
\begin{equation}
 n_{h,*}(r)=(\rho_h/m_*) (r_{\odot}/r)^{3},
 \label{rhosh}
\end{equation}
where $\rho_h=1.4\times10^{-5}~M_\odot/$pc$^3$ and $r_{\odot}=8.5$~kpc. In these
calculations, we neglect the dark matter halo non-sphericity
and the stellar halo flatness [210].

The rough criterion in (90), which characterizes the clump
mass loss, yields the clump survival probability as shown in
Fig.~7. We see that the disc makes the largest contribution to
the clump destruction, and there are almost no clumps in the
Galaxy bulge. The sharp jump in the figure is due to the use of
the bulge model with a sharp boundary.

Clump destruction by stars was also studied in [47,203,211], by numerical calculations in particular.

In some papers, the disc is regarded as a collection of stars,
and the tidal interaction is calculated in the same way as when
the clump flies through an infinite medium consisting of
point-like masses (stars), but during the finite time intervals
that the clump crosses the disc. It can be shown that in typical
January 2014 Small-scale clumps of dark matter 23
cases, the collective field of the disc gives the leading
contribution to the clump destruction. The clump internal
energy increase in one disc crossing due to interaction with
individual stars relative to that caused by the gravitational
impact with the collective disc field is
\begin{equation}
\frac{\Delta E^{\rm stars}}{\Delta E^{\rm
disk}}\sim0.6\left(\frac{R}{0.015\mbox{~pc}}\right)^{-2}e^{(r-r_{\odot})/r_0}\cos\theta,
\label{fracstardisk}
\end{equation}
where $r_0=4.5$~kpc and $R=0.015$~pc is the radius of an Earth-mass clump, $3\times10^{-6}M_{\odot}$, formed from the peak with $\nu=2$ in
the case of the density perturbation spectrum with $n_s=0.96$.
That is, only for clumps with minimal masses at $r\ge r_{\odot}$ are
both contributions comparable, and when the mass and,
accordingly, the radius of the clump increase, the collective
field effect becomes dominant.

%%%%

\subsection{Remnants of clumps}

We now turn to the question of clump core survival. We
describe the clump destruction processes in the disc, bulge,
and halo by a single equation for the clump mass loss,
\begin{equation}
 \frac{dM}{dt}=
 \left(\frac{dM}{dt}\right)_d+\left(\frac{dM}{dt}\right)_s
 \label{mmtmain}.
\end{equation}
In [82], Eqn~(102) was solved numerically for the time $t_0-t_G$
elapsed from the galaxy formation epoch until the present
instant $t_0$.

The dark matter particle annihilation is the most
important observational signature of the clumps. The dominance of the central core in the clump annihilation signal is
critical for clumps with sufficiently sharp density profiles.
Namely, the central core dominates in the annihilation signal
in the clump with power-law density profile (32) if $\beta>3/2$
and $x_c=R_c/R\ll1$. More precisely, the value $\dot N\propto\int_{r_0}^{r}4\pi r'^2dr'\rho_{\rm int}^2(r')$ is virtually independent of $r$ if $r \gg r_0$. As
a result, the annihilation luminosity of a clump with a
nearly isothermal density profile ($\beta\simeq 2$) is almost constant
during the tidal stripping process until the clump radius
becomes as small as the core radius. In other words, in the
modern Galaxy, the clump remnants stripped by tidal
interactions with $x_c<\mu(t_0)\ll1$, where $\mu(t)=M(t)/M_i$,
$t_0\simeq10^{10}$~yrs is the age of the Galaxy, satisfy Eqn~(102)
and have the same annihilation luminosity as their progenitors at $\mu=1$.

\begin{figure}[t]
\begin{center}
\includegraphics[angle=0,width=0.45\textwidth]{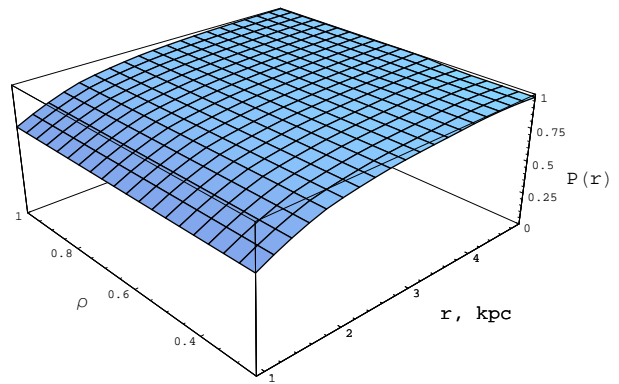}
\end{center}
\caption{Survival probability $P(r,\rho)$ as a function of the distance to the
galactic center $r$ and the mean inner clump density $\rho$ for $x_c=R_c/R=0.05$. This yields the normalized fraction of clumps in the halo $P$ that escaped tidal destructions by the disc and halo. The clump density is normalized to $7.3\times10^{-23}$~g~cm$^{-3}$ of clumps with the mass $M=10^{-6}M_\odot$ that originated from $2\sigma$ peaks, assuming a power-law perturbation spectrum with the
exponent $n_s=1$.}
 \label{ani1f}
\end{figure}

Using the solution of Eqn~(102), we can calculate the
clump survival probability $P$ over the age of the Galaxy.
Calculations show that for clumps with $x_c\ll 0.05$, $P\sim1$
everywhere (see Fig.~8). Clumps from outer regions can even
fly inside the bulge. This means that the clump remnants
mostly survive in the tidal interactions in the Galaxy. The
probability decreases, $P<1$, near the galactic center for
clumps with $x_c>0.05$. This can be easily understood
because as $x_c\to1$, we return to the old clump destruction
criterion (90) with the corresponding survival probabilities
[43,48].

The aggregate surviving clump fraction, taking early
hierarchical clustering destructions and destructions in the
halo into account, is shown in Fig.~9 as a function of the clump
mass and the perturbation spectral exponent for typical
clumps produced by perturbations with $\nu=2$.

\begin{figure}[t]
\begin{center}
\includegraphics[angle=0,width=0.45\textwidth]{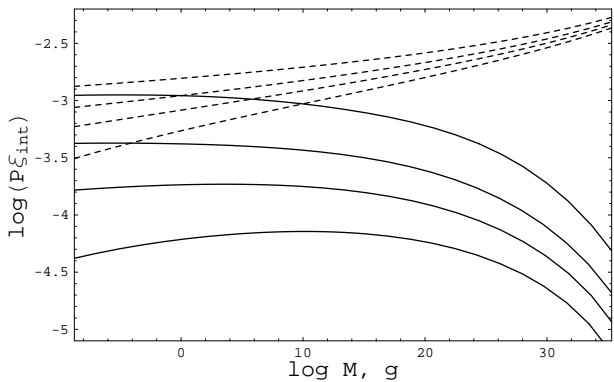}
\end{center}
\caption{Initial fraction (dashed curves) and the present-day fraction
(solid curves) of the surviving dark matter clumps in a unit logarithmic
mass interval $\delta M\sim M$ as a function of the clump mass $M$ for the spectral exponents $n_s=0.949$, $0.963$, $0.977$
and $1$  (from top to down).} \label{gr2}
\end{figure}

The clump mass function in the Galaxy is modified due to
the gradual mass loss by the clumps. According to the
theoretical model in [39] and numerical modeling in [77], the
differential number density of small-scale clumps in the
comoving frame $n(M)\,dM \propto dM/M^2$. This distribution is
shown in Fig.~10 by the solid line. In the case of flattening of
small-scale perturbations with $M<M_{\rm min}$, the additional
factor $\exp[-(M/M_{\min})^{2/3}]$ arises, which is responsible for
the perturbation decay at low masses $M$.

\begin{figure}[t]
\begin{center}
\includegraphics[angle=0,width=0.45\textwidth]{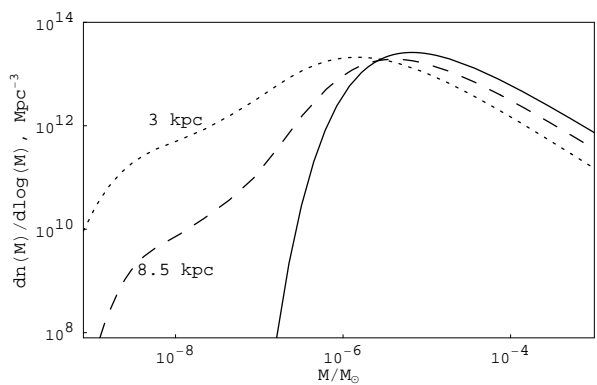}
\end{center}
\caption{The modified clump remnant mass function calculated
numerically at the galactocentric distances $3$ and $8.5$~kpc. The solid
curve shows the initial mass function.}  \label{mf1}
\end{figure}

Due to the disc and stellar destructions, the initial
(cosmological) clump mass function is transformed into the
final (modern) clump mass function, which is shown in Fig.~10
for two distances from the galactic center. The calculations
assume the clump core radius to be very small, such that all
clump remnant masses are available. With a finite core radius,
the mass function would be cut off at the clump core masses.
It can be seen from Fig. 10 that the clump remnants exist for
$\sim M_{\rm min}$. In the bulge near the galactic center, the clump
remnants are more numerous because of intensive clump
destructions in high-density stellar regions relative to the
rarefied stellar halo. The leading contribution to the formation of the mass function tail at low masses is provided by
clumps with orbits close to the disc, which suffer stronger
destructions. Another important point is the total destruction
of clumps whose orbits lie entirely within the bulge. Nevertheless, the clump number density in the bulge is nonzero
because most of the clumps have orbits extending beyond the
bulge. These ``transit'' clumps spend only a small fraction of
their orbital periods inside the bulge and thus avoid total
destruction.

%%%%%%%%%%%%%%%%%%%%%%%%%%%%%%%%%%%%%%%%%%%%%%%%%%%%%%%%%%%%%%%%%%%%

\section{Particle annihilation in clumps}

Particle-antiparticle interaction can result in their annihilation, accompanied by the rest-mass energy transformation
into other particles. For example, if photons are present
among the annihilation products, they can in principle be
registered by gamma telescopes, which provides information
on dark matter particles. Unlike charged particles, photons
are not deflected by cosmic magnetic fields; therefore, they
should be registered from the directions to the sources, which
can facilitate their identification.

The local annihilation rate is proportional to the particle
number density squared, and is therefore higher in the central
parts of dark matter halos. Generally, if any inhomogeneities
are present, $\bar{\rho^2}>(\bar{\rho})^2$; therefore, inhomogeneities always
enhance annihilation signals. An inhomogeneity can appear
as a smooth central density increase [for example, according
to profiles (45) and (46)] or as a collection of chaotic small-scale inhomogeneities on top of this smooth regular density
peak. Dark matter in the smooth profile is frequently referred
to as the diffuse halo component. The particle number density
in clumps is much higher than in the diffuse component, and
hence the clump annihilation signal in many models dominates over the diffuse signal (usually except in the most dense
central core of the halo), even if the clumps contain a small
fraction of the total halo dark matter.

Recently, an excess of gamma rays from the central parts
of the Galaxy has been reported, which cannot be explained
by the radiation of ordinary astrophysical sources and can be
related to dark matter annihilation. However, these results
have not been reliably confirmed, primarily due to uncertain
astrophysical backgrounds. Nevertheless, even an upper
bound on the annihilation radiation flux can provide invaluable information on possible dark matter particle parameters.
As far as we know, the idea to use such gamma-ray background constraints was first proposed in [212].

The annihilation rate (the number of annihilated particles,
which is equal to twice the number of annihilation events) in a
single clump is given by
\begin{equation}
\dot N_{\rm cl}\!=\!4\pi\int\limits_{0}^{\infty}\!r^2dr\rho_{\rm int}^2(r)
m^{-2}\langle\sigma_{\rm ann} v\rangle\! =\! \frac{3}{4\pi} \frac{\langle\sigma_{\rm ann}
v\rangle}{m^2} \frac{M^2}{R^3}\,S, \label{separ}
\end{equation}
where $v$ is the relative velocity of two dark matter particles
with mass $m$, and $\sigma_{\rm ann}$ is the annihilation cross section. The
function $S$ in (103) depends on the dark matter distribution
inside the clump. In particular, $S=1$ in the simplest case
where the clump has a constant density, i.e., $\rho_{\rm int}(r)=const$ at
$r\leq R$ and $\rho_{\rm
int}(r)=0$ at $r>R$, and $S\simeq 4/(9x_c)$ for the
isothermal profile $\rho\propto r^{-2}$ with a small core, $x_c\ll 1$.

In expression (103), Majorana particles are assumed, i.e.,
those that are identical to their own antiparticles, $\bar\chi=\chi$, as is
the case for neutralinos. For Dirac particles, $\rho_{\rm int}$ would mean
the sum of densities of particles and antiparticles, and the
extra factor $1/2$ should be incorporated in (103).

\subsection{Cross sections and spectra of neutralino annihilation products}

The quantity $\langle\sigma_{\rm ann} v\rangle$ can be expressed in terms of the relative
velocity $v$ of two dark matter particles as
\begin{equation}
\langle\sigma_{\rm ann}v\rangle= a+bv^2+cv^4+... \label{cross-sect}
\end{equation}
where $a$ includes only the $s$-wave contribution, while b
includes contributions from both the $s$- and $p$-waves. For
small velocities $v\ll1$, only the first term can be considered.

In the expression for $\langle\sigma_{\rm ann} v\rangle$, averaging over the thermal
particle distribution is assumed. As a rule, thermal distributions are realized in both the early Universe and the typical
dark matter halo models. Therefore, the expression for
$\langle\sigma_{\rm ann} v\rangle$ is sufficiently universal, and it can be used to connect
particle creation in the early cosmological epochs and their
annihilation rate in the modern Universe. In the case of a
strongly anisotropic or nonthermal distribution function, an
extra factor of the order of unity appears in the expression for
$\langle\sigma_{\rm ann} v\rangle$.

The most probable annihilation channel is into quarks
with the subsequent hadronization of the reaction products
and the formation of an almost universal continuum
annihilation product spectrum. Gamma quanta in this case
are generated basically due to neutral pi-meson decays:
\begin{eqnarray}
x+{\bar x} \to \pi^0+\mbox{all}, \quad \pi^0 \to \gamma+\gamma, \nonumber
\end{eqnarray}
where ``all'' means all other annihilation products. The
fragmentation function $dN_i/dx$ for the differential particle
number density of type $i$ generated in a single annihilation
event with the energy $E=xm$ is calculated in [213,214], where
various analytic approximations are also presented. An
important role in the formation of the annihilation spectrum
can be played by the intrinsic bremssrahlung radiation from
charged particles that can be present among the intermediate
annihilation products. In specific calculations, the Dark-SUSY computer package [215] is frequently used, which
allows obtaining the cross sections and spectra of the
annihilation products in different channels under different
assumptions about the supersymmetry model parameters.

Some supersymmetry models allow neutralino annihilation predominantly into the lepton channel. These models
were especially invoked in connection with data of the
PAMELA (Payload for Antimatter Matter Exploration and
Light-nuclei in Astrophysics) experiment, which is discussed
in detail in Section~9.

In most of the simplest models, the main annihilation
signal is generated in the continuum, and spectral lines, even if
they are present, are suppressed. Nevertheless, the generation
of lines, even if of a very low intensity, is possible by
neutralino annihilation into two photons, into a photon and
a $Z^0$ boson, or into a photon and a Higgs boson. The
discovery of a narrow spectral line would provide the unique
opportunity to extract the signal from the background,
because the usual astrophysical sources can hardly generate
high-energy spectral lines. In this connection, of great interest
have been the recent claims of observations of a possible line
at 130~GeV by the Fermi-LAT (Fermi Large Area Telescope)
instrument from the central region of the Galaxy [216, 217], as
well as a possible indication of the presence of this line from
galaxy clusters [218] and from some unidentified Fermi-LAT
sources (see a detailed discussion of these issues, which are
highly relevant today, in [219]). Intriguing is the $\simeq1.5^\circ$ shift
of the maximum of the emission from the dynamical center of
the Galaxy. However, the shift is difficult to accommodate in
the annihilation scenario, because the off-center dark matter
cusp location would lead to its tidal destruction in interactions with the baryonic matter cusp [220].

We now consider the Sommerfeld enhancement effect,
which can significantly boost the particle annihilation cross
section at low relative particle velocities, especially inside
clumps. The Sommerfeld enhancement became very relevant
and widely discussed in connection with the PAMELA
experimental results considered in Section 9 because this
effect only proved to be capable of producing the required
boost factor of several orders of magnitude if clumps with
masses $\ge10^5M_\odot$ are present in the Galaxy. That the
Sommerfeld enhancement plays an important role in the
annihilation of supersymmetric dark matter particles was
first argued in [221, 222]. The Sommerfeld enhancement,
which is an increase in the annihilation cross section, is due
to multiple exchanges of intermediate bosons between the
annihilating particles. This exchange, which occurs according
to ladder Feynman diagrams, corresponds to a certain
distortion of the wave function of the annihilating particles
due to the presence of an additional interaction potential in
the Schrodinger equation [223]. For supersymmetric particles, it is important here that the neutralino belongs to a
multiplet of states with close masses, between which co-annihilation occurs. Therefore, for example, for a bino,
which is a singlet, the Sommerfeld enhancement is absent.
The Sommerfeld enhancement for superheavy neutralinos
was considered in [181].

The Sommerfeld enhancement is most frequently modeled by the Yukawa potential $V(r)=-(\alpha/r)e^{-m_Vr}$,
where $m_V$ is the mass of the force-carrier particle. In the
Coulomb limit $m_V\to0$, the boost coefficient ${\cal R}$ (the ratio of
the perturbative to the nonperturbative annihilation cross
sections), which is determined by the relation $\langle\sigma v\rangle={\cal R}\langle\sigma v\rangle_0$,
has the form
\begin{equation}
 \label{SF}
   {\cal R}=\frac{\pi\alpha}{\beta}(1-e^{-\pi\alpha/\beta})^{-1},
\end{equation}
where $\beta=v/c$. For small $\beta$, ${\cal R}\propto1/v$. However, at very small
$\beta$, of the order of $10^{-4}$ (in the model with particle masses
around 1 TeV considered in [223]), saturation occurs, and no
cross-section boost occurs upon a further velocity decrease.
When changing the ratio between the masses of the dark
matter particle and the interaction carrier particle, resonances
appear in the annihilation cross section, in which the
enhancement coefficient ${\cal R}$ can be as high as ${\cal R} \sim10^4-10^5$ [223].
In addition, as shown in [223], due to the Sommerfeld effect,
the cross section of the neutralino annihilation into electron-positron pairs is of the same order of magnitude as the cross
section of the annihilation into intermediate bosons, whereas
in the absence of the Sommerfeld effect, the former cross
section is suppressed by the factor $(m_e/m)^2$.

The annihilation products can also be observed indirectly. For example, the synchrotron radiation of charged particles created in neutralino annihilations can explain
the diffuse background radio emission observed by the
ARCADE-2 (Absolute Radiometer for Cosmology, Astrophysics, and Diffuse Emission 2) experiment and by other detectors [224].

%%%

\subsection{Determination of astrophysical backgrounds that
are not connected with annihilation}

In searches for dark matter particle annihilation, the
complicated problem arises of extracting the signal from
different backgrounds produced by cosmic rays and usual
astrophysical sources. It is most difficult to determine the
interstellar cosmic ray contribution, because the rays propagate far away from their sources and generate diffuse gamma-
ray backgrounds when interacting with the interstellar gas
and radiation field.

The observed fluxes of cosmic ray charged particles are
compared with charge particle fluxes in the secondary
generation model developed by Ginzburg and Syrovatsky
[225, 226], and its modern version is conveniently realized in
the GALPROP computer code [227]. This model is key in
searching for different spectral anomalies that can suggest
dark matter particle annihilation.

Presently, searches for dark matter particle annihilation
are mainly carried out by Fermi Space Gamma-ray Observatory (the former Gamma-ray Large Area Space Telescope,
GLAST), launched in 2008, and the LAT (Large Area
Telescope) is the main instrument aboard this space gamma-ray observatory. In the GeV energy range, Fermi-LAT has a
resolution of about 1 degree, which corresponds to its point-spread function at a level of 68% [228]. Objects with a large
angular size, observed by Fermi-LAT, cannot be considered
point-like any more, and the telescope sees only part of their
gamma-ray emission in each direction. Fermi-LAT enables
detection of gamma-ray quanta with energies above 100~MeV
and has the sensitivity $\Phi(E>100\mbox{~MeV})=6.0\times10^{-9}$~cm$^2$~s$^{-1}$ for point-like sources. Not all of the point-like sources
observed have been identified, and some of them could be
produced by annihilation of large dark matter clumps.
Bounds on the number of clumps and their progenitor
perturbation spectrum in different mass ranges were
obtained in [153] by comparing the Fermi-LAT point-like
sources and the calculated annihilation signal from clumps,
both galactic and extragalactic. In addition, in [153], constraints from diffuse emission, reionization, and gravitational
microlensing were considered. A comparison of the Fermi-LAT data with the annihilation signal from clumps calculated
for a power-law dependence of the annihilation cross section
on the relative particle velocity $\sigma_{\rm ann}\propto v^{-\beta}$ (taking the
Coulomb enhancement into account), where $\beta$ is a free
parameter, was made in [72]; constraints on $\beta$ and other
parameters were obtained.

%%%

\subsection{Parameterization of the annihilation signal}

When calculating the annihilation signal, it is convenient to
separate factors related to the dark matter particle properties
from `astrophysical' factors caused by dark matter distribution and clumping. The latter include, in particular, the
character of dark matter clustering and the presence of
clumps. Following [53], we write the observed flux from the
direction $\psi$ averaged over the solid angle $\Delta\Omega$ as
\begin{equation}
J_{\gamma}(E,\psi,\Delta\Omega)=9.4\times10^{-11}\frac{dS}{dE}\langle
J(\psi)\rangle_{\Delta\Omega}, \label{igam}
\end{equation}
where
\begin{equation}
\frac{dS}{dE}=\left(\frac{100\mbox{~GeV}}{m}\right)^2\sum
\limits_{F}\frac{\langle\sigma_Fv\rangle}{10^{-26}\mbox{~cm$^3$s$^{-1}$}}
\frac{dN_{\gamma}^F}{dE}, \label{dsde}
\end{equation}
$dN_{\gamma}^F$ is the number of photons per single annihilation event
(per pair of annihilating particles), and the astrophysical
factor (in the direction c, averaged over the solid angle DO) is

$\Delta\Omega$)
\begin{equation}
\langle J(\psi)\rangle_{\Delta\Omega}=\frac{1}{8.5\mbox{~kpc}}\frac{1}{\Delta\Omega} \int
d\Omega'\int dL\left(\frac{\rho(r)}{0.3\mbox{~GeV~cm$^{-3}$}}\right)^2. \label{jpsi}
\end{equation}
The integration is carried out along the line of sight. For
Dirac particles, an extra factor $1/2$ emerges in (106), and
$\rho=\rho_++\rho_-$ denotes the total dark matter density (the sum of
particles and antiparticles).

In the case of an ordinary $\sim100$~GeV neutralino, the
generation rate of gamma-ray photons in a clump can be
approximately represented as $2\eta_{\pi^0}\dot N_{\rm cl}$, where $\eta_{\pi^0}\sim10$ is the
multiplicity of neutral pions. Here, it is assumed that the
neutralino annihilates to produce $\pi^0$ with subsequent decays
$\pi^0\to2\gamma$. The cumulative flux at the angle $\psi$ to the Galaxy
center is
\begin{eqnarray}
&~&J_{\gamma}(E>m_{\pi^0}/2,\psi)= \label{igam-pi}
\\
&~&1.9\times 10^{-10}\left(\frac{m}{100\mbox{~GeV}}\right)^{-2}\frac{\langle\sigma
v\rangle}{10^{-26}\mbox{~cm$^3$s$^{-1}$}}\langle J(\psi)\rangle_{\Delta\Omega}.\nonumber
\end{eqnarray}

The astrophysical factor for particle annihilation in clumps
has the form

\begin{eqnarray}
&~&\langle J(\psi)\rangle_{\Delta\Omega}= \label{jpsi-pi}
\\
&~&\int d\xi_{\rm cl}\left(\frac{\rho_{cl}}{0.3\mbox{~GeV~cm$^{-3}$}}\right)
\int\limits_{l.o.s.}\frac{dL}{8.5\mbox{~kpc}}\left(\frac{\rho_H(r)}
{0.3\mbox{~GeV~cm$^{-3}$}}\right),\nonumber
\end{eqnarray}
where $\xi_{\rm cl}$ is assumed to include the survival probability $P$
calculated in Section~7.3, and l.o.s. means that the integration
is performed along the line of sight.

%%%

\subsection{Enhancement of the annihilation signal}

Annihilation signal enhancement is often referred to as the
boost factor. As a rule, the enhancement is considered
relative to the model in which the Galaxy halo or another
dark matter object does not contain small-scale substructures. For example, a boost factor of 10 is required in the
ordinary neutralino annihilation model to explain the
gamma-ray excess observed by EGRET (Energetic Gamma
Ray Experiment Telescope) [229]. The boost factor in the
direction $\psi$ is defined as
\begin{equation}
B(\psi)=\frac{J^{\rm cl}(\psi)+J^{\rm hom}(\psi)}{J^{\rm hom}(\psi)}, \label{eta}
\end{equation}
where $J^{\rm hom}(\psi)$ is the signal from nonclumped dark matter in
the halo. For $\langle\sigma_{\rm ann} v\rangle=const$, the enhancement $\eta$ does not depend on the annihilation cross section and is determined by dark matter clumping only.

To calculate the total signal integrated over all directions,
instead of $\Delta\Omega^{-1}\int d\Omega'$, integration over angles is performed:
\begin{eqnarray}
\int\limits_{0}^{\pi}\!d\zeta\sin\zeta\!\!\! \int\limits_{0}^{2\pi}d\phi ~ ...
\end{eqnarray}
where $\zeta$ is the angle between the line of sight and the direction
to the Galaxy center. The distance $l$ between the clump and
the Galaxy center can be expressed through $r$ (in this case, it is
simply the distance from Earth to the clump), $r_{\odot}$ (the distance
from the Sun to the galactic center), and $\zeta$ as $l(\zeta,r) = (r^2+r_{\odot}^2-2rr_{\odot}\cos\zeta)^{1/2}$. The maximum distance from the
Sun to the outer halo boundary in the direction $\zeta$ is
$\zeta$, $r_{\rm max}(\zeta)
=(R_H^2-r_{\odot}^2\sin^2\zeta)^{1/2}$, where $R_H \sim 100$~kpc is the
virial radius of the galactic halo and $r_{\odot}=8.5$~kpc.

Under the assumption that the clump number density is
proportional to the halo density, i.e., $n_{\rm
cl}(l)=\xi\rho_{\rm DM}(l)/M$, the
boost factor can be conveniently estimated as
\begin{equation}
B \approx 1+\xi S(x_c,\beta)\frac{\bar\rho_{\rm int}} {\tilde\rho_{\rm DM}}, \label{etasimp}
\end{equation}
where $\tilde\rho_{\rm DM}=4.26\times10^{-23}$~g~cm$^{-3}$. For example, for the parameters $\beta\simeq 1.8$, $x_c\simeq0.05$, $S(x_c,\beta)\simeq5$,
$\tilde\rho_H \sim \rho_{\rm DM}(r_{\odot})\sim0.3$~GeV~cm$^{-3}$, $\bar\rho_{\rm
int}\sim2\times10^{-22}$~g~cm$^{-3}$, and $\xi\sim0.001$,
Eqn (113) yields the numerical estimate $B\sim 3$. A close
estimate, $B\sim 4$, for annihilation in the galactic halo was
recently obtained in [51].

%%%

\subsection{Annihilation in galaxies and galaxy clusters}

Properties of dark matter particles and their clustering can be
constrained by comparing the results of model calculations
with observations of the gamma-ray background and point-like gamma-ray sources. Fermi-LAT annihilation limits in
galaxy clusters were analysed in [230-234]. As noted in [235],
observations of the Virgo, Fornax, and Coma clusters likely
demonstrate some gamma-ray excess, which can be explained
by dark matter particle annihilation. Fermi-LAT annihilation limits in the Milky Way were studied, for example, in
[236-238]. A direct comparison of signals from the Galaxy
and galaxy clusters was performed in [239] (see Fig. 7 and
Section~5.3 there) and [240] (see Table~1 there). The signal of
unseen galactic satellites was analysed in [241].

The tightest bounds on the annihilation cross section are
presently obtained from dwarf galactic satellite spheroids
[230]. The Fermi-LAT limits for these objects are already
close to the thermal cross section $<\sigma v>\simeq3\times10^{-26}$~cm$^3$~s$^{-1}$,
even in the most conservative case without any boost factor
[242, 243]. In the case of the thermal cross section, dark matter
particles were produced in the early Universe in an amount
just sufficient to explain the dark matter abundance. Dwarf
galaxies seem to be very suitable for dark matter particle
annihilation searches, insofar as they contain a small amount
of gas and have a low star formation rate. For this reason, the
cosmic rays must produce a low gamma-ray background. On
the other hand, dwarf spheroids exhibit high mass--luminosity
ratios ($\sim10^3M_\odot/L_\odot$), i.e., they contain a relatively large
amount of dark matter. Using the bound $\langle\sigma v\rangle<3\times10^{-26}$~cm$^3$~s$^{-1}$, the Fermi-LAT collaboration [242] inferred a
minimum possible dark matter particle mass in the hadronic
and leptonic annihilation channels, $\simeq 27$~GeV and $\simeq 37$~GeV
respectively. An alternative analysis of signals and back-
grounds [243] suggests that the Fermi-LAT observations of
dwarf spheroids exclude, at the 95\% level, dark matter
particles with masses below 40~GeV in the hadronic annihilation channel. In addition, searches for an annihilation signal
from globular clusters are being carried out [244]. Some of
these objects may be remnants of small satellite galaxies
stripped off by tidal forces and may contain an appreciable
amount of dark matter. However, no signals from globular
clusters have been detected yet.

Details of annihilation signal calculations from dwarf
galaxies can be found in the papers cited above; here, as an
example, we calculate the signals from the Milky Way and
galaxy clusters in detail. In the Virgo cluster, the distance
from Earth to the cluster center exceeds the virial cluster
radius. Therefore, the integration over $L$ in (108) is bounded
by the values $L_{\rm max, min}(\psi) = \pm (R^2-r_V^2\sin^2\psi)^{1/2}+ r_V\cos\psi$, and
$\sin\psi<\sin\psi_{\rm max}=R/r_V$, where the distance to the Virgo
center is $r_V=16.5$~Mpc [245]. Assuming the NFW profile for
the Virgo cluster, we have $a^V=0.58$~Mpc, $R^V=2.4$~Mpc, and $\rho_0^V=1.0\times10^5M_\odot$/kpc$^3$.

We first compare the signals from the diffusive dark
matter component without clumps. Within the angular size
$\delta\psi=1^\circ$, signals from the Virgo cluster center and from the
galactic center are calculated to be

\begin{equation}
\langle J(\psi)\rangle_{\Delta\Omega}(\mbox{ Milky Way}) \simeq1.4\times10^3,\label{s01}
\end{equation}
\begin{equation}
\langle J(\psi)\rangle_{\Delta\Omega}(\mbox{ Virgo}) \simeq5\times10^{-2}.\label{s02}
\end{equation}
Signals integrated over the solid angle $\Delta\Omega=4\pi$ for the Milky
Way and $\Delta\Omega=2\pi(1-\cos\psi_{\rm max})=0.067$ for the Virgo
cluster are

\begin{equation}
\langle J(\psi)\rangle_{\Delta\Omega}(\mbox{ Milky Way}) \simeq3,
\end{equation}
\begin{equation}
\langle J(\psi)\rangle_{\Delta\Omega}(\mbox{ Virgo}) \simeq9\times10^{-4}.
\end{equation}

The centers of both the Galaxy and Virgo clusters are
fairly poor targets for annihilation signal searches due to a
strong background contamination from astrophysical
sources, including the cosmic-ray induced gamma-ray flux.
In addition, the M87 galaxy with an active nucleus is close to
the Virgo center. Chances of detecting the annihilation signal
crucially depend on the assumed signal-to-noise ratio, and
hence the above estimates remain very uncertain.

We next consider the possible contribution from small-scale clumps with distribution (88). The integration over the
clump distribution in Eqn (110) can be performed analytically,

\begin{equation}
{\displaystyle \langle J(\psi)\rangle_{\Delta\Omega}\simeq
7.01\left(\frac{S(x_c,\beta)}{S(0.01;1.8)}\right)\left(\frac{\sigma_{\rm eq}(M_{\rm
min},n_s)}{\sigma_{\rm eq}(10^{-6}M_{\odot};0.963)}\right)^3\times } \atop {\displaystyle
\times \frac{1}{\Delta\Omega} \int d\Omega'\int \frac{dL\rho(r)}{0.3\mbox{~GeV~cm$^{-3}$}}, }
\label{jpsiclan}
\end{equation}
where $S(0.01;1.8)\simeq14.5$, $\sigma_{\rm eq}(10^{-6}M_{\odot};0.963)\simeq8.76\times10^{-3}$. With the contribution from the clump taken
into account, for the central parts of the Galaxy and Virgo
clusters with the angular size $1^\circ$, we find

\begin{equation}
\langle J(\psi)\rangle_{\Delta\Omega}(\mbox{ Milky Way}) \simeq1.4\times10^2,\label{s1}
\end{equation}
\begin{equation}
\langle J(\psi)\rangle_{\Delta\Omega}(\mbox{ Virgo}) \simeq13,\label{s2}
\end{equation}
and for signals integrated over angles, we obtain

\begin{equation}
\langle J(\psi)\rangle_{\Delta\Omega}(\mbox{ Milky Way}) \simeq15,
\end{equation}
\begin{equation}
\langle J(\psi)\rangle_{\Delta\Omega}(\mbox{ Virgo}) \simeq1.3.
\end{equation}
Hence, clump boosting in the Virgo cluster can reach three
orders of magnitude.

It is also interesting to compare the signals from clumps
with $M_{\rm min}\sim10^{-6}M_\odot$ generated from the standard power-law perturbation spectrum normalized to the WMAP data
and from diffusive dark matter in the Galaxy. We assume that
the clump distribution function due to hierarchical clustering
has form (88) and that the clump survival probability can be
taken from the results in Section 7.3. The results of these
calculations are presented in Fig. 11. It is seen that depending
on the angle, the enhancement of the signal due to dark matter
clumping reaches 2.5 orders of magnitude.

\begin{figure}[t]
\begin{center}
\includegraphics[angle=0,width=0.45\textwidth]{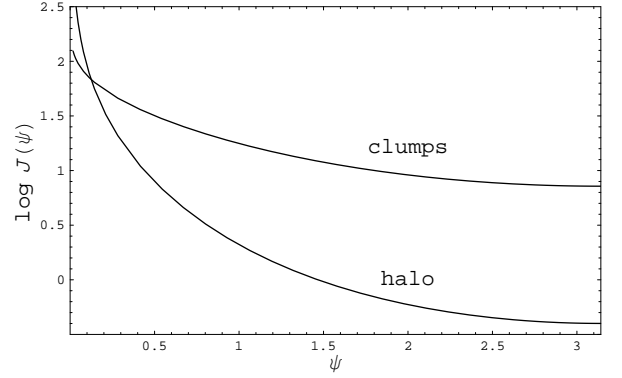}
\end{center}
\caption{The astrophysical factor $J(\psi)$ (in conventional units) as a
function of the angle c from the line of sight and the direction to the
Galactic center.} \label{grpsi_1}
\end{figure}

%%%

\subsection{Anisotropy of annihilation signals}

The annihilation signal is usually calculated by assuming a
spherically symmetric galactic halo. In this case, the anisotropy of the annihilation gamma-ray remission can only be
due to the non-central galactic location of the Sun. Nevertheless, as demonstrated in [63], the halo nonsphericity can be
fundamentally important for the annihilation emission.
According to observations, axes of the galactic ellipsoidal
halo differ by less than 10\%-20\%; however, a higher
difference, up to a factor of two, cannot be excluded [246,
247]. This leads to an order-of-magnitude uncertainty in the
predicted amplitude of the annihilation signal from the center
and anticenter of the Galaxy [63].

\begin{figure}[t]
\begin{center}
\includegraphics[angle=0,width=0.45\textwidth]{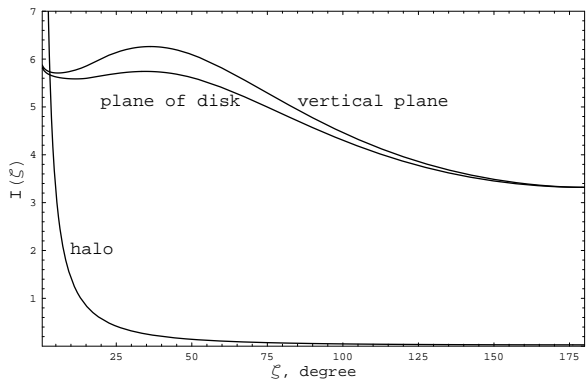}
\end{center}
\caption{The annihilation signal in the galactic disc plane and in the plane normal to the galactic disc as a function of the angle $\zeta$ between the line of sight and the direction to the galactic center. For comparison, the annihilation signal from the galactic halo without clumps is also shown.}
\label{ani2f}
\end{figure}

The anisotropy relative to the galactic plane was discussed
in [48,203]. The tidal destruction of a clump is generally
anisotropic and depends on the clump orbital inclination to
the disc plane. Accordingly, dark matter annihilation in the
halo (in surviving clumps) should also be anisotropic. In
Fig.~12, the annihilation signal calculated in [48] is shown in
the galactic plane and in the normal plane passing through the
Galaxy center. For comparison, Fig.~12 also shows the signal
from a spherically symmetric halo without clumps. This signal
is the same in both planes, and hence it can in principle be
extracted from observations. The difference between signals in
two orthogonal planes measured for the same angle $\zeta$ to the
center can be viewed as a measure of the anisotropy. We
consider the quantity $\delta=(I_2-I_1)/I_1$. At $\zeta\simeq39^{\circ}$ it takes the maximum
value $\delta \simeq 0.09$, which, however, is significantly
higher than the Fermi-LAT resolution.

The annihilation signal from the galactic center depends
on the dark matter central density profile. If there is a density
cusp [176], a bright source in the center of the Galaxy should
be present. However, this cusp in the diffuse dark matter
distribution can be smeared out by baryons [248]. Small-scale
clumps are destroyed more efficiently inside the stellar bulge
around the galactic center. It is also possible to observe dark
``gamma-ray'' circles in other galaxies, appearing due to the
absence of clumps in the central regions of these galaxies [203].

The intrinsic anisotropy of the annihilation signal due to
dark matter clustering should also be noted. The corresponding angular spectrum of the annihilation signal fluctuations at
small angular scales is related to the dark matter clump
spectrum [249-252]. In principle, nearby clumps can appear
in gamma-ray observations as point-like sources [63,72].
Some evidence of small-scale anisotropy produced by a
population of point-like sources was found in Fermi-LAT
observations [253], and it is too early to rule out that some of
these sources can be dark matter clumps. Another secondary
source of annihilation anisotropy can be the dipole anisotropy due to the motion of the Sun in the Galaxy; this effect
can easily be observed.

%%%

\subsection{Annihilation in ultradense clumps}

The annihilation signal enhancement in clumps, on the one
hand, increases the chances of dark matter annihilation
detection; on the other hand, in some models, the enhance-
ment turns out to be so strong that the predicted signal exceeds
the observed background constraints, which imposes constraints on the parameters of dark matter particles and
clustering.

We first discuss standard neutralinos with a mass of the
order of the electroweak scale, which are generated according
to the thermal scenario. We show that if not all of these
particles were assembled into superdense clumps at the RD
stage, the annihilation flux would exceed the observed
background, and therefore this scenario (superheavy clumps
from standard neutralinos) is ruled out. We choose the
parameters that minimize the annihilation flux and assume
the mean neutralino density in the clumps to be the minimal
value corresponding to the clump formation at the end of the
RD stage, $\bar{\rho}_{\rm int}= 178 \rho_{eq}$. Then the minimal gamma-ray flux averaged over angles is
\begin{equation}
J_{\gamma}^{\rm tot} = 4.3 \langle\sigma v\rangle_{26} m_{100}^{-2} ~~~ \mbox{\rm cm}^{-2}
\mbox{\rm s}^{-1} \mbox{\rm sr}^{-1}, \label{flux-num1}
\end{equation}
where $m_{100}$ is the neutralino mass $m$ in units of 100~GeV and
$\langle\sigma v\rangle_{26} =
\langle\sigma v\rangle/(10^{-26}~\mbox{\rm cm}^3 \mbox{\rm s}^{-1})$. Integral flux (123) is larger
than the observed one by five orders of magnitude. However,
the decrease in the number of clumps being formed due to the
initial nonsphericity of perturbations turns out to be sufficient to reconcile the theory with observations [131]. It is
possible to consider more massive neutralinos and to choose
the minimum possible annihilation cross section [177] for the
strongly suppressed $s$-wave annihilation channel, $\langle\sigma v\rangle= 1.7\times 10^{-30}m_{100}^{-2}$~cm$^3$/s. However, for the dark matter
clump fraction fcl51, the annihilation signal would still be
several orders of magnitude higher than the diffusive flux
measured by EGRET and Fermi-LAT. This excess was noted
in [177] in the particular case of clumps in the form of
``neutralino stars''.

The annihilation upper bounds (in the model of usual
neutralinos) were used in [152, 158, 159] to conclude that the
mass fraction of clumps in dark matter is much less than units.
A similar conclusion was reached in [149] from the analysis of
the early annihilation effect on the recombination process
from CMB measurements. In the standard model of the
neutralino ($\sim 100$~GeV), the simple possibility of avoiding
the discrepancy with a strong annihilation signal is provided
by a small amplitude of perturbations, when no superheavy
clumps can be formed. In particular, the absence of observed
point-like gamma-ray sources was used in [103] to conclude
that in the $e^+e^-$-annihilation epoch, the amplitude of the
perturbation entering the horizon was less than $10^{-3}$. In [159],
additional constraints on clumps were derived from the
analysis of dark matter annihilation effects on the reionization of the Universe.

We now discuss the annihilation of superheavy particles in
superdense clumps. Although the existence of superheavy dark matter particles is theoretically admissible and well
justified, their detection appears to be a very difficult
problem. Indeed, the creation of such particles in accelerator
experiments and their direct registration go far beyond the
existing capabilities and those planned for the foreseeable
future. The annihilation rate of such heavy particles depends
on mass approximately as $\dot N_{\rm ann}\propto m^{-4}$. Because the back-ground radiation, such as cosmic rays from astrophysical
sources and the diffusive flux of gamma-ray photons,
decreases with energy only as $1/E^{\alpha}$ with $\alpha\le 3$ as the
particle mass increases, the indirect detection of dark
matter particles seems to be more and more an impossible
task. One possibility enabling us to overcome this difficulty
is the formation of a superdense core in clumps [255]. In
[255], with the aim of explaining the origin of ultra-high-energy cosmic rays [256], superheavy dark matter particle
annihilation in the dense central part of the clumps was
discussed as an alternative to scenarios with superheavy
particle decays [20, 21].

Another possibility is the annihilation in superdense
clumps considered in [164, 181]. In sufficiently dense
clumps, relaxation due to two-particle gravitational scattering can initiate the ``gravothermal catastrophe'', and the
clump density profile can be transformed into the isothermal profile $\rho\propto r^{-2}$ with a very small core radius. This is the
case for the superheavy bino. Due to the density increase,
dark matter annihilation is substantially enhanced. Unlike
the bino, the vino and Higgsino are strongly coupled to the
thermal plasma; therefore, no gravothermal catastrophe is
occurs. Virial velocities of dark matter particles in superdense clumps are very small, which leads to the Sommerfeld
enhancement of the annihilation rate of the vino and
Higgsino. With this effect taken into account, the annihilation flux from superheavy dark matter particles was shown
in [164, 181] to be at the observable level for all types of
superheavy neutralinos.

Neutralino annihilation in superdense clumps around
cosmic string loops was analyzed in [141]. The formation of
such clumps was discussed in Section 3.5. A comparison of
the calculated signal with Fermi-LAT data treated as the
upper bound allows constraining the properties of cosmic
strings and dark matter particles.

To conclude this section, we mention three other possible
annihilation effects. Charged particles created by the particle
annihilation in clumps can form additional inhomogeneities
in interstellar magnetic fields, which can affect the cosmic ray
diffusion coefficient [257]. The annihilation of neutralinos at
redshifts $z\sim10-30$ can substantially change the thermal
balance of gas in the Universe, whereas this process without
clumps is ineffective [258]. The annihilation in clumps can
also affect the reionization history of the Universe [259].

%%%

\section{Charge particle fluxes in PAMELA,
ATIC, and other experiments}

%%%

\subsection{Observational data}

As early as 1994, the balloon experiment HEAT (High
Energy Antimatter Telescope) discovered a small excess of
$e^+$ in the energy range 6-10~GeV, possibly suggesting dark
matter particle annihilation. The excess of $e^+$ with the same
energies was detected in the AMS-01 (Alpha Magnetic
Spectrometer 01) experiment. This positron excess in the
energy range 6-10~GeV was not confirmed by later experiments, but ``instead'' new results suggested an excess of
positrons at higher energies.

The PAMELA detector aboard the Russian satellite
Resurs-DK1 for studying the charged component of cosmic
rays was launched into orbit on 15 June 2006. It can register
positrons and protons in cosmic rays. The 2008 data release
revealed an excess of $e^+$ in the energy range 10-60~GeV, where
no solar modulation is expected. In a short time, this excess
was reliably confirmed up to 90~GeV.

Over 850 days of observations from July 2006 to
December 2008, the PAMELA detector registered $\simeq1500$
antiprotons with energies from 60 to 180 GeV; this antiproton
flux, up to model and experimental errors, is in good
correspondence with the secondary generation of protons in
cosmic ray interactions with the interstellar gas [260]. In
particular, the PAMELA data reproduce the expected shape
of the spectrum with a maximum at about 2~GeV. Thus, the
excess of $e^+$ observed by PAMELA is not associated with the
excess of $\bar p$.

The ATIC (Advanced Thin Ionization Calorimeter)
balloon-borne experiments revealed the presence of an excess
of electrons with energies $300-800$~GeV, with a sharp cut-off at
higher energies. Because electrons can easily be absorbed by
the interstellar gas, such an excess can be produced only by a
nearby source.

The results of measurements in different experiments
significantly differ in some energy ranges. The flux of
electrons measured by ATIC somewhat exceeds the standard
model prediction, while the PAMELA measurements below
625~GeV are in agreement with the standard generation
model [261]. The PAMELA data look somewhat more
convincing because the PAMELA detector provides good
magnetic separation of particles, has a larger volume
calorimeter than ATIC, and is equipped with a neutron
detector; in addition, the PAMELA data are visually
scrutinized at the final stage.

PAMELA also provided the best constraints to date on
antideuterium and antihelium fluxes, except in some energy
ranges where more stringent bounds are provided by the
BESS (Balloon-borne Experiment with a Superconducting
Spectrometer) experiment. These constraints are interesting
from the dark matter standpoint, because $d\bar d$ and $He\bar{He}$ pairs
can, with some probability, be created by dark matter particle
annihilation.

Recently, in the AMS-02 experiment aboard the International Space Station, an excess of the positron fraction at
energies up to 350~GeV was reported [262], and thus the
PAMELA results were confirmed and extended toward
higher energies.

%%%

\subsection{Annihilation scenario and its problems}

Dark matter particle annihilation provides an explanation for
the observed e? excess in the PAMELA experiment. This
hypothesis has been widely discussed, suggesting indirect
registration of dark matter. Discussions of the annihilating
dark matter in connection with the PAMELA data made
these studies very topical and stimulated detailed investigations of the neutralino models, accurate calculations of the
annihilation cross sections and the annihilation product
spectra, etc. Models with decaying dark matter particles
have also been discussed. It should be noted from the very
beginning that simpler astrophysical explanations have been
immediately suggested (see Section~9.3), and the dark matter
model is presently not considered to be the lead one.

Explaining the PAMELA results by the annihilation of
ordinary neutralinos generated by the thermal mechanism
requires invoking a significant boost factor, because the
annihilation cross section is determined by the neutralino
generation model, and this cross section is too small to create
the observed charged particle flow. As we know, high-density
clumps can increase the boost factor [263]. Because the
particle velocity dispersion in the clumps is low, the
Sommerfeld enhancement and radiation corrections, which
can increase the annihilation cross section by several orders of
magnitude, are equally important. They can increase gamma-ray and positron fluxes due to annihilation [264]. Earlier
explanations of the HEAT results by neutralino annihilation
also required a boost factor of the order of 30.

Thus, apparently, neutralino annihilation can easily
explain the PAMELA data. However, this scenario runs
into a serious difficulty. As noted in Section 9.1, PAMELA
observations do not show the antiproton excess, whereas
typical annihilation models predict the generation of positron
and antiproton excesses simultaneously. This problem is
pertinent to the boost factor produced due to both clumps
and other possible sources, i.e., the problem is inherent to the
annihilation scenario in general, irrespective of the clump
models.

Another major problem arises in relation to the Fermi-LAT flux bounds from dwarf galaxies and galaxy clusters.
These bounds strongly constrain or (in the most typical
models) exclude the PAMELA data interpretation of the
neutralino annihilation. The point is that for the positron
flux registered by PAMELA to be produced, the annihilation
cross section must be so large that the gamma-ray signal from
dwarf galaxies and clusters should necessarily be observed.
This problem is to a great extent dependent on the annihilating
particle model, namely, on the predominant annihilation
channels. In the simplest case of the usual neutralino, the
dark matter annihilation model as an explanation of the
PAMELA results can already be excluded due to the absence
of antiprotons and disagreement with Fermi-LAT observations, but hypothetical variants with suppressed annihilation
into gamma-ray photons and antiprotons are still possible.
For example, the annihilation into lepton channels only was
considered in [265], and a model invoking the dark sector was
studied in [266].

A sufficiently good correspondence of the calculations
with the ATIC data can be obtained in the annihilation
model for Kaluza-Klein particles with masses $\simeq620$~GeV
[267]. It can be supposed that ATIC discovered a
sufficiently close clump in which annihilation occurs [268],
or the annihilation is under way in a dense density peak
around an intermediate-mass black hole. We note, however, that dark matter particles of one sort are unlikely to
simultaneously explain both PAMELA and ATIC results,
although attempts have been made to develop a unified
model.

%%%

\subsection{Alternative explanations}

The simplest explanation of the positron excess observed by
PAMELA is provided by the assumption that cosmic ray
propagation and secondary generation models are incomplete [269]. The corrected model in [269] corresponds to the observed positron excess if the additional positrons result
from the annihilation of vino-like neutralinos with masses of
180~GeV, which were produced in the early Universe by a
nonthermal mechanism [270].

The $e^+$ excess registered by PAMELA together with the
absence of the corresponding antiproton excess can be
explained by the generation of electrons and positrons in
pulsars. However, these models encounter difficulties because
reproducing different parts of the $e^+$ spectrum requires
assuming different pulsar model parameters; in other words,
to reach agreement with observations, the summation of
fluxes from several pulsars with different properties is
required. A similar explanation of the ATIC results also
meets with difficulties, since the electron spectra generated
by neutron stars and micro-quasars are different from the
observed one.

An interesting solution can be found in the framework of
the model of generation of an $e^+$ excess in the processes
$\pi^+\to e^++...$ during flares on dwarf main-sequence stars,
which are the most numerous stars in the Galaxy. It is known
from astronomical observations that low-mass stars have
quite unsteady photospheres, in which very powerful flares,
much more powerful than solar flares, occur. Calculations
[271,272] indicate that stellar flares could reproduce the
observed $e^+$ excess. This model is attractive because it does
not postulate any new phenomena and is based only on the
known astrophysical processes.

Finally, a simple and elegant solution of the $e^+$ excess
problem is given in [273,274], where the secondary positrons
are assumed to be generated and accelerated by cosmic-ray
sources themselves. This model successfully reproduces the
observed $e^+$ spectrum with no antiproton excess and well-studied energy ranges. A concern may arise only with nuclei
and antiprotons at very high energies ($>100$~GeV), although
the reality of such concerns has not been confirmed by specific
calculations, and the observational accuracy at these energies
is still poor.

%%%

\section{Other possible observational manifestations
of clumps}

%%%

\subsection{Direct detection of dark matter particles. Ministreams}

The probability that Earth is now inside a dark matter clump
is estimated to be from 0.0001\% [211] to 0.1\% [68], depending
on the clump mass and the assumed perturbation spectrum. If
such a rare event is indeed taking place now, the probability of
direct and indirect dark matter particle detection should be
much higher [68].

It was concluded in [40, 203] that almost all clumps in the
Galaxy are tidally disrupted in interactions with stars. As
shown in Section~7.3, for a significant fraction of clumps, tidal
interactions with stars indeed lead to stripping off their outer
layers, i.e., to a substantial mass loss, but this does not imply
total clump destruction, because the clump cores can survive
[83, 263]. The stripped mass is transformed into ``ministreams''
of dark matter [40, 203, 263]. For the direct experimental
detection of dark matter, the ministreams are interesting
because dark matter particles in them move along several
different discrete directions corresponding to the progenitor
clumps before the destruction. According to [211], presently
some $10^2-10^4$ ministreams may be crossing Earth. The
ministream formation and evolution during clump destructions by stars were also addressed in [41]. Dark matter in the
streams spreads in the space like a broad tail, and therefore
the probability of a ministream colliding with Earth is much
higher than that of the entire clump crossing.

%%%

\subsection{Registration of clumps by gravitational wave
detectors}

It has been argued in some papers that the planned
gravitational-wave detectors like LISA (Laser Interferometer Space Antenna) can measure small variations of the
gravitational field caused by nearby flying compact objects.
LISA was supposed to search for PBHs [275], asteroids [276],
or compact dark matter objects of an unknown nature [277].
Clumps can be included into this list. A detectable signal is
produced by the tidal gravitational force, which changes the
interferometer arm length and, accordingly, causes a phase
shift. Using LISA, it is possible to search for compact objects
in the mass range $10^{16}$~g$\le M\le10^{20}$~g according to [275], or
$10^{14}$~g$\le M\le10^{20}$~g according to the calculations in [277]. The
signal is expected to consist of single pulses with the
characteristic frequency near the low-frequency sensitivity
limit of LISA, and the event rate of these signals can be
$\sim1$ per decade, assuming that such objects constitute most of
the dark matter. In the usual scenario, the clump fraction is
only 1\%-10\% of all dark matter and hence the clump
detection rate should be 1-2 orders of magnitude lower. In
addition, the radii of such clumps, as a rule, are much larger
than the LISA arm length $L\simeq5\times10^{11}$~cm, and therefore the
tidal force amplitude is smaller. Hence, the detection of
standard clumps by LISA seems to be unlikely.
The next generation of gravitational-wave detectors
opens more possibilities (see [275] for more details). Superdense clumps, due to their compactness, easily meet the
condition $R<L\simeq5\times10^{11}$~cm in the mass range $10^{14}$~g$\le M\le10^{20}$~g, and these clumps, if they exist, can be detected by LISA.

%%%

\subsection{Neutralino stars and microlensing}

To explain microlensing events observed in the Galaxy halo,
Gurevich, Zybin, and Sirota were the first to propose the
model of gravitationally bound noncompact clumps (self-gravitating clumps) with a mass $\sim(0.01\div1)M_{\odot}$ consisting of weakly interacting dark matter particles [166].
Because the main dark matter particle candidate was the
neutralino, such clumps were dubbed ``neutralino stars''. It
was also noted that the neutralino stars produce a very
strong annihilation signal even for a very large core size,
$R_c\sim0.1R$, and in order to not violate the observational
limits, the neutralino was postulated to annihilate with a
small cross section in a p-wave. Neutralino stars with a very
small core radius complying with the annihilation limits
were then considered in [177]; it was shown that in this case,
too, the annihilation signal is expected to significantly
exceed the observational constraints, even by assuming the
minimum possible annihilation cross section available in
supersymmetry models.

The criterion for clump formation from adiabatic density
perturbations with PBH constraints taken into account was
improved in [88]. In particular, it was shown there that to
make compact gravitational lenses, the clumps should be
produced as early as at the radiation-dominated stage
(although at its end). Hence, neutralino stars can be qualified
as the superdense clumps considered in Section 3.3. For their
formation, a maximum should exist in the standard power-law density perturbation cosmological spectrum. The results
in Section~3.3 suggest that in the case of adiabatic perturbations, the existence of such a maximum would inevitably lead
to the creation of a significant number of PBHs with masses
$\sim10^5M_{\odot}$ at the RD stage.

The authors of [166] also noted that the baryonic core, if
formed inside clumps, should modify the microlensing light
curve by the clumps. In later studies [157, 159, 160], the
clumps were also regarded as microlensing objects. Two types
of clumps were analyzed in [157]:minihalos around PBHs and
clumps without central PBHs. The substellar mass clumps in
other galaxies can in theory be detected in gravitational
lensing observations of quasars in these galaxies [279, 280].
Gravitational lensing of type-Ia supernovae also yielded
constraints on the dark matter clumping [281]. Some
clumping constraints can also be derived from pulsar timing
observations, because the gravitational field of clumps
located near the line of sight must produce delays in the time
of arrival of periodical signals from pulsars [282]. Figure 13
displays microlensing constraints obtained by the MACHO
(Massive Astrophysical Compact Halo Object) project, as
well as femto- and picolensing constraints obtained from
gamma-ray burst observations [118].

\begin{figure}[t]
\begin{center}
\includegraphics[angle=0,width=0.45\textwidth]{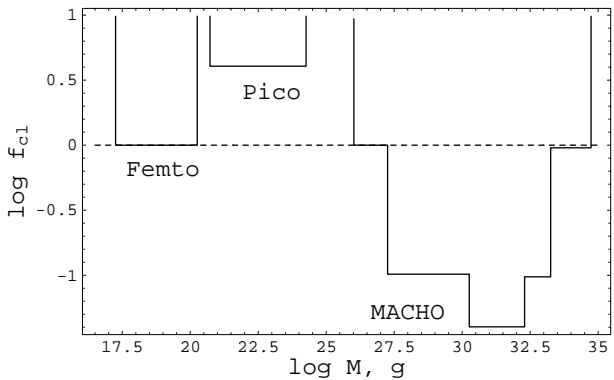}
\end{center}
\caption{Upper bounds on the relative clumped dark matter fraction
$f_{\rm cl}=\Omega_{\rm cl}/\Omega_m$ from microlensing MACHO observations, as well as from
femto- and picolensing observations of cosmic gamma-ray bursts [118].} \label{grmicro}
\end{figure}

We note that the problem of microlenses has become less
relevant now because all microlensing events can be explained
by stars. However, the microlensing observations can be used
to impose upper bounds on the number of PBHs and
superdense dark matter clumps in some mass ranges, thus
constraining the primordial perturbation spectrum on the
corresponding scales.

As shown in [166], a clump can serve as a gravitational
lens and explain the observed microlensing events only if its
radius is less than 10 times the Einstein radius for this object.
This condition implies very stringent constraints on the
nonlinear compression of clumps. It turned out [88] that
clumps can be gravitational lenses only for the high PBH
formation threshold $\delta_{\rm th}>0.5$ obtained in the model of the
critical gravitational collapse, whereas the clump model of
microlensing objects is excluded for lower values of $\delta_{\rm th}$.

%%%

\subsection{Baryons in clumps}

Dissipative processes in the baryonic gas inside dark matter
clumps were studied in [166] and in the clumps around PBHs,
in [283]. It was concluded in [166] that the settling of baryons
toward the clump center causes the dark matter density to
increase as well, which enhances its particle annihilation rate.

In addition, it was noted that the baryonic core modifies the
microlensing light curve of the clumps.
However, we note that it is hard to expect a noticeable
baryonic flux toward substellar-mass clumps, because virial
temperatures in the clumps are too low for the effective
baryon cooling to operate. Additionally, due to Compton
scattering of CMB photons, baryons cannot be accelerated to
virial velocities in clumps; therefore, it is hard to expect the
enhancement of baryonic accretion at high redshifts. In
particular, baryons from almost homogeneous low-density
surroundings can be accreting onto the clumps around
PBHs [283].

Sufficiently heavy clumps with masses $\sim10^5-10^6M_\odot$
that form potential wells into which baryons are collected to
form early population-III stars were studied in [157,284].
These stars can be responsible for the reionization of the
Universe at $z\geq10$. The annihilation of dark matter particles
captured by a star in the central part of a clump can
significantly contribute to the energy balance of the star and
affect its evolution [285,286].

Clumps with masses $M\geq10^{-3}M_\odot$ can be manifested in
21~cm atomic hydrogen line observations by producing
temperature fluctuations in the baryonic gas.

%%%%%%%%%%%

   \subsection{Motion of clumps on the celestial sphere}

If a signal from nearby clumps is detected, it will be possible to
measure the proper motions of the clump in the sky, as in
astrometric observations of nearby stars [69,72,288,289].
However, as shown in [289], the possibility of observing
proper motions is limited by annihilation gamma-ray constraints from the galactic center and other sources. The
nearby clumps and their proper motion can be detected only
for a large annihilation cross section. But then the signal from
other sources would violate the existing observational
bounds. However, we note that this effect is strongly
dependent on the sensitivity and angular resolution of
gamma-ray telescopes, and the calculation in [289] was done
using the Fermi-LAT parameters. Clearly, with instrumental
and observational progress, it may be possible to simultaneously measure the proper motion of clumps in the sky and
annihilation signals from different sources.

%%%

\section{Conclusion}

Small-scale dark matter clumps are interesting to study for
several reasons.
These early formed objects can be the densest dark matter
objects in the Universe; therefore, the dark matter annihilation in these small-scale clumps can be very effective. The
clumps enable the annihilation signal to be enhanced in
galactic halos by several times or even orders of magnitude.
High densities like those in the clumps can appear only in the
central dark matter cusps in galactic centers, if the density
growth is not stopped at some large radius, or in cusps around
central black holes. Thus, the clumps, if they exist, open
additional prospects for indirect dark matter detection via the
annihilation channel, which may help solve one of the biggest
modern mysteries: the nature of dark matter.

If a signal from annihilating dark matter is detected, it will
be possible to study the dark matter distribution in greater
detail and to obtain information on the primordial perturbation spectrum, from which different-scale structures were
formed, as well as to clarify the physics (shape of the scalar
field potential or other characteristics) of the inflation stage,
when these perturbations were generated. The clumps could
shed light on the small scales (in comparison with the galactic
scale) at which the perturbations were generated at the end of
the inflation stage.

Annihilation in clumps could change the thermal balance
in the gas in the pre-galactic epoch and dramatically affect the
evolution of the first stars and hence the chemical evolution of
matter in galaxies, the number of PBHs, etc.
The physics of dark matter clumps is also concerned with
the problem of small-scale perturbation generation, gravitational dynamics, dark matter particle properties, and the
annihilation product effects in the clumps; it is a truly
interdisciplinary field of modern astrophysics and cosmology.

This paper was supported by the Ministry of Education
and Science of the Russian Federation (contract No. 8525)
and by grants OFN-17 from the RAS and NSh-871.2012.2.

%%%


\begin{thebibliography}{300}

\bibitem{} Gorbunov D S, Rubakov V A Introduction to the Theory of the Early
Universe: Hot Big Bang Theory (Singapore: World Scientific, 2011)
[Translated from Russian: Vvedenie v Teoriyu Rannei Vselennoi:
Teoriya Goryachego Bol'shogo Vzryva (Moscow: LKI, 2008)]

\bibitem{} Gorbunov D S, Rubakov V A Introduction to the Theory of the Early
Universe: Cosmological Perturbations and Inflationary Theory
(Singapore: World Scientific, 2011) [Translated from Russian:
Vvedenie v Teoriyu Rannei Vselennoi: Kosmologicheskie Vozmushcheniya. Inflyatsionnaya Teoriya (Moscow: KRASAND, 2010)]

\bibitem{} Weinberg S Cosmology (Oxford: Oxford Univ. Press, 2008)
[Translated into Russian: Kosmologiya (Moscow: URSS, 2012)]

\bibitem{} Lukash V N, Mikheeva E V Fizicheskaya Kosmologiya (Physical
Cosmology) (Moscow: Fizmatlit, 2010)

\bibitem{} Bisnovatyi-Kogan G S Relyativistskaya Astrofizika i Fizicheskaya
Kosmologiya (Relativistic Astrophysics and Physical Cosmology)
(Moscow: KRASAND, 2011)

\bibitem{} Lukash V N, Mikheeva E V, Malinovsky A M Phys. Usp. 54 983
(2011) [Usp. Fiz. Nauk 181 1017 (2011)]

\bibitem{} Cherepashchuk A M Phys. Usp. 56 509 (2013) [Usp. Fiz. Nauk 183
535 (2013)]

\bibitem{} Chernin A D Phys. Usp. 51 253 (2008) [Usp. Fiz. Nauk 178 267
(2008)]

\bibitem{} Lukash V N, Rubakov V A Phys. Usp. 51 283 (2008) [Usp. Fiz. Nauk
178 301 (2008)]

\bibitem{} Bolotin Yu L, Erokhin D A, Lemets O A Phys. Usp. 55 876 (2012)
[Usp. Fiz. Nauk 182 941 (2012)]

\bibitem{} Massey R et al. Nature 445 286 (2007)

\bibitem{} Ryabov V A, Tsarev V A, Tskhovrebov A M Phys. Usp. 51 1091
(2008) [Usp. Fiz. Nauk 178 1129 (2008)]

\bibitem{} Jungman G, Kamionkowski M, Griest K Phys. Rep. 267 195 (1996)

\bibitem{} Troitsky S V Phys. Usp. 55 72 (2012) [Usp. Fiz. Nauk 182 77 (2012)]

\bibitem{} Dodelson S, Widrow L M Phys. Rev. Lett. 72 17 (1994); hep-ph/9303287

\bibitem{} Gorbunov D, Khmelnitsky A, Rubakov V JCAP (10) 041 (2008)

\bibitem{} Sikivie P, arXiv:0909.0949

\bibitem{} Berezinsky V S Phys. Lett. B 261 71 (1991)

\bibitem{} Bolz M, Brandenburg A, Buchmuller W Nucl. Phys. B 606 518
(2001); Nucl. Phys. B 790 336 (2008)

\bibitem{} Berezinsky V, Kachelrieu M, Vilenkin A Phys. Rev. Lett. 79 4302
(1997)

\bibitem{} Kuzmin V A, Rubakov V A Phys. Atom. Nucl. 61 1028 (1998) [Yad.
Fiz. 61 1122 (1998)]

\bibitem{} Berezinsky V, Kachelrieu M, Solberg M A Phys. Rev. D 78 123535
(2008)

\bibitem{} Dubrovich V K, Khlopov M Yu JETP Lett. 77 335 (2003) [Pis'ma
Zh. Eksp. Teor. Fiz. 77 403 (2003)]

\bibitem{} Ivanov P, Naselsky P, Novikov I Phys. Rev. D 50 7173 (1994)

\bibitem{} Lacki B C, Beacom J F Astrophys. J. Lett. 720 L67 (2010)

\bibitem{} Add G et al. (ATLAS Collab.) Phys. Lett. B 716 1 (2012)

\bibitem{} Chatrchyan S et al. (CMS Collab.) Phys. Lett. B 716 30 (2012)

\bibitem{} Rubakov V A Phys. Usp. 55 949 (2012) [Usp. Fiz. Nauk 182 1017
(2012)]

\bibitem{} Nath P, arXiv:1210.0520

\bibitem{} Belli P et al. Phys. Rev. D 84 055014 (2011)

\bibitem{} Aalseth C E et al. (and CoGeNT Collab.) Phys. Rev. Lett. 107
141301 (2011)

\bibitem{} Hinshaw G et al., arXiv:1212.5226

\bibitem{} Ade P A R et al. (Planck Collab.), arXiv:1303.5062

\bibitem{} Lacey C, Cole S Mon. Not. R. Astron. Soc. 262 627 (1993)

\bibitem{} Schmid C, Schwarz D J, Widerin P Phys. Rev. Lett. 78 791 (1997)

\bibitem{} Wasserman I, in Second Intern. A.D. Sakharov Conf. on Physics
Moscow, Russia 20-24 May 1996 (Eds I M Dremin, A M Semikhatov) (Singapore: World Scientific, 1997) p. 191; astro-ph/
9608012

\bibitem{} Schmid C, Schwarz D J, Widerin P Phys. Rev. D 59 043517 (1999)

\bibitem{} Schwarz D J, Hofmann S Nucl. Phys. B Proc. Suppl. 87 93 (2000)

\bibitem{} Berezinsky V, Dokuchaev V, Eroshenko Y Phys. Rev. D 68 103003
(2003)

\bibitem{} Zhao H et al., astro-ph/0502049

\bibitem{} Green A M, Hofmann S, Schwarz D J JCAP (08) 003 (2005)

\bibitem{} Moore B et al., astro-ph/0502213

\bibitem{} Berezinsky V, Dokuchaev V, Eroshenko Yu Phys. Rev. D 73 063504
(2006)

\bibitem{} Diemand J, Kuhlen M, Madau P Astrophys. J. 649 1 (2006)

\bibitem{} GreenAM, Goodwin S P Mon. Not. R. Astron. Soc. 375 1111 (2007)

\bibitem{} Bertschinger E Phys. Rev. D 74 063509 (2006)

\bibitem{} Angus G W, Zhao H S Mon. Not. R. Astron. Soc. 375 1146 (2007)

\bibitem{} Berezinsky V S, Dokuchaev V I, Eroshenko Yu N JCAP (07) 011
(2007)

\bibitem{} Giocoli C, Pieri L, Tormen G Mon. Not. R. Astron. Soc. 387 689
(2008)

\bibitem{} Kamionkowski M, Koushiappas S M, Kuhlen M Phys. Rev. D 81
043532 (2010)

\bibitem{} Anderhalden D, Diemand J JCAP (04) 009 (2013); arXiv:1302.0003

\bibitem{} Koushiappas S M New J. Phys. 11 105012 (2009)

\bibitem{} Bergstrom L et al. Phys. Rev. D 59 043506 (1999)

\bibitem{} Schwarz D J Ann. Physik 12 220 (2003)

\bibitem{} Gurevich A V, Zybin K P Sov. Phys. JETP 67 1 (1988) [Zh. Eksp.
Teor. Fiz. 94 3 (1988)]

\bibitem{} Gurevich A V, Zybin K P Sov. Phys. JETP 67 1957 (1988) [Zh.
Eksp. Teor. Fiz. 94 (4) 5 (1988)]

\bibitem{} Gurevich A V, Zybin K P Phys. Usp. 38 687 (1995) [Usp. Fiz. Nauk
165 723 (1995)]

\bibitem{} Profumo S, Sigurdson K, Kamionkowski M Phys. Rev. Lett. 97
031301 (2006)

\bibitem{} Bringmann T New J. Phys. 11 105027 (2009)

\bibitem{} Weinberg S Astrophys. J. 168 175 (1971)

\bibitem{} Hofmann S, Schwarz D J, Stocker H Phys. Rev. D 64 083507 (2001)

\bibitem{} Loeb A, Zaldarriaga M Phys. Rev. D 71 103520 (2005)

\bibitem{} Calcaneo-Roldan C, Moore B Phys. Rev. D 62 123005 (2000)

\bibitem{} Nieto D et al., arXiv:1110.4744

\bibitem{} Zechlin H-S et al., arXiv:1110.6868

\bibitem{} Pieri L, Branchini E, Hofmann S Phys. Rev. Lett. 95 211301 (2005)

\bibitem{} Oda T, Totani T, Nagashima M Astrophys. J. 633 L65 (2005)

\bibitem{} Kamionkowski M, Koushiappas S M Phys. Rev.D77 103509 (2008)

\bibitem{} Pieri L, Bertone G, Branchini E Mon. Not. R. Astron. Soc. 384 1627
(2008)

\bibitem{} Pinzke A, Pfrommer C, Bergstrom L Phys. Rev. Lett. 103 181302
(2009)

\bibitem{} Baxter E J et al. Phys. Rev. D 82 123511 (2010)

\bibitem{} Belotsky K M, Kirillov A A, Khlopov M Yu, arXiv:1212.6087

\bibitem{} Zhang D Mon. Not. R. Astron. Soc. 418 1850 (2011)

\bibitem{} Yang Y, Yang G, Zong H Phys. Rev. D 87 103525 (2013)

\bibitem{} Anderhalden D, Diemand J JCAP (04) 009 (2013)

\bibitem{} Bertschinger E Astrophys. J. Suppl. 58 (5) 39 (1985)

\bibitem{} Diemand J, Moore B, Stadel J Nature 433 389 (2005)

\bibitem{} Mikheeva E, Doroshkevich A, Lukash V Nuovo Cimento B 122 1393
(2007)

\bibitem{} Doroshkevich A G, Lukash V N, Mikheeva E V Phys. Usp. 55 3
(2012) [Usp. Fiz. Nauk 182 3 (2012)]

\bibitem{} Kaplinghat M Phys. Rev. D 72 063510 (2005)

\bibitem{} Strigari L E, Kaplinghat M, Bullock J S Phys. Rev. D 75 061303(R)
(2007)

\bibitem{} Ishiyama T, Makino J, Ebisuzaki T Astrophys. J. Lett. 723 L195
(2010)

\bibitem{} Berezinsky V, Dokuchaev V, Eroshenko Y Phys. Rev. D 77 083519
(2008)

\bibitem{} Kolb E W, Tkachev I I Phys. Rev. Lett. 71 3051 (1993)

\bibitem{} Kolb E W, Tkachev I I Phys. Rev. D 50 769 (1994)

\bibitem{} Khlopov M Yu , Sakharov A S, Sokoloff D D Nucl. Phys. B Proc.
Suppl. 72 105 (1999)

\bibitem{} Khlopov M J. Phys. Conf. Ser. 66 012032 (2007)

\bibitem{} Dokuchaev V I, Eroshenko Yu N JETP 94 1 (2002) [Zh. Eksp. Teor.
Fiz. 121 5 (2002)]

\bibitem{} Bardeen J M et al. Astrophys. J. 304 15 (1986)

\bibitem{} Green A M, Hofmann S, Schwarz D J Mon. Not. R. Astron. Soc. 353
L23 (2004)

\bibitem{} Press W H, Schechter P Astrophys. J. 187 425 (1974)

\bibitem{} Bond J R et al. Astrophys. J. 379 440 (1991)

\bibitem{} Bond J R, Myers S T Astrophys. J. Suppl. 103 1 (1996)

\bibitem{} Ade P A R et al. (Planck Collab.), arXiv:1303.5082

\bibitem{} Green A M, Liddle A R Phys. Rev. D 56 6166 (1997)

\bibitem{} Starobinskii A A JETP Lett. 55 489 (1992) [Pis'ma Zh. Eksp. Teor.
Fiz. 55 477 (1992)]

\bibitem{} Blais D et al. Phys. Rev. D 67 024024 (2003)

\bibitem{} Yokoyama J Astron. Astrophys. 318 673 (1997)

\bibitem{} GarcoAa-Bellido J, Linde A, Wands D Phys. Rev. D 54 6040 (1996)

\bibitem{} Cline J M, Crotty P, Lesgourgues J JCAP (09) 010 (2003)

\bibitem{} Chung D J H et al. Phys. Rev. D 62 043508 (2000)

\bibitem{} Gelmini G B, Gondolo P JCAP (10) 002 (2008)

\bibitem{} Scott P, Sivertsson S Phys. Rev. Lett. 103 211301 (2009)

\bibitem{} Kolb E W, Tkachev I I Astrophys. J. Lett. 460 L25 (1996)

\bibitem{} Gott J R (III) Astrophys. J. 201 296 (1975)

\bibitem{} Gunn J E Astrophys. J. 218 592 (1977)

\bibitem{} Silk J, Stebbins A Astrophys. J. 411 439 (1993)

\bibitem{} Zel'dovich Ya B, Novikov I D Sov. Astron. 10 602 (1967) [Astron.
Zh. 43 758 (1966)]

\bibitem{} Hawking S Mon. Not. R. Astron. Soc. 152 75 (1971)

\bibitem{} Carr B J Astrophys. J. 201 1 (1975)

\bibitem{} Nadezhin D K, Novikov I D, Polnarev A G Sov. Astron. J. 22 129
(1978) [Astron. Zh. 55 216 (1978)]

\bibitem{} Novikov I D et al. Astron. Astrophys. 80 104 (1979)

\bibitem{} Polnarev A G, Khlopov M Yu Sov. Phys. Usp. 28 213 (1985) [Usp.
Fiz. Nauk 145 369 (1985)]

\bibitem{} Bugaev E V, Konishchev K V Phys. Rev. D 65 123005 (2002)

\bibitem{} Choptuik M W Phys. Rev. Lett. 70 9 (1993)

\bibitem{} Niemeyer J C, Jedamzik K Phys. Rev. D 59 124013 (1999)

\bibitem{} Yokoyama J Phys. Rev. D 58 107502 (1998)

\bibitem{} Carr B J et al. Phys. Rev. D 81 104019 (2010)

\bibitem{} Sikivie P, Tkachev I I, Wang Y Phys. Rev. D 56 1863 (1997)

\bibitem{} Navarro J F, Frenk C S, White S D M Astrophys. J. 462 563 (1996)

\bibitem{} Fukushige T, Makino J Astrophys. J. Lett. 477 L9 (1997)

\bibitem{} Moore B et al. Astrophys. J. 524 L19 (1999)

\bibitem{} Jing Y P, Suto Y Astrophys. J. 529 L69 (2000)

\bibitem{} White S D M, astro-ph/9410043

\bibitem{} Knobel C, arXiv:1208.5931

\bibitem{} Tolman R C Phys. Rev. 35 875 (1930)

\bibitem{} McCreaW H Proc. R. Soc. London A 206 562 (1951)

\bibitem{} Peebles P J E The Large-Scale Structure of the Universe (Princeton,
N.J.: Princeton Univ. Press, 1980) [Translated into Russian:
Struktura Vselennoi v Bol'shikh Masshtabakh (Moscow: Mir, 1983)]

\bibitem{} Padmanabhan T, Subramanian K Astrophys. J. 417 3 (1993)

\bibitem{} Eisenstein D J, Loeb A Astrophys. J. 439 520 (1995)

\bibitem{} Berezinsky V S, Dokuchaev V I, Eroshenko Yu N JCAP (11) 059
(2013); arXiv:1308.6742

\bibitem{} Doroshkevich A G Astrophysics 6 320 (1970) [Astrofizika 6 581
(1970)]

\bibitem{} Sheth R K, Mo H J, Tormen G Mon. Not. R. Astron. Soc. 323 1
(2001)

\bibitem{} Shandarin S F, Doroshkevich A G, Zel'dovich Ya B Sov. Phys. Usp.
26 46 (1983) [Usp. Fiz. Nauk 139 83 (1983)]

\bibitem{} Gurbatov S N, Saichev A I, Shandarin S F Phys. Usp. 55 223 (2012)
[Usp. Fiz. Nauk 182 233 (2012)]

\bibitem{} Shlaer B, Vilenkin A, Loeb A, arXiv:1202.1346

\bibitem{} Vilenkin A, Shellard E P S Cosmic Strings and Other Topological
Defects (Cambridge: Cambridge Univ. Press, 1994)

\bibitem{} Vilenkin A, in Inflating Horizons of Particle Astrophysics and
Cosmology: Proc. of the Yamada Conf. LIX, June 20-24, 2005,
Tokyo, Japan (Frontiers Science Ser., No. 46, Eds H Suzuki et al.)
(Tokyo: Universal Acad. Press, 2006) p. 159; hep-th/0508135

\bibitem{} Vanchurin V, Olum K D, Vilenkin A Phys. Rev. D 74 063527 (2006)

\bibitem{} Blanco-Pillado J J, Olum K D, Shlaer B Phys. Rev. D 83 083514
(2011); arXiv:1101.5173

\bibitem{} Berezinsky V S, Dokuchaev V I, Eroshenko Yu N JCAP (12) 007
(2011)

\bibitem{} Olum K D, Vilenkin A Phys. Rev. D 74 063516 (2006)

\bibitem{} Lynden-Bell D Mon. Not. R. Astron. Soc. 136 101 (1967)

\bibitem{} White S D M, in Gravitational Dynamics: Proc. of the 36th
Herstmonceux Conf., in Honour of Professor D. Lynden-Bell's 60th
Birthday, Cambridge, UK, August 7-11, 1995 (Eds O Lahav,
E Terlevich, R J Terlevich) (Cambridge: Cambridge Univ. Press,
1996) p. 121; astro-ph/9602021

\bibitem{} Henriksen R N, Widrow L M Mon. Not. R. Astron. Soc. 302 321
(1999)

\bibitem{} Dokuchaev V I, Eroshenko Yu N Astron. Lett. 27 759 (2001) [Pis'ma
Astron. Zh. 27 883 (2001)]

\bibitem{} Dokuchaev V I, Eroshenko Yu N Astron. Astrophys. Trans. 22 727
(2003)

\bibitem{} Mack K J, Ostriker J P, Ricotti M Astrophys. J. 665 1277 (2007)

\bibitem{} Yang Y et al. Phys. Rev. D 84 043506 (2011); arXiv:1109.0156

\bibitem{} Saito R, Shirai S Phys. Lett. B 697 95 (2011)

\bibitem{} Yang Y et al. Eur. Phys. J. Plus 126 123 (2011)

\bibitem{} Yang Y et al. JCAP(12) 020 (2011)

\bibitem{} Carr B J, Rees M J Mon. Not. R. Astron. Soc. 206 801 (1984)

\bibitem{} Bertone G, Zentner A R, Silk J Phys. Rev. D 72 103517 (2005)

\bibitem{} Baushev A Inter. J. Mod. Phys. D 18 1195 (2009)

\bibitem{} Fillmore J A, Goldreich P Astrophys. J. 281 1 (1984)

\bibitem{} Ricotti M, Gould A Astrophys. J. 707 979 (2009)

\bibitem{} Josan A S, Green AMPhys. Rev. D 82 083527 (2010)

\bibitem{} Bringmann T, Scott P, Akrami Y Phys. Rev. D 85 125027 (2012)

\bibitem{} Li F, Erickcek A L, Law N M Phys. Rev. D 86 043519 (2012)

\bibitem{} Yang Y et al. Phys. Rev. D 87 083519 (2013); arXiv:1206.3750

\bibitem{} Yang Y-P, Yang G-L, Zong H-S, arXiv:1210.1409

\bibitem{} Yang Y-P, Yang G-L, Zong H-S Europhys. Lett. 101 69001 (2013)

\bibitem{} Berezinsky V et al. Phys. Rev. D 81 103529 (2010)

\bibitem{} Saslaw W C Gravitational Physics of Stellar and Galactic Systems
(Cambridge: Cambridge Univ. Press, 1985) [Translated into Rus-
sian: Gravitatsionnaya Fizika Zvezdnykh i Galakticheskikh Sistem
(Moscow: Mir, 1989)]

\bibitem{} Gurevich A V, Zybin K P, Sirota V A Phys. Usp. 40 869 (1997) [Usp.
Fiz. Nauk 167 913 (1997)]

\bibitem{} Iocco F et al. JCAP (11) 029 (2011)

\bibitem{} Tremaine S, Gunn J E Phys. Rev. Lett. 42 407 (1979)

\bibitem{} Taylor J E, Navarro J F Astrophys. J. 563 483 (2001)

\bibitem{} Tremaine S, Henon M, Lynden-Bell D Mon. Not. R. Astron. Soc.
219 285 (1986)

\bibitem{} Ryden B S Astrophys. J. 329 589 (1988)

\bibitem{} Ryden B S, Gunn J E Astrophys. J. 318 15 (1987)

\bibitem{} Hiotelis N Astron. Astrophys. 382 84 (2002)

\bibitem{} Ascasibar Y et al. Mon. Not. R. Astron. Soc. 352 1109 (2004)

\bibitem{} Ullio P et al. Phys. Rev. D 66 123502 (2002)

\bibitem{} Berezinsky V S, Gurevich A V, Zybin K P Phys. Lett. B 294 221
(1992)

\bibitem{} Berezinsky V, Bottino A, Mignola G Phys. Lett. B 391 355 (1997)

\bibitem{} Chung D J H, Kolb E W, Riotto A Phys. Rev. D 59 023501 (1999)

\bibitem{} Kuzmin V A, Tkachev I I JETP Lett. 68 271 (1998) [Pis'ma Zh.
Eksp. Teor. Fiz. 68 255 (1998)]

\bibitem{} Lyth D H, Roberts D, Smith M Phys. Rev. D 57 7120 (1998)

\bibitem{} Berezinsky V et al. Phys. Rev. D 81 103530 (2010)

\bibitem{} Bilic N, Munyaneza F, Viollier R D Phys. Rev. D 59 024003 (1998)

\bibitem{} Moore B et al. Astrophys. J. 499 L5 (1998)

\bibitem{} Moore B et al. Mon. Not. R. Astron. Soc. 310 1147 (1999)

\bibitem{} Gao L et al. Mon. Not. R. Astron. Soc. 387 536 (2008)

\bibitem{} Burkert A Astrophys. J. 447 L25 (1995)

\bibitem{} Syer D, White S D M Mon. Not. R. Astron. Soc. 293 337 (1998)

\bibitem{} Springel V et al. Mon. Not. R. Astron. Soc. 391 1685 (2008)

\bibitem{} Bullock J S et al. Mon. Not. R. Astron. Soc. 321 559 (2001)

\bibitem{} Lifshitz E M, Pitaevskii L P Physical Kinetics (Oxford: Pergamon
Press, 1981) [Translated from Russian: Fizicheskaya Kinetika
(Moscow: Nauka, 1979)]

\bibitem{} ZybinKP, Vysotsky M I, Gurevich A V Phys. Lett. A 260 262 (1999)

\bibitem{} Widrow LMet al. Mon. Not. R. Astron. Soc. 397 1275 (2009)

\bibitem{} Arhipova N A et al. Grav. Cosmol. Suppl. 8 (Suppl. I) 66 (2002)

\bibitem{} Giocoli C et al. Mon. Not. R. Astron. Soc. 395 1620 (2009)

\bibitem{} Angulo R E, White S D M Mon. Not. R. Astron. Soc. 401 1796
(2010)

\bibitem{} Oguri M, Lee J Mon. Not. R. Astron. Soc. 355 120 (2004)

\bibitem{} Gnedin O Y, Hernquist L, Ostriker J P Astrophys. J. 514 109 (1999)

\bibitem{} Fall S M, Rees M J Mon. Not. R. Astron. Soc. 181 37P (1977)

\bibitem{} Gnedin O Y, Ostriker J P Astrophys. J. 513 626 (1999)

\bibitem{} Taylor J E, Babul A Astrophys. J. 559 716 (2001)

\bibitem{} Diemand J, Kuhlen M, Madau P Astrophys. J. 667 859 (2007)

\bibitem{} Goerdt T et al. Mon. Not. R. Astron. Soc. 375 191 (2007)

\bibitem{} Zhao H et al. Astrophys. J. 654 697 (2007)

\bibitem{} Ostriker J P, Spitzer L (Jr.), Chevalier R A Astrophys. J. 176 L51
(1972)

\bibitem{} Weinberg M D Astron. J. 108 1403 (1994)

\bibitem{} Eddington A S Mon. Not. R. Astron. Soc. 76 572 (1916)

\bibitem{} Widrow L M Astrophys. J. Suppl. 131 39 (2000)

\bibitem{} D'Onghia E et al. Astrophys. J. 709 1138 (2010)

\bibitem{} Launhardt R, Zylka R, Mezger P G Astron. Astrophys. 384 112
(2002)

\bibitem{} Bell E F et al. Astrophys. J. 680 295 (2008)

\bibitem{} Schneider A, Krauss L, Moore B Phys. Rev. D 82 063525 (2010)

\bibitem{} Silk J, Bloemen H Astrophys. J. 313 L47 (1987)

\bibitem{} Berezinsky V, Kachelrieu M Phys. Rev. D 63 034007 (2001)

\bibitem{} Aloisio R, Berezinsky V, Kachelrieu M Phys. Rev. D 69 094023
(2004)

\bibitem{} Gondolo P et al. JCAP (07) 008 (2004)

\bibitem{} Bringmann T et al. JCAP (07) 054 (2012)

\bibitem{} Weniger C JCAP (08) 007 (2012)

\bibitem{} Hektor A, Raidal M, Tempel E, arXiv:1207.4466

\bibitem{} Bringmann T, Weniger C Phys. Dark Universe 1 194 (2012);
arXiv:1208.5481

\bibitem{} Gorbunov D, Tinyakov P Phys. Rev. D 87 081302(R) (2013);
arXiv:1212.0488

\bibitem{} Hisano J, Matsumoto S, Nojiri M M Phys. Rev. Lett. 92 031303
(2004)

\bibitem{} Profumo S Phys. Rev. D 72 103521 (2005)

\bibitem{} Lattanzi M, Silk J Phys. Rev. D 79 083523 (2009)

\bibitem{} Hooper D et al. Phys. Rev. D 86 103003 (2012); arXiv:1203.3547

\bibitem{} Ginzburg V L, Syrovatskii S I The Origin of Cosmic Rays ((Oxford:
Pergamon Press, 1964)) [Translated from Russian: Proiskhozhdenie
Kosmicheskikh Luchei (Moscow: Izd. AN SSSR, 1963)]

\bibitem{} Berezinskii V S et al. Astrophysics of Cosmic Rays (Ed. V L Ginzburg) (Amsterdam: North-Holland, 1990) [Translated from Russian: Astrofizika Kosmicheskikh Luchei (Ed. V L Ginzburg) 2nd ed.
(Moscow: Nauka, 1990)]

\bibitem{} The GALPROP code for cosmic-ray transport and diffuse emission
production, http://galprop.stanford.edu

\bibitem{} Atwood W B et al. Astrophys. J. 697 1071 (2009)

\bibitem{} de Boer W et al. Astron. Astrophys. 444 51 (2005)

\bibitem{} Ackermann M et al. (The Fermi-LAT Collab.) JCAP (05) 025
(2010)

\bibitem{} Yuan Q et al. Phys. Rev. D 82 023506 (2010)

\bibitem{} Pinzke A, Pfrommer C, Bergstrom L Phys. Rev. D 84 123509 (2011)

\bibitem{} Huang X, Vertongen G, Weniger C JCAP (01) 042 (2012)

\bibitem{} Ando S, Nagai D JCAP (07) 017 (2012)

\bibitem{} Han J et al., arXiv:1201.1003

\bibitem{} Vitale V, Morselli A (for the Fermi/LAT Collab.), arXiv:0912.3828

\bibitem{} Hooper D, Goodenough L Phys. Lett. B 697 412 (2011)

\bibitem{} Ellis J, Olive K A, Spanos V C JCAP (10) 024 (2011)

\bibitem{} Sanchez-Conde M A et al. JCAP (12) 011 (2011)

\bibitem{} Gao L et al. Mon. Not. R. Astron. Soc. 419 1721 (2012)

\bibitem{} Ackermann M et al. (The Fermi LAT Collab.) Astrophys. J. 747 121
(2012)

\bibitem{} Ackermann M et al. (The Fermi-LAT Collab.) Phys. Rev. Lett. 107
241302 (2011)

\bibitem{} Geringer-Sameth A, Koushiappas SM Phys. Rev. Lett. 107 241303
(2011)

\bibitem{} Feng L et al. JCAP (04) 030 (2012)

\bibitem{} Mei S et al. Astrophys. J. 655 144 (2007)

\bibitem{} OllingRP, Merrifield M R Mon. Not. R. Astron. Soc. 311 361 (2000)

\bibitem{} OllingRP, Merrifield M R Mon. Not. R. Astron. Soc. 326 164 (2001)

\bibitem{} Mashchenko S, Couchman H M P, Wadsley J Nature 442 539 (2006)

\bibitem{} Ando S, Komatsu E Phys. Rev. D 73 023521 (2006)

\bibitem{} Ando S Phys. Rev. D 80 023520 (2009)

\bibitem{} FornasaM et al. Phys. Rev. D 80 023518 (2009)

\bibitem{} Fornasa M et al. Mon. Not. R. Astron. Soc. 429 1529 (2012);
arXiv:1207.0502

\bibitem{} Gomez-Vargas G (on behalf of the Fermi-LAT Collab.), Komatsu E
Nuovo Cimento C 034 327 (2011)

\bibitem{} Hooper D, Serpico P D JCAP (06) 013 (2007)

\bibitem{} Blasi P, Dick R, Kolb EWAstropart. Phys. 18 57 (2002)

\bibitem{} Troitsky S V Phys. Usp. 56 304 (2013) [Usp. Fiz. Nauk 183 323
(2013)]

\bibitem{} de Boer W, Zhukov V, arXiv:0709.4576

\bibitem{} Myers Z, Nusser A Mon. Not. R. Astron. Soc. 384 727 (2008)

\bibitem{} Natarajan A, Schwarz D J Phys. Rev. D 78 103524 (2008)

\bibitem{} Adriani O et al. Phys. Rev. Lett. 105 121101 (2010)

\bibitem{} Adriani O et al. (PAMELA Collab.) Phys. Rev. Lett. 106 201101
(2011)

\bibitem{} Aguilar M et al. (AMS Collab.) Phys. Rev. Lett. 110 141102 (2013)

\bibitem{} Diemand J et al. Nature 454 735 (2008)

\bibitem{} Bergstrom L, Bringmann T, Edsjo J Phys. Rev. D 78 103520 (2008)

\bibitem{} Cirelli M et al. Nucl. Phys. B 813 1 (2009)

\bibitem{} Cholis I et al. Phys. Rev. D 80 123518 (2009)

\bibitem{} Chang J et al. Nature 456 362 (2008)

\bibitem{} Cumberbatch D, Silk J Mon. Not. R. Astron. Soc. 374 455 (2007)

\bibitem{} Kane G, Lu R, Watson S Phys. Lett. B 681 151 (2009)

\bibitem{} Adriani O et al. Nature 458 607 (2009)

\bibitem{} Stozhkov Yu I, Galper A M ``Iternational PAMELA experiment'',
Report on G T Zatsepin Seminar ``Neutrino and nuclear astrophysics'' February 18, 2011

\bibitem{} Stozhkov Yu I Bull. Russ. Acad. Sci. Phys. 75 323 (2011) [Izv. Ross.
Akad. Nauk Fiz. 75 (3) 352 (2011)]

\bibitem{} Blasi P Phys. Rev. Lett. 103 051104 (2009)

\bibitem{} Blasi P, Serpico P D Phys. Rev. Lett. 103 081103 (2009)

\bibitem{} Seto N, Cooray A Phys. Rev. D 70 063512 (2004)

\bibitem{} Tricarico P Class. Quantum Grav. 26 085003 (2009)

\bibitem{} Adams A W, Bloom J S, astro-ph/0405266

\bibitem{} Lewis G F, Gil-Merino R Astrophys. J. 645 835 (2006)

\bibitem{} Chen J, Koushiappas SMAstrophys. J. 724 400 (2010)

\bibitem{} Zackrisson E et al. Mon. Not. R. Astron. Soc. 431 2172 (2013)

\bibitem{} Metcalf R B, Silk J Phys. Rev. Lett. 98 071302 (2007); Phys. Rev.
Lett. 98 099903(E) (2007)

\bibitem{} Siegel E R, Hertzberg M P, Fry J N Mon. Not. R. Astron. Soc. 382
879 (2007)

\bibitem{} Ricotti M, Ostriker J P, Mack K J Astrophys. J. 680 829 (2008)

\bibitem{} Freese K et al. Astrophys. J. 693 1563 (2009)

\bibitem{} Casanellas J, Lopes I, arXiv:1002.2326

\bibitem{} Smith R J et al. Astrophys. J. 761 154 (2012); arXiv:1210.1582

\bibitem{} Zurek K M, Hogan C J Phys. Rev. D 76 063002 (2007)

\bibitem{} Koushiappas S M AIP Conf. Proc. 921 142 (2007); astro-ph/0703778

\bibitem{} Ando S et al. Phys. Rev. D 78 101301(R) (2008)

\end{thebibliography}
\end{document}